\documentclass[11pt]{article}
\usepackage[margin=1in]{geometry}
\usepackage[utf8]{inputenc}

\usepackage{amsmath,amssymb,amsfonts,amsthm}

\usepackage[font=small,labelfont=bf]{caption}

\usepackage{breqn}

\usepackage{bold-extra}

\usepackage{bm}

\usepackage{bbm}

\usepackage{graphicx}
\usepackage[dvipsnames,svgnames]{xcolor}

\usepackage{url}
\usepackage{hyperref}
\hypersetup{
  colorlinks   = true, 
  urlcolor     = blue, 
  linkcolor    = blue, 
  citecolor   = blue 
}
\usepackage[numbers,sort&compress]{natbib}
\let\OLDthebibliography\thebibliography
\renewcommand\thebibliography[1]{
  \OLDthebibliography{#1}
  \setlength{\parskip}{4pt}
  \setlength{\itemsep}{0pt plus 0.3ex}
}

\allowdisplaybreaks
\usepackage{titlesec}
\titleformat*{\section}{\bfseries\boldmath}
\titleformat*{\subsection}{\bfseries\boldmath}
\titleformat*{\subsubsection}{\bfseries\boldmath}

\usepackage{multicol}
\usepackage{enumitem}

\usepackage{soul}

\newcommand\citere[1]{Ref.~\cite{#1}}
\newcommand\citeres[1]{Refs.~\cite{#1}}

\newcommand{\TB}[1]{{\color{black} #1}}

\newcommand{\gev}{\ \mathrm{GeV}}

\newcommand{\ii}{\textrm{i}}
\newcommand{\ee}{\textrm{e}}

\newcommand{\ca}{\cos\alpha}
\newcommand{\sa}{\sin\alpha}

\newcommand{\la}{\lambda}

\usepackage[capitalise]{cleveref}

\begin{document}

\thispagestyle{empty}

\def\thefootnote{\fnsymbol{footnote}}

\begin{flushright}
  \footnotesize
  KA-TP-25-2025 ~~~
  IFT-UAM/CSIC-25-90
\end{flushright}

\vspace*{0.4cm}

\begin{center}

{\Large
  \textbf{Towards a Unified Framework for Pseudo-Nambu-Goldstone\\[0.4em]
  Dark Matter and Electroweak Baryogenesis}
}

\vspace{1.8em}

Thomas Biek\"otter$^1$\footnote{thomas.biekoetter@desy.de},
Pedro Gabriel$^{2,3}$\footnote{ptgabriel@fc.ul.pt},
Milada Margarete M\"uhlleitner$^2$\footnote{milada.muehlleitner@kit.edu},
and
Rui Santos$^{3,4}$\footnote{rasantos@fc.ul.pt}

\vspace*{0.8em}

\textit{
$^1$Instituto de F\'isica Te\'orica UAM/CSIC,\\
Calle Nicolás Cabrera 13-15,
Cantoblanco, 28049, Madrid, Spain\\[0.4em]
$^2$Institute for Theoretical Physics,
Karlsruhe Institute of Technology,\\
Wolfgang-Gaede-Str.~1, 76131 Karlsruhe, Germany\\[0.4em]
$^3$Centro de F\'isica T\'eorica e Computacional,
Faculdade de C\^iencias,
Universidade de Lisboa, Campo Grande, Edif\'icio C8 1749-016 Lisboa, Portugal\\[0.4em]
$^4$ISEL – Instituto Superior de Engenharia de Lisboa,\\
Instituto Polit\'ecnico de Lisboa, 1959-007 Lisboa, Portugal
}

\vspace*{0.2cm}

\begin{abstract}
We propose the complex singlet-extended 2-Higgs-Doublet
Model (cS2HDM), a spin-0 
Dark Matter (DM) model with a Higgs sector consisting
of two Higgs doublets and a complex singlet, as a benchmark 
for LHC DM searches. The model predicts a
pseudo-Nambu-Goldstone DM candidate whose 
interactions with nuclei are naturally suppressed,
while allowing for all sources of CP-violation 
under the assumption of flavour alignment in
the Yukawa sector,
which enables CP-violating interactions of the Higgs bosons even
in the alignment limit. 
This feature makes the model attractive for studies of
electroweak baryogenesis while accommodating 
a Higgs-portal DM candidate with standard thermal freeze-out.
We confront the model with a comprehensive set of
theoretical and experimental constraints,
including Higgs-boson signal strength measurements, searches for 
additional Higgs bosons, DM relic abundance and
direct detection, as well as electroweak precision 
observables and the electron EDM, with emphasis on the
impact of the new CP-violating sources. 
For DM direct detection, we perform a one-loop computation
of DM-nucleon scattering including CP-violating effects. 
We provide a public software package to facilitate
future phenomenological studies of the cS2HDM.
\end{abstract}

\end{center}

\renewcommand{\thefootnote}{\arabic{footnote}}
\setcounter{footnote}{0} 

\newpage

{
  \hypersetup{linkcolor=black}
  \tableofcontents
}

\section{Introduction}

The Standard Model~(SM) contains a minimal
prescription of electroweak~(EW) symmetry breaking
postulating a single fundamental scalar particle.
In 2012 the ATLAS and the CMS collaborations
reported the discovery of a new particle at the
Large Hadron Collider~(LHC) in agreement
with the predictions of
the SM~\cite{ATLAS:2012yve,CMS:2012qbp}.
After the Run~2 of the LHC operating at 13~TeV,
the coupling measurements of the detected
Higgs boson have reached a precision of
better than 
10\% in the experimentally most sensitive
channels~\cite{ATLAS:2022vkf,CMS:2022dwd}.
So far, no significant discrepancies from the
SM predictions have been observed, such that
any realistic model has to contain a spin-0
particle that behaves -- within the current
experimental uncertainties -- according to
the predictions of the~SM.

With the Higgs boson the last missing fundamental
particle contained in the SM has been observed.
However, cosmological and astronomical observations
are in disagreement with the~SM, calling for
extensions of
the SM particle content.
For instance, the SM does not contain a
particle that can contribute to the observed
relic density of cold Dark Matter~(DM),
and it does not explain the origin of the
observed asymmetry between matter and antimatter
in the Universe.
These shortcomings can be addressed in theories
beyond the Standard Model~(BSM) that feature
extended scalar sectors. These models generically
predict additional spin-0 particles.
The search for additional Higgs bosons is
one of the prime tasks of the current and
future LHC programme~\cite{ATLAS:2024itc}.

An important class of models with extended
Higgs sectors consists of the so-called
\textit{Higgs-portal}
DM models~\cite{Silveira:1985rk,Patt:2006fw},
where the particles contained in the
dark sector interact with the visible/SM sector
exclusively via the exchange of Higgs bosons.
In fact, in the SM the dimension-two
operator $|H|^2$, where $H$ is the Higgs doublet
field, is the only 
gauge- and Lorentz-invariant operator 
to which a dark sector with fields
that are not charged under the
SM gauge interactions can be coupled
to in a renormalizable way~\cite{Falkowski:2015iwa}.
Given the fact that this operator is
super-renormalizable,
from an effective-field-theory perspective a Higgs-portal
coupling between the dark sector and the
visible sector is in general
a renormalizable operator that can
be relevant even if
it is generated at energy scales that
are much larger than the electroweak scale.

In the context of models with extended scalar sectors,
Higgs-portal DM can be realized most economically
by extending the SM with a real gauge singlet
scalar field acting as the DM~\cite{Barger:2007im}.
However, this minimal possibility
is excluded by DM direct detection experiments
apart from a small mass interval where the
DM mass is approximately equal to half
the mass of the 125~GeV Higgs
boson or for 
rather heavy DM masses of more than about 2~TeV,
leaving only little discovery potential for the
LHC~\cite{Biekotter:2022ckj,Arcadi:2024wwg}.
Phenomenologically more interesting models
arise under the presence of a complex
gauge singlet field, where (depending on
the applied symmetry on the singlet 
field)
one component of the singlet field can
act as the DM particle and the other
component can mix with the Higgs field
contained in the SM, giving rise to
the presence of a second
Higgs boson~\cite{Barger:2008jx}.
Notably, this construction allows for the
presence of so-called pseudo-Nambu-Goldstone~(pNG)
DM~\cite{Gross:2017dan} which is
naturally faint to direct-detection
experiments as a consequence of a
softly broken global $U(1)$ symmetry (see discussion below).
This feature makes the model
especially appealing for DM searches
at particle colliders~\cite{Huitu:2018gbc},
whereas in other
models a possible detection at colliders
is largely ruled out by DM direct-detection
experiments.
However, in the minimal pNG DM model
it is not possible to accommodate
additional sources of CP-violation, and
one cannot account
for a strong first-order electroweak (EW) phase transition~\cite{Kannike:2019wsn},
both required to realize EW
baryogenesis~\cite{Sakharov:1967dj,Kuzmin:1985mm}
(see Ref.~\cite{vandeVis:2025efm} for a recent review).
As such, the complex-singlet extended SM is
not able to explain the observed asymmetry between
matter and antimatter in the Universe, while at
the same time containing a pNG DM state.

In light of this, pNG DM models have been
studied that -- in addition to a complex
gauge singlet scalar field -- augment the
Higgs sector of the SM by a second
Higgs doublet~\cite{Jiang:2019soj,
Zhang:2021alu,
Biekotter:2021ovi,
Muhlleitner:2021cci,
Biekotter:2022bxp,Darvishi:2022wnd,Biekotter:2023jld}.
In a 2-Higgs-Doublet Model~(2HDM) the additional
quartic couplings in the potential allow for
a strong first-order EW phase transition
by means of sizable radiative and thermal
corrections to the effective
potential~\cite{Cline:1996mga}.
Additionally,
the 2HDM in general also contains additional
sources of CP-violation in the Higgs
sector~\cite{Weinberg:1990me}.\footnote{An exception is
the so-called inert
2HDM in which the two doublets have opposite
parity under an exact $\mathbb{Z}_2$ symmetry.}
Consequently, 2HDM + complex singlet
models are the minimal framework for pNG
DM models that
in addition have the potential to dynamically
generate the
matter-antimatter asymmetry of the Universe
via EW baryogenesis.\footnote{The DM and collider
phenomenology of more general 2HDM models
extended with a complex scalar singlet,
where no natural suppression mechanism of
direct-detection scattering rates is present
and without considering CP-violation in the scalar sector,
has been studied, for example, in Refs.~\cite{Baum:2018zhf,
Heinemeyer:2021msz,Dutta:2022qeq,Dutta:2025nmy}.}

The fact that EW baryogensis may 
be realized without spoiling the cancellation
mechanism that suppresses the scattering rates
at DM direct-detection experiments
distinguishes the model from models
in which the stringent constraints
from DM direct-detection experiments are
evaded by assuming that the Higgs portal
is confined to the CP-odd sector.
The minimal UV-complete realization of
such a pseudoscalar Higgs-portal is the
2HDM+a~\cite{Ipek:2014gua,No:2015xqa}.
The 2HDM+a has been used most commonly as
one of the benchmark models for LHC DM
searches during 
Run~2~\cite{LHCDarkMatterWorkingGroup:2018ufk}.
There, due to the pseudoscalar nature of
the portal couplings, the
spin-independent interactions between the
DM and nucleons are
momentum-suppressed~\cite{Boehm:2014hva,
Alves:2014yha}.
However, in general this attractive feature
of the 2HDM+a is lost if there
are explicit CP-violating phases in the
scalar potential, in contrast to the
cancellation mechanism that protects
pNG DM which is related to the global
$U(1)$ symmetry and not the
CP-symmetry.\footnote{In the 2HDM+a,
the cancellation can be maintained if the
CP-violation necessary for EW baryogenesis is
generated spontaneously at finite temperature
and vanishes at zero temperature~\cite{Huber:2022ndk,
Liu:2023sey}.
In our paper, we restrict the discussion to
explicit CP violation in the scalar sector.}
Another important difference between the model
put forward here and the 2HDM+a
is that in our model
the DM particle and its portal couplings
are contained in the scalar
sector, whereas in the 2HDM+a  the
DM particle and its interactions with the
visible sector
are in principle yet undefined
(typically the DM particle in the 2HDM+a
is chosen to be a Dirac fermion with
Yukawa-like portal couplings).

According to the discussion above, we propose
in this paper the CP-violating singlet-extended
2HDM~(cS2HDM) as a benchmark model for future
LHC DM searches and LHC searches for additional
Higgs bosons motivated by the simultaneous description
of both DM and EW baryogenesis.\footnote{The
same model has recently been discussed
in \citere{Darvishi:2022wnd}, focusing on the
predictions for the eEDM and the indirect detection
of DM in the numerical analysis.}
To this end, we provide the necessary computational
tools in order to study the phenomenology of
the cS2HDM via the software package
\texttt{cs2hdmtools} available at:
\begin{center}
  \href{https://gitlab.com/thomas.biekoetter/cs2hdmtools}
    {https://gitlab.com/thomas.biekoetter/cs2hdmtools}
\end{center}
This tool includes predictions for the
branching ratios of the Higgs bosons via
a modified version of the public code
\texttt{HDECAY}~\cite{Djouadi:1997yw,Djouadi:2018xqq},
cross section predictions
and methods to confront the model with
measurements from the LHC via an interface
to the public code \texttt{HiggsTools}~\cite{Bahl:2022igd},
and predictions for the EW $\rho$-parameter
and the electron electric dipole moment~(eEDM)
that can be used to indirectly constrain the
parameter space of the model with
precision measurements.
Furthermore, we provide functions to apply
theoretical constraints from vacuum stability
and from perturbative unitarity that give rise
to important bounds on the parameters
in the scalar potential.
Finally, we provide an interface to the
public code \texttt{micrOMEGAs}~\cite{Belanger:2006is,Belanger:2018ccd}
to compute the DM relic abundance assuming
thermal freeze-out in the early universe,
and we provide routines to compute the
one-loop induced DM-nucleon scattering interactions
that can be used to confront the cS2HDM
with current and future constraints
from DM direct detection experiments.

The outline of the paper is as follows.
In \cref{sec:model} we introduce the model,
with a special focus on the various new sources
of explicit CP-violation and the Higgs-portal
interactions of the DM scalar.
In \cref{sec:constraints} we discuss the
theoretical and experimental constraints that are
relevant in the cS2HDM, and we discuss how we
confront the model with these constraints in
our numerical analysis.
In \cref{sec:numdis} we present a selection of
representative benchmark scenarios that demonstrate
how the model can manifest itself at the LHC
with novel signatures and CP-violating couplings
of the 125~GeV Higgs boson, while satisfying all
theoretical and experimental constraints and,
in addition, predicting a large fraction or all
of the observed DM relic abundance.
We summarize and conclude our findings in
\cref{sec:conclu}.

\section{The Model cS2HDM}
\label{sec:model}

The cS2HDM 
extends the particle content of the
SM by a second Higgs doublet field and a complex
gauge singlet scalar.
In this section we introduce all elements of the
model that are relevant for our numerical analysis,
and we fix the notation used throughout the paper.

\subsection{Higgs Sector}
\label{sec:higgssector}

The scalar potential
is defined such that the singlet field $\Phi_S$
respects a global $U(1)$ symmetry which is only
softly broken via a dimension-two term.
Under this assumption, the most general scalar
potential can be written as
\begin{align}
V &=
m_{11}^2 |\Phi_1|^2 +
m_{22}^2 |\Phi_2|^2 -
\left(
  m_{12}^2 \Phi_1^\dagger \Phi_2 + \mathrm{h.c.}
\right) +
\dfrac{1}{2} m_S^2 |\Phi_S|^2 -
\dfrac{1}{4} \left(
  m_\chi^2 \Phi_S^2 + \mathrm{h.c.}
\right) \notag \\
&+ \dfrac{1}{2} \lambda_1 |\Phi_1|^4 +
\dfrac{1}{2} \lambda_2 |\Phi_2|^4 +
\lambda_3 |\Phi_1|^2 |\Phi_2|^2 +
\lambda_4 (\Phi_1^\dagger \Phi_2) (\Phi_2^\dagger \Phi_1)
\notag \\
& +\left(
  \dfrac{1}{2} \lambda_5 (\Phi_1^\dagger \Phi_2)^2 +
  \lambda_6 (\Phi_1^\dagger \Phi_2) |\Phi_1|^2 +
  \lambda_7 (\Phi_1^\dagger \Phi_2) |\Phi_2|^2 +
  \, \mathrm{h.c.} \right) \notag \\
&+ \dfrac{1}{2} \lambda_8 |\Phi_S|^4 +
\lambda_9 |\Phi_S|^2 |\Phi_1|^2 +
\lambda_{10} |\Phi_S|^2 |\Phi_2|^2 +
\left(
  \lambda_{11} (\Phi_1^\dagger \Phi_2) |\Phi_S|^2 +\, \mathrm{h.c.}
\right) \ ,
\label{eq:scapot}
\end{align}
where $\Phi_1$ and $\Phi_2$ are $SU(2)$ doublet fields
with hypercharge~1/2, and $\Phi_S$ is the gauge singlet
complex scalar field.
As pointed out in \citere{Darvishi:2022wnd},
the term proportional to $\lambda_{11}$ is redundant and
could be removed via a field redefinition of $\Phi_1$
and $\Phi_2$.
However, we write the potential including this terms since
it allows us to use a physically more convenient basis for
the input parameters using 6 independent angles that
parametrize the mixing of the neutral scalars. The parameters $m_{12}^2,m_\chi^2,\lambda_5,\lambda_6,
\lambda_7,\lambda_{11}$ can be complex, whereas the remaining parameters are real.

The electroweak vacuum is defined in such a way that
the neutral components of the Higgs doublets $\Phi_1$
and $\Phi_2$ and the singlet field have non-zero
real vacuum expectation values (vevs) $v_1$, $v_2$ and $v_S$, respectively.
In this case the scalar fields can be expanded around
their vevs and written as
\begin{equation}
\Phi_1 =
\left(
\begin{array}{c}
 \varphi_1^+ \\
 (v_1 + \varphi_1 + \ii \sigma_1) / \sqrt{2} \\
\end{array}
\right) \ , \quad
\Phi_2 =
\left(
\begin{array}{c}
 \varphi_2^+ \\
 (v_2 + \varphi_2 + \ii \sigma_2) / \sqrt{2} \\
\end{array}
\right) \ , \quad
\Phi_S = (v_S + \varphi_S + \ii \chi) / \sqrt{2} \ .
\label{eq:vevsew}
\end{equation}
We assume $v_1$ and $v_2$ to be real since we do not
consider the possibility of spontaneous
CP-violation.
We leave the exploration of spontaneous CP-violation
in the model for future work and focus here on
explicit CP-violation.
A complex vev $v_S$ would spoil the stability
of the DM candidate $\chi$, see discussion below.
The stationary conditions with respect to the three
fields that obtain a vev are then given by
\begin{align}
\frac{\partial V}{\partial \varphi_1} = 0 &\quad \Rightarrow \notag \\
m_{11}^2 =& \frac{v_2}{v_1} (m_{12}^2)^{\rm Re} -
  \frac{v_1^2}{2} \la_1 -
  \frac{v_2^2}{2} \left( \la_3 + \la_4 + \la_5^{\rm Re} \right) -
  \frac{3 v_1 v_2}{2} \la_6^{\rm Re} -
  \frac{v_2^3}{2 v_1} \la_7^{\rm Re} -
  \frac{v_S^2}{2} \la_9 -
  \frac{v_2 v_S^2}{2 v_1} \la_{11}^{\rm Re} \ , \\
\frac{\partial V}{\partial \varphi_2} = 0 &\quad \Rightarrow \notag \\
m_{22}^2 =& \frac{v_1}{v_2} (m_{12}^2)^{\rm Re} -
  \frac{v_2^2}{2} \la_2 -
  \frac{v_1^2}{2} \left( \la_3 + \la_4 + \la_5^{\rm Re} \right) -
  \frac{v_1^3}{2 v_2} \la_6^{\rm Re} -
  \frac{3 v_1 v_2}{2} \la_7^{\rm Re} -
  \frac{v_S^2}{2} \la_{10} -
  \frac{v_1 v_S^2}{2 v_2} \la_{11}^{\rm Re} \ , \\
\frac{\partial V}{\partial \varphi_S} = 0 &\quad \Rightarrow \notag \\
m_S^2 =& (m_\chi^2)^{\rm Re} -
  v_S^2 \la_8 -
  v_1^2 \la_9 -
  v_2^2 \la_{10} -
  2 v_1 v_2 \la_{11}^{\rm Re} \ ,
\end{align}
where the superscript 'Re' refers to the real component of the respective parameter.
In order to make connection to the 2HDM with
natural flavour conservation, we define the
parameter $\tan\beta = v_2 / v_1$. However, it
should be kept in mind that without a discrete
symmetry applied on the Higgs doublets that acts
differently on $\Phi_1$ and $\Phi_2$, the parameter
$\tan\beta$ is not basis invariant and has no direct
physical meaning in the general case~\cite{Haber:2006ue}.

All terms in \cref{eq:scapot} that contain the singlet field $\Phi_S$ respect
a global $U(1)$ symmetry, with the exception of the dimension-two term
proportional to  $m_\chi^2$ in the first line. If $v_S >0$,
the imaginary component of $\Phi_S$ acts as a pseudo-Nambu-Goldstone
boson under this symmetry. The mass of $\chi$ is then
given by the $U(1)$-breaking parameter $m_\chi$.
A discrete symmetry $\chi \to -\chi$
remains as a remnant of the $U(1)$ symmetry, such that $\chi$ is stable
and can play the role of the DM candidate of the model.

As stated above, the scalar potential contains six potentially
complex parameters: $m_{12}^2,m_\chi^2,\lambda_5,\lambda_6,
\lambda_7,\lambda_{11}$.
In the following, we will denote with $a^{\rm Re}$
and $a^{\rm Im}$ the real and imaginary parts of
a complex parameter $a$, respectively.
The imaginary components of the complex parameters
in the scalar potential
can in general be the source of explicit CP-violation.
However, assuming an electroweak vacuum as given
in \cref{eq:vevsew}, the imaginary part of $m_\chi^2$ has to vanish
according to the stationary condition,
\begin{equation}
\frac{\partial V}{\partial \chi}
= (m_\chi^2)^{\rm Im} = 0 \ .
\end{equation}
Moreover, using the stationary conditions with respect
to either of the two CP-odd doublet 
components\footnote{Since the theory
is CP-violating there are no CP-odd fields. We call them CP-odd to make the connection with the CP-conserving version of the model, noting that they do not correspond to a physical mass eigenstate.} $\sigma_1$ and
$\sigma_2$, one can express the imaginary part of~$m_{12}^2$
in dependence of the imaginary parts of the complex
quartic couplings, such that
\begin{equation}
(m_{12}^2)^{\rm Im} = \frac{v_1 v_2}{2} \lambda_5^{\rm Im} +
\frac{v_1^2}{2} \lambda_6^{\rm Im} +
\frac{v_2^2}{2} \lambda_7^{\rm Im} +
\frac{v_S^2}{2} \lambda_{11}^{\rm Im} \ .
\end{equation}
Hence, only four independent CP-violating phases
remain in the Higgs potential. As will be discussed
in \cref{sec:parabasis}, one can use three (of a total of six)
mixing angles in the
neutral scalar sector in order
to parametrize the mixing of gauge eigenstates
with opposite CP charges. In this
case the imaginary parts of $\lambda_6$,
$\lambda_7$ and $\lambda_{11}$ are dependent parameters,
and only the imaginary part of $\lambda_5$ remains as
free input parameter.

After decomposing the neutral scalar fields
in real and imaginary parts, there are a total
of six real neutral scalar fields
$\varphi_1$, $\varphi_2$, $\sigma_1$, $\sigma_2$,
$\varphi_S$ and $\chi$, see \cref{eq:vevsew}.
As a consequence of the
global $U(1)$ symmetry and since
$v_S$ is real, the field $\chi$ does not mix with
the other fields. Furthermore, a rotation of the
imaginary components of the doublet fields via
the angle $\beta$ in the form $\sigma_1 = s_\beta A_0
+ c_\beta G_0$ and $\sigma_2 = c_\beta A_0 - s_\beta G_0$
removes the mixing between the unphysical Goldstone
boson $G_0$ and the physical states. The field $A_0$
corresponds to a CP-odd state that in general
is not a mass eigenstate but mixes with the
CP-even fields $\varphi_1$, $\varphi_2$ and $\varphi_S$
under the presence of CP-violation.
Hence, there are a total of four neutral scalar
Higgs bosons in the model. The mass matrix of the neutral
scalar sector can be written in the field basis
$\phi = (\varphi_1, \varphi_2, \varphi_S, A_0)$
as
\begin{equation}
- \mathcal{L} = \phi^T \mathcal{M}_H^2 \phi \ , \quad \textrm{with }
\left(\mathcal{M}_H^2\right)_{ij}  =
\dfrac{\partial^2 V}{\partial \phi_i \phi_j} \ ,
\end{equation}
where the elements of the mass matrix are not
written here for the sake of brevity.
In order to obtain the mass eigenstates,
we define the $4 \times 4$ orthogonal matrix $R$
that diagonalizes $\mathcal{M}_H^2$ such that
\begin{equation}
\left(
\begin{matrix}
m_{H_1}^2 & 0 & 0 & 0 \\
0 & m_{H_2}^2 & 0 & 0 \\
0 & 0 & m_{H_3}^2 & 0 \\
0 & 0 & 0 & m_{H_4}^2 \\
\end{matrix}
\right)
=
R \ \mathcal{M}_H^2 R^T \ ,
\label{eq:mneudiag}
\end{equation}
with $m_{H_i}$ being the masses
of the neutral Higgs bosons.
We parametrize the mixing matrix $R$
in term of six mixing angles $\alpha_{i=1,\dots,6}$ as
\begin{align}
R_{11} &=
\ca_1 \ca_2 \ca_6 \ , \label{eq:R11} \\ 
R_{12} &=
\ca_2 \ca_6 \sa_1 \ , \label{eq:R12} \\
R_{13} &=
\ca_6 \sa_2 \ , \label{eq:R13} \\
R_{14} &= - \sa_6 \ , \label{eq:R14} \\
R_{21} &=
- \sa_1
(
  \ca_3 \ca_4 + \sa_3 \sa_4 \sa_5
)
+ \ca_1
(
  - \sa_2
    (
      \ca_4 \sa_3
      \notag \\ &-
      \ca_3 \sa_4 \sa_5
    )
  - \ca_2 \ca_5 \sa_4 \sa_6
) \ , \label{eq:R21} \\
R_{22} &=
\ca_1
(
  \ca_3 \ca_4 + \sa_3 \sa_4 \sa_5
)
+ \sa_1
(
  - \sa_2
  (
    \ca_4 \sa_3
    \notag \\ &-
    \ca_3 \sa_4 \sa_5
  )
  - \ca_2 \ca_5 \sa_4 \sa_6
) \ , \label{eq:R22} \\
R_{23} &=
\ca_2
(
  \ca_4 \sa_3 - \ca_3 \sa_4 \sa_5
)
- \ca_5 \sa_2 \sa_4 \sa_6 \ , \label{eq:R23} \\
R_{24} &=
- \ca_5 \ca_6 \sa_4 \ , \label{eq:R24} \\
R_{31} &=
\ca_5 \sa_1 \sa_3 -
\ca_1
(
  \ca_3 \ca_5 \sa_2 + \ca_2 \sa_5 \sa_6
) \ , \label{eq:R31} \\
R_{32} &=
- \ca_1 \ca_5 \sa_3 - \sa_1
(
  \ca_3 \ca_5 \sa_2 + \ca_2 \sa_5 \sa_6
) \ , \label{eq:R32} \\
R_{33} &=
\ca_2 \ca_3 \ca_5 - \sa_2 \sa_5 \sa_6 \ , \label{eq:R33} \\
R_{34} &=
- \ca_6 \sa_5 \ , \label{eq:R34} \\
R_{41} &=
- \sa_1
(
  \ca_3 \sa_4 - \ca_4 \sa_3 \sa_5
)
+ \ca_1
(
  - \sa_2
  (
    \sa_3 \sa_4
    \notag \\ &+ \ca_3 \ca_4 \sa_5
  )
  + \ca_2 \ca_4 \ca_5 \sa_6
) \ , \label{eq:R41} \\
R_{42} &=
\ca_1
(
  \ca_3 \sa_4 - \ca_4 \sa_3 \sa_5
) + \sa_1
(
  - \sa_2
  (
    \sa_3 \sa_4
    \notag \\
    &+ \ca_3 \ca_4 \sa_5
  ) + \ca_2 \ca_4 \ca_5 \sa_6
) \ , \label{eq:R42} \\
R_{43} &=
\ca_2
(
  \sa_3 \sa_4 + \ca_3 \ca_4 \sa_5
)
\ca_4 \ca_5 \sa_2 \sa_6 \ , \label{eq:R43} \\
R_{44} &=
\ca_4 \ca_5 \ca_6 \ . \label{eq:R44}
\end{align}
We note that the model features a so-called
alignment limit in which one of the states
$H_i$ has couplings to fermions and gauge
bosons of the same form as the ones predicted
for the Higgs boson in the SM.
The couplings of the mass eigenstate $H_1$
only depend on the mixing angles
$\alpha_1$, $\alpha_2$ and $\alpha_6$
as can be seen in \cref{eq:R11,eq:R12,eq:R13,eq:R14}.
A convenient choice to parametrize the
alignment limit is therefore
\begin{equation}
\alpha_1 = \beta \ , \quad
\alpha_2 = 
\alpha_6 = 0 \ ,
\label{eq:alignlim}
\end{equation}
in which case the state $H_1$ is the one
resembling a SM Higgs boson of mass $m_{H_1}$.
\TB{For instance, the couplings of the neutral
Higgs bosons $H_i$ to the massive
gauge bosons, $\Gamma_{H_i VV}$, with $V = Z,W^\pm$,
normalized to the corresponding ones of the
SM Higgs boson, $\Gamma^{\rm SM}_{HVV}$, are
in general given by
\begin{equation}
  \kappa^V_i = 
  \frac{\Gamma_{H_i VV}}{{\Gamma^{\rm SM}_{HVV}}} =
    \frac{v_1 R_{i1} + v_2 R_{i2}}{v} =
    c_\beta R_{i1} + s_\beta R_{i2} \, .
\end{equation}
In the limit shown in \cref{eq:alignlim}, we
find $R_{11} = c_\beta$ and $R_{12} = s_\beta$,
and thus $\kappa_1^V = 1$. In the same way, one
can show that the couplings of $H_1$ to fermions
become equal to the ones of the SM Higgs boson.
This is discussed in detail in \cref{sec:cpviolationferm}.}

One should note that in this alignment limit
the singlet field is in general not decoupled
since the remaining states $H_2$, $H_3$ and $H_4$
can still mix with each other as long as
$\alpha_3$, $\alpha_4$ and/or $\alpha_5$
are not equal to zero
(see the discussion in \cref{sec:relic} for details).

The charged scalar sector of the model
consists of a pair of physical charged
Higgs bosons $H^\pm$ and the unphysical charged
Goldstone bosons $G^\pm$.
As in the 2HDM, the mass matrix in the
charged scalar sector is diagonalized via
a rotation with angle $\beta$, giving rise
to the massive mass eigenstates $H^\pm$
with mass
\begin{equation}
m_{H^\pm} = - \dfrac{v_1^2 + v_2^2}{2 v_1 v_2}
\left(
  -2 m_{12}^{2\, \rm Re} + v_1 v_2 \left(\lambda_4 + \lambda_5^{\rm Re}\right) +
  v_1^2 \lambda_6^{\rm Re} + v_2^2 \lambda_7^{\rm Re} +
  v_S^2 \lambda_{11}^{\rm Re} 
\right)\ .
\label{eq:mhpmdef}
\end{equation}
In the absence of CP-violation one finds
the relation
$m_{H^\pm}^2 - m_A^2 = v^2 (\lambda_5 - \lambda_4) / 2$
that is known from the real 2HDM,
where $m_A$ is the mass of the CP-odd mass
eigenstate $A_0$.

\subsection{Yukawa Sector}
\label{sec:yukawasector}

The most general Yukawa interactions can be written as
\begin{equation}
\mathcal{L}_{\mathrm{Yuk}} = - \sum_{a=1}^2\Big(
(Y_u^{(a)})_{ij} \Phi_a Q_i \bar u_j +
(Y_d^{(a)})_{ij} \tilde \Phi_a Q_i \bar d_j +
(Y_\ell^{(a)})_{ij} \tilde \Phi_a L_i \bar e_j
\Big) + \mathrm{h.c.} \ ,
\end{equation}
where the sum runs over the two Higgs doublets,
with $\tilde \Phi_a = (\Phi_a^T \varepsilon)^\dagger$
and $\epsilon_{21} = -1$,
and the indices $i,j=1,2,3$ run over the fermion flavours.\footnote{We
do not consider the possible presence of right-handed
neutrinos. The Yukawa sector and the flavour alignment (see
discussion below) can be trivially extended to take into
account right-handed neutrinos.}
The Yukawa couplings $Y^{(1,2)}_{u,d,\ell}$ are $3 \times 3$
matrices in flavour spaces. Their elements can
be complex, which would in general give rise to
CP-violation in the Yukawa sector.
After electroweak symmetry breaking, one finds the
following mass matrices for the quarks and the charged leptons,
\begin{equation}
- \mathcal{L} = f_{iL}
\left(
  \mathcal{M}_{f}
\right)_{ij}
f_{jR}^* +
\mathrm{h.c.} \ , \qquad
\left(
  \mathcal{M}_{f}
\right)_{ij} =
\dfrac{1}{\sqrt{2}}
\left(
  v_1 ( Y^{(1)}_{f} )_{ji} +
  v_2 ( Y^{(2)}_{f} )_{ji}
\right)
\ , \quad
f_i = u_i,d_i,\ell_i \ .
\label{eq:fermmassmatrix}
\end{equation}
If the matrices $Y^{(1)}_f$ and $Y^{(2)}_f$ are not
diagonalizable simultaneously, the diagonalization of
the fermion mass matrices would yield flavour-changing
couplings of the neutral Higgs bosons to fermions,
and thus to flavour-changing neutral currents (FCNC)
which are tightly constraint by experiments.
In order to avoid FCNC, we apply the so-called
flavour-alignment scenario~\cite{Pich:2009sp}.
Here it is assumed that the Yukawa matrices
$Y^{(1)}_f$ and $Y^{(2)}_f$ are proportional to
each other, in which case they can be diagonalized
by means of the same fermion rotation matrices
which then also diagonalize the mass matrices.
We parametrize the flavour alignment
in terms of the flavour alignment parameters $\xi_u$, $\xi_d$
and $\xi_\ell$ for the up-type quarks, the down-type
quarks and the charged leptons, respectively, such that
\begin{equation}
(Y^{(1)}_u)_{ij} = \xi_u \, (Y^{(2)}_u)_{ij} \, , \quad
(Y^{(1)}_d)_{ij} = \xi_d \, (Y^{(2)}_d)_{ij} \, , \quad
(Y^{(1)}_\ell)_{ij} = \xi_\ell \, (Y^{(2)}_u)_{ij} \, .
\end{equation}
The flavour alignment parameters are in general
complex. Their imaginary parts constitute three
sources of CP violation in the Yukawa sector.
Without loss of generality, we assume the
Yukawa matrices $Y^{(2)}_f$ to be
diagonal,
\begin{equation}
Y^{(2)}_u = \mathrm{diag}(Y_u, Y_c, Y_t) \, , \quad
Y^{(2)}_d = \mathrm{diag}(Y_d, Y_s, Y_b) \, , \quad
Y^{(2)}_\ell = \mathrm{diag}(Y_e, Y_\mu, Y_\tau) \, ,
\end{equation}
where the Yukawa couplings on the right-hand side of
the equations are the Yukawa coupling of each fermion
contained in the SM.\footnote{For simplicity, we
neglect the quark-flavour mixing in terms of the
CKM matrix in this discussion, but it is taken into
account in the numerical analysis if relevant.}
These Yukawa couplings can be written in dependence
of the physical masses $m_f$ of the fermions,
the vevs of the Higgs doublets
and the flavour alignment parameters,
\begin{equation}
Y_f = \dfrac{\sqrt{2} \, m_f}{\xi_f v_1 + v_2}
\, , \quad
f = u,c,t,d,s,b,e,\mu,\tau \, .
\label{eq:yukfrommass}
\end{equation}
We note that although the Yukawa coupling $Y_f$
is complex if $\xi_f$ is complex, the elements
of the fermion mass
matrices shown in \cref{eq:fermmassmatrix}
are always real.

In general, the flavour alignment is not protected
by a symmetry. The only exception is
the case in which all fermions with the
same electric charge only couple
to one of the two Higgs doublets, leading
to the traditional Yukawa types of the 2HDM,
see discussion below~\cite{Ferreira:2010xe}.
Thus, radiative corrections give
rise to a breaking of the flavour alignment.
As a consequence, the flavour alignment can
be set at one particular energy scale
(here assumed to be the EW scale), but the
renormalization group running would induce
a breaking of the flavour alignment at different
energy scales. 
This being said,
if flavour alignment is assumed at the EW scale,
the impact of the radiatively induced breaking
of the flavour alignment at other experimentally
accessible scales is typically too
small to generate observable
FCNCs at the current experimental
precision~\cite{Penuelas:2017ikk}.

\begin{table}
\renewcommand{\arraystretch}{1.2}
\centering
\begin{tabular}{l||ccc|cc}
\textbf{Type} & $\xi_u^{\rm Re}$ & $\xi_d^{\rm Re}$ & \
  $\xi_\ell^{\rm Re}$ & $\Phi_1$ & $\Phi_2$ \\
\hline
\hline
\textbf{I} & 0 & 0 & 0 & - & $u,d,\ell$ \\
\textbf{II} & 0 & $\infty$ & $\infty$ & $d,\ell$ & $u$ \\
\textbf{LS/III} & 0 & 0 & $\infty$ & $\ell$ & $u,d$ \\
\textbf{FL/IV} & 0 & $\infty$ & 0 & $d$ & $u,\ell$
\end{tabular}
\caption{Correspondence between special choices
of the flavour
alignment parameters and the four different
Yukawa types if natural flavour alignment
is assumed.
The imaginary parts $\xi_{u,d,\ell}^{\rm Im}$
can be chosen to be vanishing in the
Yukawa types since a complex
phase of the flavour alignment parameters can
be removed via field redefinitions.
The last two columns indicate
which of the two Higgs doublets $\Phi_1$
and $\Phi_2$ is coupled to the different
kind of fermions, where $u$, $d$ and $\ell$
corresponds to up-type quarks, down-type quarks
and charged leptons, respectively.}
\label{ta:flavalign}
\renewcommand{\arraystretch}{1.0}
\end{table}

In order to make the flavour alignment
stable under the renormalization group,
one can impose the so-called natural
flavour alignment, where a discrete
$\mathbb{Z}_2$ symmetry is responsible for the
invariance of the flavour alignment.
In this case, one of the Higgs doublets
changes the sign while the other transforms
trivially, such that only one Higgs doublet
can be coupled to one kind of fermion, but
not both at the same time.
There are four different ways to extend
the $Z_2$ symmetry to the fermions, giving
rise to the so-called Yukawa types~I, II,
lepton-specific~(LS)/III and
Flipped~(FL)/IV~\cite{Glashow:1976nt}.
The four Yukawa types correspond to
particular choices of the flavour parameters
as summarized in \cref{ta:flavalign}.
Imposing such a $\mathbb{Z}_2$ symmetry
is a significant restriction of the model
and its phenomenology at the LHC (and other
experiments) since it constrains the
Higgs-boson couplings to fermions.
Moreover, the symmetry removes the possibility
of CP violation in the scalar sector
in the Higgs alignment limit of the model.
The current LHC measurements are
well compatible with the predictions for a
Higgs boson predicted by the SM, thus driving
the model towards the alignment limit.
As a consequence, the presence of CP violation
in the Higgs sector, and in particular in the
couplings of the Higgs boson at 125~GeV, are
severely restricted by the LHC measurements
if a $\mathbb{Z}_2$ is imposed
(see \citere{Biekotter:2024ykp} for a recent analysis in
the complex 2HDM).
Consequently, in order to allow for the
presence of CP-violation in the alignment limit
we use in our analysis the
less restrictive flavour alignment scenario.

\subsection{Parameter Basis}
\label{sec:parabasis}

To investigate the phenomenology of the model,
it is convenient to choose a basis of free input
parameters which are more closely related to
physical observables than the bare parameter
appearing in the Lagrangian.
In the scalar sector we make use of the
masses and the mixing angles as input
parameters. The mixing angles are suitable input
parameters since they are related to the couplings
of the scalars to fermions and gauge bosons.
Using the stationary coditions of the
scalar potential and the mass matrix relations
shown in \cref{eq:mneudiag} for the neutral
and in \cref{eq:mhpmdef}
for the charged scalars, one can trade 11
Lagrangian parameters
for the physical masses and
the mixing angles according to the relations
\begin{align}
\la_1 &= \frac{1}{c_\beta^2 v^2} \bigg[
    \sum_{i=1}^4 R_{i1}^2 m_{H_i}^2 -
    s_\beta^2 M^2 \bigg] -
  \frac{3}{2} t_\beta \la_6^{\rm Re} +
  \frac{1}{2} t_\beta^2 \la_7^{\rm Re} +
  \frac{v_S^2}{2 v^2} \frac{t_\beta}{c_\beta^2} \la_{11}^{\rm Re} \ , \\
\la_2 &= \frac{1}{s_\beta^2 v^2} \bigg[
    \sum_{i=1}^4 R_{i2}^2 m_{H_i}^2 -
    c_\beta^2 M^2 \bigg] +
  \frac{1}{2} \frac{1}{t_\beta^3} \la_6^{\rm Re} -
  \frac{3}{2} \frac{1}{t_\beta} \la_7^{\rm Re} +
  \frac{v_S^2}{2 \, v^2} \frac{1}{s_\beta^2 t_\beta} \la_{11}^{\rm Re} \ , \\
\la_3 &= \frac{2}{v^2} m_{H^\pm}^2 +
  \frac{1}{c_\beta s_\beta v^2} \bigg[
    \sum_{i=1}^4 R_{i1} R_{i2} m_{H_i}^2 - c_\beta s_\beta M^2
  \bigg] -
  \frac{1}{2} \frac{1}{t_\beta} \la_6^{\rm Re} -
  \frac{1}{2} t_\beta \la_7^{\rm Re} +
  \frac{1}{s_{2\beta}} \frac{v_S^2}{v^2} \la_{11}^{\rm Re} \ , \\
\la_4 &= - \frac{2}{v^2} m_{H^\pm}^2 +
  \frac{1}{v^2} \bigg[
    \sum_{i=1}^4 R_{i4}^2 m_{H_i}^2 + M^2 \bigg] -
  \frac{1}{2} \frac{1}{t_\beta} \la_6^{\rm Re} -
  \frac{1}{2} t_\beta \la_7^{\rm Re} -
  \frac{1}{s_{2\beta}} \frac{v_S^2}{v^2} \la_{11}^{\rm Re} \ , \\
\la_5^{\rm Re} &= \frac{1}{v^2} \bigg[
    \sum_{i=1}^4 R_{i4}^2 m_{H_i}^2 - M^2 \bigg] -
  \frac{1}{2} \frac{1}{t_\beta} \la_6^{\rm Re} -
  \frac{1}{2} t_\beta \la_7^{\rm Re} -
  \frac{1}{s_{2\beta}} \frac{v_S^2}{v^2} \la_{11}^{\rm Re} \ , \\
\la_6^{\rm Im} &= -
  \frac{1}{c_\beta v^2} \sum_{i=1}^4 R_{i1} R_{i4} m_{H_i}^2 -
  \frac{1}{2} t_\beta \la_5^{\rm Im} \ , \\
\la_7^{\rm Im} &= -
  \frac{1}{s_\beta v^2} \sum_{i=1}^4 R_{i2} R_{i4} m_{H_i}^2 -
  \frac{1}{2} \frac{1}{t_\beta} \la_5^{\rm Im} \ , \\
\la_8 &=
  \frac{1}{v_S^2} \sum_{i=1}^4 R_{i3}^2 m_{H_i}^2 \ ,
  \label{eq:lam8} \\
\la_9 &=
  \frac{1}{c_\beta v v_S} \sum_{i=1}^4 R_{i1} R_{i3} m_{H_i}^2 -
  t_\beta \la_{11}^{\rm Re} \ ,
  \label{eq:lam9} \\
\la_{10} &=
  \frac{1}{s_\beta v v_S} \sum_{i=1}^4 R_{i2} R_{i3} m_{H_i}^2 -
  \frac{1}{t_\beta} \la_{11}^{\rm Re} \ ,
 \label{eq:lam10} \\
\la_{11}^{\rm Im} &= -
  \frac{1}{v v_S} \sum_{i=1}^4 R_{i3} R_{i4} m_{H_i}^2 \ ,
  \label{eq:lam11}
\end{align}
where we have introduced the parameter
\begin{equation}
M^2 = \frac{(m_{12}^2)^{\rm Re}}{c_\beta s_\beta} \ ,
\end{equation}
and the scalar mixing matrix elements $R_{ij}$ are
expressed in terms of the six mixing angles $\alpha_i$
as shown in
\cref{eq:R11,eq:R12,eq:R13,eq:R14,eq:R21,eq:R22,eq:R23,eq:R24,eq:R31,eq:R32,eq:R33,eq:R34,eq:R41,eq:R42,eq:R43,eq:R44}.
In the fermion sector, we will use the fermion
masses and the flavour alignment parameters as
input. The Yukawa couplings can then be derived
from \cref{eq:yukfrommass}.
This leaves us with the following set of input parameters
to specify a parameter point
\begin{align}
v = 246.2\gev \ , \quad
m_{H_1} = 125.1\gev \ , \quad
m_{H_{2,3,4}} \ , \quad
m_{H^\pm} \ , \quad
m_{\chi} \ , \quad
\tan\beta \ , \quad
M \ , \notag \\
v_S \ , \quad
\alpha_{1,2,3,4,5,6} \ , \quad
\la_5^{\rm Im} \ , \quad
\la_6^{\rm Re} \ , \quad
\la_7^{\rm Re} \ , \quad
\la_{11}^{\rm Re} \ , \quad
\xi_{u,d,\ell}^{\rm Re} \ , \quad
\xi_{u,d,\ell}^{\rm Im} \ .
\label{eq:freeparas}
\end{align}
Since the number of free parameters is sizable,
we will make different simplifying assumptions
throughout our numerical analysis.
This also facilitates the comparison to simpler
models that have been extensively studied
in the literature.
This concerns the parameters $\la_6^{\rm Re}$
and $\la_7^{\rm Re}$ and $\la_{11}^{\rm Re}$
which we set to zero in order to eliminate
FCNC at tree level which are tightly constrained
experimentally.
In addition, since we assume $H_1$ to play the role of the
detected Higgs boson at about 125~GeV,
we will make use of the conditions shown
in \cref{eq:alignlim} if we impose the alignment limit.
Moreover, if we consider natural flavour conservation (NFC) in terms of the
well-known Yukawa types, we set $\xi_{u,d,\ell}^{\rm Im} = 0$
and, depending on the considered type,
the parameters $\xi_{u,d,\ell}^{\rm Re}$ as depicted
in \cref{ta:flavalign}.

\subsection{The Dark Sector}

The model predicts a neutral stable scalar particle $\chi$
that plays the role of the DM candidate.
$\chi$ interacts with the fermions and gauge bosons
only via Higgs-boson exchange.
The tree-level couplings $\chi\chi H_i$ are given by
\begin{equation}
\ii \Gamma_{\chi\chi H_i} =
  v \left[
    c_\beta \la_9 R_{i1} +
    s_\beta \la_{10} R_{i2} +
    (s_\beta R_{i1} + c_\beta R_{i2}) \la_{11}^{\rm Re} +
    \TB{\la_{11}^{\rm Im}} R_{i4} +
    \frac{v_S}{v} \la_8 R_{i3}
  \right] \TB{= \frac{m_{H_i}^2}{v_S} R_{i3}} \ .
\label{eq:chichihcpl}
\end{equation}
\TB{where in the last step
we used the parameter relations
for $\lambda_8$, $\lambda_9$, $\lambda_{10}$ and
$\lambda_{11}$ shown above
and the orthogonality of the mixing matrix~$R$
to express the couplings in terms of the masses
of the neutral Higgs bosons.}
The coupling of $\chi$ to the doublet components
is generated via the terms proportional to the portal
couplings $\la_9$, $\la_{10}$ and $\la_{11}$, whose
corresponding potential terms, as shown in \cref{eq:scapot}, contain
both the doublet fields $\Phi_1$ or $\Phi_2$ and the
gauge singlet field $\Phi_S$. On the other hand,
the coupling of $\chi$ to the singlet components of
the states $H_i$ is generated via the coupling $\la_8$
that appears in the $|\Phi_S|^4$ term in the scalar
potential. One important consequence is that
the detected Higgs boson at 125~GeV can be coupled
to $\chi$ only if either at least one of the portal
couplings is non-zero, or if it has a non-vanishing
singlet admixture and $\la_8 > 0$.
The $\mathbb{Z}_2$-breaking couplings $\la_6$ and $\la_7$
do not appear in the $\chi\chi H_i$ couplings
at leading order and thus do not play a major
role for the DM phenomenology.
The DM particle
$\chi$ is in general coupled to all four neutral Higgs bosons.
This gives rise to an important distinction to
the CP-conserving version of the model as studied in
\citere{Biekotter:2021ovi}, where
the DM state is only coupled to the three
CP-even scalars but not to the CP-odd pseudoscalar.
This additional coupling of $\chi$ can play an important
role for the prediction of the DM relic abundance,
as will be discussed in \cref{sec:relic}.

In the alignment limit, as defined in
\cref{eq:alignlim},
and additionally setting $\alpha_3 = \alpha_4
= \alpha_5 = 0$,
the expression shown in
\cref{eq:chichihcpl}
can be written as
\begin{equation}
\ii \Gamma_{\chi\chi H_{1,2,4}} = 0 \ , \quad
\ii \Gamma_{\chi\chi H_3} = \frac{m_{H_3}^2}{v_S} \ .
\end{equation}
\TB{Thus, in this limit the DM particle 
$\chi$ only interacts with the singlet state
$H_3$. Since a pure gauge-singlet scalar does not
couple directly to SM fermions or gauge bosons and
acquires such couplings only via mixing with the
Higgs doublet fields, the absence of doublet
admixtures in $H_3$ implies that both 
$\chi$ and $H_3$ form a decoupled dark sector
whose interactions with the visible sector are
switched off (see also \cref{sec:cpviolationferm}
for a more detailed discussion of the $H_i f \bar f$
couplings).}
As such, departures from
this parameter configuration
have to be considered in order to
enable the standard freeze-out scenario
for the production of the DM relic abundance
in agreement with observations.
Staying in the alignment limit,
these can be generated by mixing the singlet
state~$H_3$ with the second CP-even state $H_2$
and/or by mixing the singlet state $H_3$
with the CP-odd state $H_4$ under the presence
of CP-violation.
For instance, setting $\alpha_3 \neq 0$ and keeping
the other mixing angles as before,
both $H_2$ and $H_3$ share a non-zero component
of the singlet field, and one finds
\begin{equation}
\ii \Gamma_{\chi\chi H_{1,4}} = 0 \ , \quad
\ii \Gamma_{\chi\chi H_{2}} = \frac{m_{H_2}^2 s_{\alpha_3}}
  {v_S} \ , \quad
\ii \Gamma_{\chi\chi H_{3}} = \frac{m_{H_3}^2 c_{\alpha_3}}
  {v_S} \ .
\end{equation}
Since in this case both
CP-even states $H_2$ and $H_3$ couple to
fermions and gauge bosons, the freeze-out mechanism
can efficiently proceed via these two states
without modifications to the couplings of the
SM-like Higgs boson $H_1$.
On the other hand, if 
mixing is considered
via $\alpha_5 \neq 0$ (with $\alpha_3 = \alpha_4 =  0$), then
the states $H_3$ and $H_4$ are CP-odd
and both share a singlet
admixture, and one finds for the couplings
to the DM state
\begin{equation}
\ii \Gamma_{\chi\chi H_{1,2}} = 0 \ , \quad
\ii \Gamma_{\chi\chi H_{3}} = \frac{m_{H_3}^2 c_{\alpha_5}}
  {v_S} \ , \quad
\ii \Gamma_{\chi\chi H_{4}} = \frac{m_{H_4}^2 s_{\alpha_5}}
  {v_S} \ .
\end{equation}
This opens up the interesting possibility that
the DM annihilation proceeds mainly via a
CP-odd Higgs-portal
which is not possible
for pseudo-Nambu-Goldstone (pNG) DM if the Higgs
sector respects the CP symmetry.

The special feature of the cS2HDM 
is that the DM state $\chi$ acts as a pNG
boson under a softly-broken global U(1)
symmetry. As a pNG boson, the on-shell interactions
of $\chi$ with the neutral Higgs bosons $H_i$
are proportional to the momentum squared of
$H_i$. In \citere{Gross:2017dan} it was shown
that this leads to a cancellation of the
scattering process of $\chi$ on nucleons
in the limit of vanishing momentum transfer.
As a consequence, for non-relativistic
pNG DM the scattering rates at direct detection
experiments are highly suppressed.
This opens up parameter space regions and
DM masses that can be probed by the LHC,
where in other models the possibility of a
detection of DM at the LHC is largely excluded
by direct-detection experiments.
Thus, the model put forward here is especially
motivated to serve as an LHC benchmark model
for future DM searches.
Still, since the global $U(1)$ symmetry acting
on the singlet field is softly broken,
the cancellation mechanism is not exact,
and radiative corrections from diagrams
with $\chi$ in the loops generate
contributions to the DM-nucleon scattering
that do not vanish in the limit of vanishing
momentum exchange.\footnote{Finite-momentum
contributions have been shown to be too small
to generate detectable scattering
cross sections~\cite{Azevedo:2018exj}
and are therefore not considered here.}
The impact of the radiative corrections, and
how we take them into account in order to confront
the model with the constraints from direct-detection
experiments, is discussed in \cref{sec:directdetection}.

\subsection{CP-Violation}
\label{sec:cpviolation}

The model contains different sources of CP-violation
both in the scalar and in the Yukawa sector.
The CP-violation can physically manifest itself
at low-energy experiments, most notably in the form
of electromagnetic dipole moments~(EDMs),
or at higher energies at collider experiments
like the LHC or LEP in the production and
decay of the Higgs bosons.

\subsubsection{CP-Violation in the Scalar Potential}
\label{sec:cpviolationscalar}

In the scalar sector, and using the set of input
parameters introduced in \cref{sec:parabasis},
one can impose CP-violation in two different ways.
The first possibility is to choose the mixing angles
$\alpha_i$ in such a way that there is a mixing between
at least one of the CP-even fields $\varphi_1$
or $\varphi_2$
and the CP-odd field $A_0$
(see discussion in \cref{sec:higgssector}).
If a CP-violating mixing is present in the scalar sector,
the CP-violation manifests itself phenomenologically
in various ways. For instance, at least two neutral scalars $H_i$
possesses couplings to fermions $f$ that contain both
a CP-even component $H_i f \bar f$ and a CP-odd
component $\ii H_i f \gamma_5 \bar f$,
giving rise
to charge-conserving but parity~(P)-violating
interactions~\cite{Haber:2022gsn}.
In addition, purely bosonic CP-violating interactions
are generated. For instance, one finds in general
that the couplings $H_i H_j Z$
exist for all four neutral Higgs bosons
$i=1,2,3,4$ with $j \neq i$, which is a
P-even but CP-violating phenomenon~\cite{Haber:2022gsn}.
On the other hand, one can remove the CP-violating
mixing of the scalar fields by choosing the mixing
angles in such a way that the CP-odd state $A_0$
is a mass eigenstate and does not mix with
the other fields. This can be achieved, for example, by setting
\begin{equation}
\alpha_4 = 0, \pi, -\pi \ , \quad
\alpha_5 = 0, \pi, -\pi \ , \quad
\alpha_6 = 0, \pi, -\pi \ \quad \Rightarrow \quad
R_{i,4} = R_{4,i} = \pm \delta_{i4} \ , \quad i=1,2,3,4 \ .
\label{eq:alphasCPconv}
\end{equation}
Then, only a mixing between the three CP-even fields
$(\varphi_1,\varphi_2,\varphi_S)$ $\to$ $(H_1,H_2,H_3)$
remains that is parametrized in terms of the
mixing angles $\alpha_1$, $\alpha_2$ and $\alpha_3$,
and the state $H_4 = A_0$ does not mix with the other
mass eigenstates.
Here it should be noted that at the loop level
CP-violating couplings give in general rise to
a mixing between the states $H_{1,2,3}$ and the
state $H_4$, such that the states $H_{1,2,3}$ and
$H_4$ should only be understood as approximate
CP-even and CP-odd eigenstates, respectively,
see also the discussion below.

\begin{figure}
\centering
\includegraphics[width=5cm]{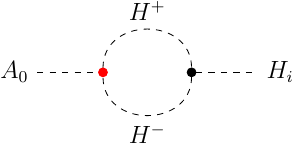}
\qquad
\includegraphics[width=4cm]{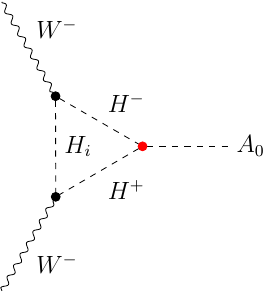}
\caption{Left: one-loop self-energy diagram that
gives rise to a CP-violating mixing between
$A_0$ and $H_i$ under the presence of a bosonic
CP-violating coupling $A_0 H^+ H^-$.
Right: one-loop diagram that gives rise to
a coupling between a seemingly CP-odd state
$A_0$ and the $W$ boson under the presence of
the same CP-violating coupling considered in
the left diagram. See text for details.}
\label{fig:CPvioFromlam5im}
\end{figure}

Even if the CP-violating mixing in the scalar sector
is removed (at the classical level)
by choosing the mixing angles according
to \cref{eq:alphasCPconv},
the model still allows for the presence of
CP-violation by setting $\la_5^{\rm Im} \neq 0$.
In this case,
using the free parameters
shown in \cref{eq:freeparas}, one finds CP-violation
only in scalar interactions, whereas all couplings
of the scalars to the fermions and gauge bosons
are in agreement with the predictions of
a CP conserving model.
For instance, the state $A_0$ acts as a pseudoscalar
in its interactions with SM particles, i.e.~it couples
to fermions with a CP-odd coupling $A_0 f \gamma_5 \bar f$,
and its couplings
to gauge bosons are absent.
Nevertheless, it also features a coupling to the charged
scalars of the form
\begin{equation}
\ii \Gamma_{A_0 H^+ H^-}^{(0)} =
\frac{v \la_5^{\rm Im}}{2 c_\beta s_\beta} \ ,
\label{eq:axxalignlim}
\end{equation}
which is a coupling that is P-even but CP-violating.
Here it should be noted that this purely
\textit{bosonic CP-violation} resulting from
$\la_5^{\rm Im} \neq 0$ leaks into the
other couplings of the Higgs bosons and in their
mixing among each other at the radiative level.
Another purely scalar CP-violating interaction that
is generated in this case are $A_0 H_i H_j$ interactions.
In general, due to the mixing of the states $H_i$
the corresponding couplings have a complicated form.
However, in the
alignment limit defined in \cref{eq:alignlim},
in combination with the absence of CP-violating
mixings by setting $\alpha_4 = \alpha_5 = 0$,
the couplings
are non-vanishing only for the states $H_2$
and $H_3$ and given by
\begin{equation}
\ii \Gamma_{A_0 H_2 H_2}^{(0)} =
\frac{c_{\alpha_3}^2 v \la_5^{\rm Im}}{s_{2\beta}} \, , \quad
\ii \Gamma_{A_0 H_2 H_3}^{(0)} =
\frac{s_{\alpha_3}c_{\alpha_3} v \la_5^{\rm Im}}{s_{2\beta}} \, , \quad
\ii \Gamma_{A_0 H_3 H_3}^{(0)} =
\frac{s_{\alpha_3}^2 v \la_5^{\rm Im}}{s_{2\beta}} \, .
\end{equation}
As a result of the CP-violating scalar vertices
$A_0 H^+ H^-$ and $A_0 H_i H_j$
at tree level, also other forms of CP-violating
interactions are generated via quantum corrections.
Already at one-loop level, the self-energy diagram depicted
in \cref{fig:CPvioFromlam5im} induces a CP-violating mixing between a seemingly
CP-even state $H_i$
and the seemingly CP-odd state $A_0$.
Moreover, the Feynman diagram on the right-hand side
induces a coupling
of $A_0$ to the $W$ boson
which would not be allowed for a pure pseudoscalar.
Taking such radiative effects into account in the
mixing between the scalar fields would
require a complete renormalization of the scalar potential,
which lies beyond the scope of this paper,
and is not considered in our numerical discussion.
However, we do take into account the CP-violating scalar
couplings in the (tree-level) computation of the
partial decay widths of the Higgs bosons, as these
couplings open up novel decay modes that would be impossible
in a CP-conserving model.

\subsubsection{CP-Violation in the Yukawa Sector}
\label{sec:cpviolationferm}

Imposing NFC forbids sources
of CP-violation in the Yukawa sector of 2HDMs.
Instead, we
impose in our analysis the less restrictive flavour alignment.
This allows for three additional CP-violating phases
in the Yukawa sector.
We parametrize the CP-violation
by treating the flavour alignment parameters
$\xi_u$, $\xi_d$ and $\xi_\ell$
as potentially complex parameters. If one or more
of these three parameters has an imaginary part,
there is CP-violation in the Yukawa sector, which
then gives rise to CP-violating $H_i f \bar f$ interactions.
The corresponding couplings can be written at leading order as
\begin{align}
\ii \Gamma_{H_i f \bar f}^{(0)} &= \Gamma_{H_i f \bar f}
  + \ii \gamma_5 \tilde\Gamma_{H_i f \bar f} \ , \notag \\
&\Gamma_{H_i f \bar f} =
  \frac
    {m_f\left[
      (|\xi_f|^2 c_\beta + \xi_f^{\rm Re} s_\beta) R_{i1} +
      (\xi_f^{\rm Re} c_\beta + s_\beta) R_{i2} -
      \xi_f^{\rm Im} R_{i4}
    \right]}
    {v\left[|\xi_f|^2 c_\beta^2 + s_\beta^2 + \xi_f^{\rm Re} s_{2\beta}\right]}
     \ , \label{eq:gamhffeven} \\
&\tilde\Gamma_{H_i f \bar f} =
  \frac
    {m_f\left[
      - \xi_f^{\rm Im} s_\beta R_{i1}
      + \xi_f^{\rm Im} c_\beta R_{i2}
      + \left(
          (1 - |\xi_f|^2) \frac{s_{2\beta}}{2} +
          \xi_f^{\rm Im} c_{2\beta}
        \right) R_{i4}
    \right]}
    {v\left[(\xi_f^{\rm Im} c_\beta)^2 + (\xi_f^{\rm Re} c_\beta + s_\beta)^2\right]}
     \ . \label{eq:gamhffodd}
\end{align}
CP-violation is present in the couplings of the
Higgs bosons to fermions if at least one scalar
$H_i$ features a coupling in which both
$\Gamma_{H_i f \bar f} \neq 0 $ and
$\tilde\Gamma_{H_i f \bar f} \neq 0$.
In our model this effect can arise from
two independent sources:
(1) if there is CP-violation in the scalar
sector such that $R_{i4} \neq 0$ for $i=1,2,3$,
see the last terms in \cref{eq:gamhffeven} and \cref{eq:gamhffodd}.
(2) if there is CP-violation in the Yukawa
sector such that $\xi_f$ is a complex number,
see the first two terms in \cref{eq:gamhffeven} and \cref{eq:gamhffodd}. 
It should also be noted that in the alignment limit 
the expression shown in \cref{eq:gamhffodd}
yields $\tilde \Gamma_{H_1 f \bar f} = 0$  for the
SM-like Higgs boson $H_1$, irrespective of
non-zero values of $\xi_f^{\rm Im}$, and only
one of the other BSM Higgs bosons obtains
CP-violating couplings to fermions.
It is useful to define so-called $\kappa$-factors
that are given by normalizing the CP-even and
CP-odd components of the $H_i f \bar f$ couplings
to the CP-even coupling of a SM Higgs boson
of the same mass, i.e.
\begin{equation}
  \Gamma_{H_i f \bar f} =
    \frac{m_f}{v} \kappa_i^f \ , \quad
  \tilde \Gamma_{H_i f \bar f} =
    \frac{m_f}{v} \tilde \kappa_i^f \ .
\end{equation}
Hence, in the alignment limit defined
in \cref{eq:alignlim} one finds
$\kappa_1^f = 1$ and $\tilde \kappa_1^f = 0$
for the SM-like state $H_1$,
$\kappa_{2,3}^f \neq 0$ and
$\tilde \kappa_{2,3}^f = 0$ for the
second CP-even doublet and the singlet
states $H_2$ and $H_3$,
and $\kappa_4^f = 0$ and
$\tilde \kappa_4^f \neq 0$ for the
CP-odd state $H_4$. 

We end this discussion
by noting that the presence of sources for CP-violation
in the scalar sector and in the Yukawa sector
allows for cancellations in the theoretical predictions of
the EDMs between the contributions from both sectors~\cite{Kanemura:2020ibp}.
As a consequence, compared to models with NFC in
which the Yukawa sector is CP conserving,
models with flavour alignment have more freedom to
accommodate sizable CP-violating
couplings of the Higgs bosons without being in
conflict with experimental upper limits on
the EDMs. We will discuss this in more detail
in \cref{sec:edms}.

\subsection{Comments on the Early Universe}
\label{sec:earlyuni}

The cS2HDM 
predicts a phenomenologically viable
DM candidate while also containing all required
ingredients to realize EW baryogensis.
Here, the
strong constraints from the non-observation of
the eEDM can be avoided 
via a cancellation
between CP-violating effects between the Higgs
sector and the Yukawa sector,
as will be illustrated in specific
parameter planes below.
The strong constraints
from DM direct detection experiments
are evaded due to the pNG nature of the DM and the
strongly suppressed DM-nucleon interactions,
see discussion above.
Although it is not the main focus of this paper, we briefly
comment on general cosmological aspects of the model
and possible phenomenological consequences from the
thermal history of the model in the early universe.

Regarding the DM sector, the most natural production
mechanism of the DM in the early universe proceeds
via the freeze-out mechanism~\cite{Lee:1977ua}.
For temperatures of
the order of the DM mass, assumed to be of the
order of the EW scale here,
the DM state $\chi$ is in
thermal equilibrium with the particles from the
visible sector by means of the Higgs-portal
interactions. After the temperature of the universe
drops significantly below the DM mass,
the portal interaction rate falls below the expansion
rate of the universe, at which moment the number
density of the DM effectively freezes out.
The resulting relic abundance of DM depends on the
mass of the DM, $m_\chi$, the masses $m_{H_i}$ of
the Higgs bosons acting as mediators between
the dark and the visible sector,
and the size of the portal couplings
$\lambda_9$, $\lambda_{10}$ and $\lambda_{11}$,
see \cref{eq:scapot}.
We note here that in our parametrization only
the real part of $\lambda_{11}$ is an input parameter,
whereas $\lambda_9$, $\lambda_{10}$ and
$\lambda_{11}^{\rm Im}$ are derived parameters
according to \cref{eq:lam9}, \cref{eq:lam10} and
\cref{eq:lam11}, respectively.

Modifications from the usual freeze-out mechanism
have to be considered in special corners of the
parameter space. For instance,
if the DM mass is approximately
equal to half the mass of one of the Higgs bosons,
the DM annihilation is resonantly enhanced.
To predict
a relic abundance in agreement with the Planck
measurement then requires strongly suppressed portal
couplings.
This can give rise to the effect of
\textit{early kinematic decoupling}
that should be taken into account for an accurate
prediction for the DM relic abundance~\cite{Binder:2017rgn}.
Early kinematic decoupling of pNG DM has been
studied recently in \citere{Abe:2021jcz}.
Parameter space regions where the DM mass is set
to be roughly equal to the mediator mass
can be used to evade the limits from
DM direct-detection experiments, because the small
portal couplings also reduce the scattering rates.
In our model, however, the scattering rates are
momentum-suppressed already due to the pNG nature,
such that it is not necessary to consider resonance
regions with $m_\chi \approx m_{H_i} / 2$ for a viable
DM phenomenology. We therefore do not take into account
the effect of early kinematic decoupling here.
Another alternative process for the production of the
DM in the early universe consists of the
freeze-in mechanism~\cite{Hall:2009bx}, where the interaction rate
between the DM and the ordinary matter is so low
that the DM particle is never in thermal equilibrium
with the visible sector. The DM is typically assumed
to be produced via the decay of a heavier particle
that is in equilibrium with the thermal bath.
In our model the portal couplings are related to the
EW interactions, and thus, in the absence of
additional symmetries restricting the Higgs-portal
interactions, are significantly too large
to accommodate the freeze-in mechanism.
We will therefore not consider the freeze-in of
DM here.

A strong first-order EW phase transition in the
early universe is necessary for the
realization of
EW baryogensis, which seeks to explain the
matter-antimatter asymmetry of the universe.
The model considered here is an extension of the 2HDM,
such that it can accommodate an EW phase transition
in a manner similar to the 2HDM.
However, compared to the most studied
$\mathbb{Z}_2$-symmetric 2HDM with CP-violation,
usually denoted C2HDM in the literature,
our model offers additional advantages.
The stringent experimental limits on CP-violation
in the scalar sector can be mitigated through
potential cancellations with CP-violating couplings
in the Yukawa sector.
Furthermore, constraints from LHC searches for additional
Higgs bosons can be relaxed
if the BSM Higgs bosons decay with a sizable
branching fraction invisibly into two DM scalars.
This allows for a lighter BSM Higgs boson spectrum,
which would strengthen the EW phase transition
and enhance the resulting gravitational wave signal
from the transitions. We leave a detailed discussion
of the thermal history of the model, the possible
patterns of cosmological phase transitions and
the associated gravitational wave signals
for future work.

\section{Constraints}
\label{sec:constraints}

The parameter space of the model is subject to
various theoretical and experimental constraints.
In the following, we will discuss the constraints
that we apply in our numerical analysis, starting
with the theoretical requirements that we have imposed.

\subsection{Vacuum Stability}
\label{sec:vacuumstability}

We verify if the scalar potential
is bounded from below in order to ensure that
the EW vacuum can be stable.
While in simpler extensions of the SM
by a second Higgs doublet and/or a
real scalar singlet
analytic conditions can be found that
can be used to determine whether the
scalar potential is bounded from below,
analytic conditions are not avalaible
in this model due to the additional terms
present in the scalar potential and the
correspondingly larger number of parameters.
We therefore apply a numerical method to
check for the behaviour of the potential
for very large field values.

To this end, we expressed the quartic part
of the potential in terms of gauge-invariant
bilinears
\begin{equation}
|\Phi_1|^2 = \frac{h_1^2}{2} \, , \quad
|\Phi_2|^2 = \frac{h_2^2}{2} \, , \quad
|\Phi_S|^2 = \frac{h_S^2}{2} \, , \quad
\Phi_2^\dagger \Phi_1 = \rho \, \ee^{- \ii \eta} \frac{h_1 h_2}{2} \ ,
\Phi_1^\dagger \Phi_2 = \rho \, \ee^{\ii \eta} \frac{h_1 h_2}{2} \ .
\end{equation}
This reduces the number of field dimensions
that have to be considered
to five, parametrized by the five real parameters
$h_1$, $h_2$, $h_S$, $\rho$ and $\eta$.
Note here that the possible ranges of the
parameters are restricted:
$0 \leq \rho \leq 1$ and $-\pi \leq \eta \leq \pi$,
and $h_1$, $h_2$ and $h_S$ are non-negative.
Using these invariants, the quartic part of
the scalar potential is given by
\begin{align}
V_{\rm quartic} &=
\frac{\lambda_1}{8} h_1^4 +
\frac{\lambda_2}{8} h_2^4 +
\frac{\lambda_3}{4} h_1^2 h_2^2 +
\frac{\lambda_4}{4} h_1^2 h_2^2 \rho^2 +
\frac{\lambda_8}{8} h_S^4 +
\frac{\lambda_9}{4} h_1^2 h_S^2 +
\frac{\lambda_{10}}{4} h_2^2 h_S^2 \notag \\ &+
\frac{1}{4} h_1^2 h_2^2 \rho^2 (c_{2\eta} \lambda_5^{\rm Re}
  - s_{2\eta} \lambda_5^{\rm Im}) +
\frac{1}{2} h_1^3 h_2 \rho (c_\eta \lambda_6^{\rm Re}
  - s_\eta \lambda_6^{\rm Im}) \notag \\ &+
\frac{1}{2} h_1 h_2^3 \rho (c_\eta \lambda_7^{\rm Re}
  - s_\eta \lambda_7^{\rm Im}) +
\frac{1}{2} h_1 h_2 h_S^2 \rho (c_\eta \lambda_{11}^{\rm Re}
  - s_\eta \lambda_{11}^{\rm Im}) \ .
\end{align}
The scalar potential is unbounded if there are
values of the invariants for which $V_{\rm quartic} < 0$,
and it is bounded from below otherwise.
We numerically verified this by minimizing $V_{\rm quartic}$
and checking the sign of $V_{\rm quartic}$ at the
deepest minimum that is found. Since the possible
ranges of the invariants is finite, we used
a constrained minimization method called
bound optimization by quadratic
approximation~\cite{10.1093/comjnl/7.2.155}
as implemented in the public Fortran package
\texttt{PowellOpt}~\cite{powellopt}.
In order to improve the reliability of this
procedure, we applied the minimizer up to ten times
with different initial seed values for the invariants
for each considered parameter point.

\begin{figure}
\centering
\includegraphics[width=0.32\textwidth]{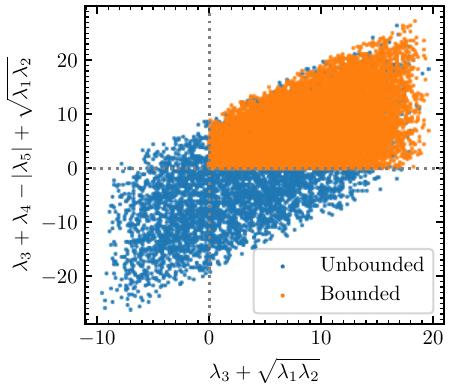}~
\includegraphics[width=0.32\textwidth]{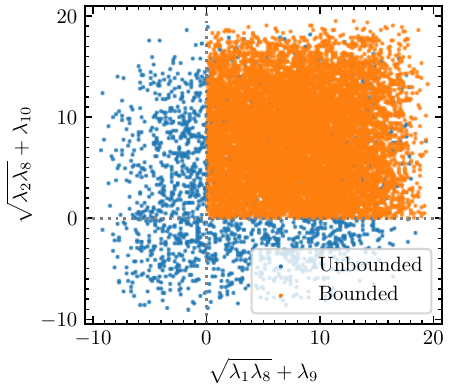}~
\includegraphics[width=0.32\textwidth]{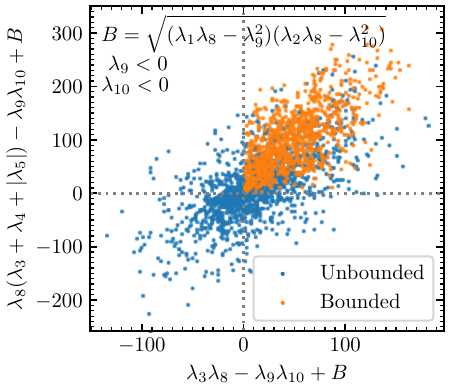}
\caption{Set of points without (with) vacuum stability constraints in blue (orange) in three different plane projections. The points are obtained in a limit where analytical conditions are available~\cite{Arhrib:2018qmw}. The plot shows that in this limit our strategy to obtain the conditions numerically works.}
\label{fig:bfb}
\end{figure}

In order to test the reliability of this numerical
approach, we generated parameter points in a
limiting case for which analytical conditions
are known to verify the boundedness of the
potential. In this limit all quartic
couplings $\lambda_i$ are set to be real, and we set
$\lambda_{6} = \lambda_7 = \lambda_{11} = 0$.\footnote{The
scalar potential then resembles
the one of the so-called Next-to
2HDM~(N2HDM)~\cite{Chen:2013jvg}.}
By scanning over the remaining $\lambda_i$
in a range from -10 to 10 (without considering
any other constraint), we randomly generated
10,000 parameter points that feature
a bounded scalar potential
according to our numerical procedure.
Subsequently, we generated 10,000 parameter
points in the same way that are predicted
to feature an unbounded potential.
The two sets of parameter points are shown in
\cref{fig:bfb}
in orange and blue colour, respectively,
in three different parameter planes.
The axes of the three plots
show different combinations of
the quartic scalar couplings that
according to~\citere{Arhrib:2018qmw} should
be positive if the potential is bounded.
One can observe that all orange points lie
within the upper right quadrant.
This demonstrates that our numerical method
is able to
reliably predict the boundedness of the
potential for all 10,000 parameter points
in agreement with the analytical conditions
in the considered limit.

\subsection{Perturbative Unitarity}
\label{sec:perturbativeunitarity}

Upper limits on the absolute values of the
quartic scalar couplings $\lambda_i$ can
be found
on the basis of unitarity.
To this end, we apply so-called perturbative
unitariy constraints at tree level
in the high-energy limit in which only
scalar self interactions have to be considered.
We require that the absolute values of
the eigenvalues of the $2 \to 2$ scalar
scattering matrices are smaller than $8 \pi$.
The scalar scattering matrices in the high-energy
limit
for neutral, charged and doubly-charged
final and initial states are matrices
of dimension 25, 12 and 3, respectively.
We do not display the elements of these
matrices here for brevity,
but they can be found in the
module \texttt{perturbative_unitarity}
of our code.

\begin{figure}
\centering
\includegraphics[width=0.48\textwidth]{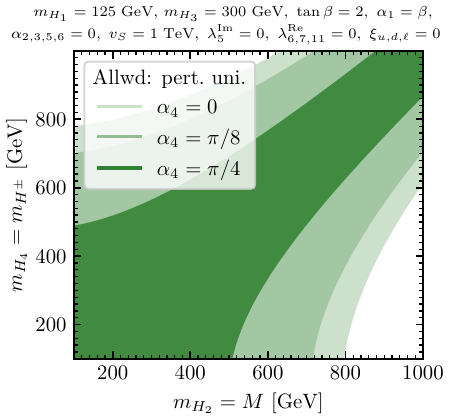}
\caption{Allowed parameter space after imposing perturbative unitarity constraints in a projection plane with $m_{H_2} = M$ on the horizontal axis and $m_{H_4} = m_{H^\pm}$ on the vertical axis, keeping all other parameters fixed as shown on top of the plot. Different tones of green correspond to three values of the CP-violating mixing angle$\alpha_4$.}
\label{fig:pertuni}
\end{figure}

The perturbative unitarity requirement
gives rise to upper limits on the
absolute values of the scalar quartic
couplings $\lambda_i$.
This, in turn, results in upper limits on
the mass splitting between neutral scalar states
that are mixed with each other, and also on the
splitting between the masses of the
scalars
and the two BSM mass scales
$M = m_{12}^2/(s_\beta c_\beta)$
and $v_S$ contained in the model.
In \cref{fig:pertuni} we show the impact of
the perturbative unitarity constraints in a
benchmark plane where we vary $m_{H_2} = M$
on the horizontal and $m_{H_4} = m_{H^\pm}$ on the
vertical axis, keeping all other parameters fixed
as shown on top of the plot.
Such a parameter plane is of interest in the context
of realizing a strong first-order EW phase transition,
in which the strengths of the transition is typically
correlated with sizable mass splitting between
two neutral scalar states~\cite{Dorsch:2014qja,
Basler:2016obg,Biekotter:2022kgf,
Biekotter:2023eil}.\footnote{Strong
first-order EW phase transition,
although of substantially
smaller strengths, can also be realized
without large mass splittings between the neutral
Higgs bosons~\cite{Dorsch:2017nza,Basler:2017uxn,
Biekotter:2025fjx}.}
The green shaded regions indicate the allowed parameter
regions for different values of the CP-violating mixing
angle $\alpha_4$. One can see that the possible amount
of mass splitting between $H_2$ and $H_4$ becomes
more constraint the larger the value of $\alpha_4$,
i.e.~the more CP-violating mixing is present
in the scalar sector (see also
Ref.~\cite{Basler:2019iuu} for a detailed
discussion).
Since the model studied here allows for
CP-violation without non-zero values for the
CP-violating mixing angles, it would be interesting
to investigate whether in this model
one can realize a first-order
EW phase transition in or close to the alignment limit
with a transition strength
and an amount of CP-violation that
are sufficient to account for the observed baryon
asymmetry of the universe via EW baryogenesis,
while at the same time being in agreement with
the experimental upper limits on the eEDM.
We leave this analysis for future work.

One should note that using the limit of $8\pi$ on
the absolute values of the eigenvalues
of the scattering matrices still allows
for large values of individual $\lambda_i$
couplings which (depending on the observables
of interest) may lie outside
of the region in which the model can
be treated perturbatively.
Additional constraints can be imposed
on the absolute values of the parameters
$\lambda_i$ themselves. A typical choice is
to require $|\lambda_i| < 4 \pi$ in order to
ensure perturbativity. It should be noted that
such conditions depend on the
numerical prefactors 
of the different terms
contained in the scalar potential, whereas
the perturbative unitarity conditions derived
from the scattering matrix are independent
of these prefactors.
Furthermore, such a limit can also be too
restrictive, excluding in principle allowed
parameter regions with potentially interesting
phenomenology. Thus, in order to be conservative
regarding the exclusion of parameter points,
and since we present a broad phenomenological study
of the model instead of focusing on a particular observable,
we do not include additional perturbativity
limits in our numerical analysis.

\subsection{Collider Constraints}

We now turn to the experimental constraints, starting
with the ones stemming from measurements at
colliders. These comprise the LHC cross-section measurements
of the detected Higgs boson at 125~GeV,
cross-section limits from searches for additional
Higgs bosons at LEP and the LHC,
measurements of flavour-physics observables
which are especially sensitive to the presence
of the charged Higgs bosons, and measurements of
electroweak precision observables~(EWPO).

\subsubsection{LHC Higgs-Boson Measurements}
\label{sec:h125}

After the LHC Run~2 at 13~TeV we have entered
the era of precision measurements of the
125~GeV Higgs boson. The Higgs boson has now
been produced and detected in several different
production and decay channels.
Models with additional Higgs bosons, as the one
considered here, have to accommodate a Higgs boson
at 125~GeV that predicts the measured signal rates
in agreement with the LHC measurements.
Since the measured cross sections are so far
compatible with the predictions for a Higgs boson
contained in the SM, extended Higgs sector models
have to feature a Higgs boson that approximately
resembles a SM Higgs boson.

In the model considered here, there are four
different neutral scalar states that can play
the role of the detected Higgs boson.\footnote{We
do not consider the possibility that there are
more than one Higgs boson with masses of about
125~GeV whose combined signal rates give rise
to the experimentally measured values at
the LHC.} We test whether a parameter point
is consistent with the current LHC data by
using 
\texttt{HiggsSignals}~\cite{Bechtle:2013xfa,
Bechtle:2014ewa,Bechtle:2020uwn}
contained in the software package
\texttt{HiggsTools v.1.2}~\cite{Bahl:2022igd}.
We perform a $\chi^2$ fit
comparing the theoretically predicted signal rates
(see \cref{sec:searches} for a discussion of the
theory predictions)
of the scalar at 125~GeV against the experimentally
measured values, taking into account total
cross section measurements and the
measurements published in the framework
of the simplified template cross sections.
We use the \texttt{HiggsSignals} data set
with version number~1.1
which contains
the currently most precise Run~2 measurements.
We reject a parameter point if the total
$\chi^2$ from \texttt{HiggsSignals} is larger
than the one obtained assuming a SM Higgs boson
by 6.18 units, i.e.~if the fit result of the
parameter point is worse compared to
the SM fit result at a confidence level of
more than $2\sigma$ in joint estimation of two parameters.

As already mentioned in \cref{sec:higgssector},
the LHC Higgs data drives the model towards
the alignment limit.
If this limit is imposed, the state $H_1$
will feature couplings that are identical
to the ones of a SM Higgs boson.
One way to enforce the
alignment limit is given in \cref{eq:alignlim}.
Applying the constraints from the Higgs-boson
measurements severely constraints the
allowed values of the six mixing angles
$\alpha_i$. Hence, a random scan over these
angles would be highly inefficient as it would
give rise to tiny fractions of allowed
parameter points.
Therefore, in our analysis we will often
start from a valid parameter point in the alignment
limit and then investigate to what extend
departures from the alignment limit are
still allowed. The modifications to the
properties of the state at 125~GeV compared
to the SM can arise in several
ways in our model, e.g.~from a misalignment
of the mixing of the doublet state away
from the so-called Higgs basis~\cite{Georgi:1978ri},
from a singlet admixture, or from CP-violating
effects that, as we remind the reader,
can be present in this model even
in the alignment limit.

\subsubsection{BSM Higgs-Boson Searches}
\label{sec:searches}

While a vast search programme for additional
scalar particles is ongoing at the LHC,
so far only one Higgs boson has been detected.
Hence, the BSM Higgs bosons predicted
by the model have to respect the cross-section limits
from the various collider searches that have been
performed at the LHC, but also at previous colliders
like LEP or Tevatron. We verify whether a parameter
point is in agreement with all existing cross section
limits by using 
\texttt{HiggsBounds}~\cite{Bechtle:2008jh,Bechtle:2011sb,
Bechtle:2013wla,Bechtle:2020pkv}
which is now incorporated in \texttt{HiggsTools.}
For the full set of experimental results
implemented in the \texttt{HiggsBounds} data set
with version number~1.4 we identify the most sensitive
search for each BSM Higgs boson based on the
expected experimental sensitivity. For the selected
searches, we apply the observed cross section limits
at 95\% confidence level.
If one or more of the BSM Higgs bosons features
a cross section that is larger than the experimentally
observed limit, we regard the corresponding parameter
point as excluded.

In order to use \texttt{HiggsBounds} (and
\texttt{HiggsSignals} for the test against the
LHC Higgs data of the detected Higgs boson)
one needs the cross sections and partial decay
widths of all the neutral and charged Higgs
bosons contained in the model.
We obtain the cross sections at the LHC and
LEP as functions of
effective coupling coefficients
by using the package \texttt{HiggsPredictions}
of \texttt{HiggsTools}
For the neutral scalars $H_i$, the effective
coupling coefficients
correspond to the so-called $\kappa$-factors which are
defined as the couplings normalized to
the prediction of a SM Higgs boson of the
same mass.
For the charged Higgs bosons, 
the process $pp \to H^\pm t b$ is currently
the only one for which LHC searches exist.
We compute the cross sections at 13 TeV
as functions of the couplings $t \bar b H^-$
and $\bar t b H^+$ using \texttt{HiggsPredictions}.
For the scalar DM state, we only take into account
its production via decays of the neutral scalars.
This can give rise to final states with large
missing transverse energy and/or mono-Higgs or
mono-jet final states.

We compute the partial decays of the Higgs bosons
using an extension of the public code
\texttt{HDECAY}~\cite{Djouadi:1997yw,Djouadi:2018xqq}.
This code was originally developed for the
SM and the minimal supersymmmetric extension of the SM and later extended to the 2HDM and other BSM models. We extended it to include the additional
particles and their interactions predicted
in the model considered here.
The resulting Fortran library \texttt{CS2HDECAY}
will be made publicly
available contained in the Fortran package
\texttt{cs2hdmTools}. With this 
package and its
interfaces to \texttt{HiggsTools} and
\texttt{C2SHDECAY} the parameter points of the
model can be readily confronted with the current
LHC measurements and cross section limits.

\subsubsection{Flavour-Physics Observables}

Flavour-physics observables tightly constrain models
with more than one Higgs doublet. In the SM FCNCs
are naturally suppressed, whereas
this suppression is lost in general under the
presence of a second Higgs doublet field.

In order to comply with the corresponding experimental
limits, we employ the flavour alignment in the
Yukawa sector as discussed in \cref{sec:yukawasector}.
Since the flavour alignment is not enforced by
a symmetry, small deviations from the flavour alignment
are generated at loop level that would ultimately
lead to FCNCs. However, it has been shown that these
radiative flavour-misalignment effects are
sufficiently small to
not be in conflict with existing
experimental limits on FCNCs~\cite{Gori:2017qwg}.

In the scalar sector,
\TB{FCNCs are generated at loop level even under the
assumption of flavour alignment}
if at least one of
the quartic scalar couplings that break the $\mathbb{Z}_2$
symmetry under which the doublet fields
transform as $\Phi_1 \to - \Phi_1$ and $\Phi_2 \to \Phi_2$
are non-zero. This mainly concerns the parameters
$\la_6$ and $\la_7$. Hence, in order to suppress FCNCs
we set $\lambda_6^{\rm Re} = \lambda_7^{\rm Re} = 0$
in our numerical analysis, and we note that their imaginary components
are typically suppressed because they violate CP.
A third $\mathbb{Z}_2$-breaking
term proportional to $\la_{11}$ is present that can
induce dangerous FCNCs at
\TB{loop level}. 
Since in the corresponding
potential term the singlet field $\Phi_S$ appears, its
contribution to FCNCs is milder, potentially only
relevant if the mixing between the
doublet fields and the singlet field are sizable
and the portal couplings are large.
As for $\la_6$ and $\la_7$, in a conservative 
approach
to avoid the appearance of 
FCNCs, we set
$\la_{11}^{\rm Re} = 0$. The imaginary part
$\la_{11}^{\rm Im}$ is a derived parameter in our
analysis and not generically zero, but its impact on
FCNCs is suppressed in a two-fold way due to the
CP breaking and the singlet admixture required
to induce FCNCs from this parameter.

Under the assumptions discussed above, the appearance
of FCNCs in this model reduces to the case of the
2HDM with softly-broken $\mathbb{Z}_2$ symmetry,
where the presence of the charged scalars
gives rise, most importantly, to radiative decays
of $B$ mesons. This leads to strong constraints
from measurements of $b \to s \gamma$ transitions,
giving rise to a lower limit on the mass of
the charged scalars of $m_{H^\pm} \gtrsim  500$~GeV
in the type~II 2HDM~\cite{Misiak:2017bgg,Misiak:2020vlo,
Biekotter:2024ykp}, 
and to a lower limit on
$\tan\beta$ that, depending on the value of $m_{H^\pm}$
ranges between $\tan\beta \gtrsim 2$ for
$m_{H^\pm} = 200$~GeV and
$\tan\beta \gtrsim 1$ for $m_{H^\pm} = 1$~TeV
for all Yukawa types.
Similar, but currently slightly weaker, limits on
$\tan\beta$ result from measurements of leptonic
$B$-meson decays, $B^0_s \to \mu^+ \mu^-$ and
$B^0 \to \mu^+ \mu^-$.
We consider these limits from the 2HDM in our
analysis whenever we consider values of the
flavour-alignment parameters that result
in one of the four Yukawa types of the
$\mathbb{Z}_2$-symmetric 2HDM, see
\cref{ta:flavalign}.
For all other cases, the application of experimental
bounds from flavour-physics observables requires
a dedicated computation which is beyond the scope
of this paper, and thus left for future work.

\subsubsection{Electroweak Precision Observables}

In the SM the $\rho$ parameter~\cite{Ross:1975fq}
\begin{equation}
\rho = \frac{M_W^2}{M_Z^2 c_w^2} \ ,
\end{equation}
with $c_w$ the cosine of the weak mixing angle,
is predicted to be equal to one at the
classical level as a consequence of the custodial
symmetry of the Higgs potential.
This prediction is confirmed by
experiment at the sub-permille
level~\cite{ParticleDataGroup:2022pth},
making the $\rho$ parameter one of the most
important probes of new physics.
In models with extended scalar sectors where
several scalar fields obtain a vev,
as considered here, the predictions for
the $\rho$ parameter are in general different
from one. However, in extensions of the SM
that contain additional scalar fields that
are either gauge-singlets or $SU(2)$ doublets
with hypercharge $\pm 1/2$, the SM prediction
$\rho = 1$ is maintained at the classical level,
and modifications to the $\rho$ parameter
are only generated via radiative corrections.
In this case, modifications to the $\rho$
parameter are typically parametrized in terms
of the parameter $\Delta \rho$, which is defined
by~\cite{Einhorn:1981cy}
\begin{equation}
\Delta \rho =
\left[
  \frac{\Pi_{WW}^T(0)}{M_W^2} - \frac{\Pi_{ZZ}^T(0)}{M_Z^2}
\right]_{\rm BSM} -
\left[
  \frac{\Pi_{WW}^T(0)}{M_W^2} - \frac{\Pi_{ZZ}^T(0)}{M_Z^2}
\right]_{\rm SM} \ ,
\end{equation}
where $\Pi_{WW}^T(0)$ and $\Pi_{ZZ}^T(0)$ are the
transverse parts of the $W$-boson and $Z$-boson self-energies
at zero external momentum, respectively, and the
first and the second bracket contain the self-energies in the BSM theory
under consideration and in the SM, respectively.
The parameter $\Delta \rho$ is related to the
oblique parameter $T$~\cite{Peskin:1990zt} via the relation
$\Delta \rho = \alpha T$, where $\alpha$ is the
fine-structure constant.

In our model the main modifications to the EWPOs
arise via weak isospin breaking generated at
loop-level, giving rise to non-zero values
of~$T$. As a consequence, the two most
precisely measured EWPO, the $W$-boson mass $M_W$
and the leptonic effective weak mixing angle, usually
denoted as $\sin^2 \Theta_{\rm eff}$, receive
corrections with respect to the SM prediction,
which can be used to constrain the parameter space
of the model. To this end, we compute $T$ at the
one-loop level and apply the limit from
the Gfitter collaboration
\cite{Haller:2018nnx}
that was obtained from a global fit
to various EWPOs.\footnote{More recent measurements
of the $W$-boson mass performed by
LHCb~\cite{LHCb:2021bjt},
ATLAS~\cite{ATLAS:2024erm} and
CMS~\cite{CMS-PAS-SMP-23-002} are
not contained in the Gfitter 2018 analysis.
However, these measurements are
in good agreement with the SM and the PDG average
value of $M_W$ on which the Gfitter 2018 result is
based. Hence, the new measurements performed at the LHC
would not yield significant modifications to
the allowed parameter space
regions discussed here. On the other hand, the CDF $M_W$
measurement published in 2022 is in significant tension
with the SM~\cite{CDF:2022hxs}.
While the large upwards shift
in $M_W$ in the direction of the CDF measurement can
be accommodated in
2HDM~\cite{Song:2022xts,Bahl:2022xzi}
and 2HDM + singlet
models~\cite{Biekotter:2022abc},
we do not consider this possibility here due to the
discrepancies between the CDF measurement and
the LHC (and LEP) measurements.}

\begin{figure}[t]
\centering
\includegraphics[width=0.48\textwidth]{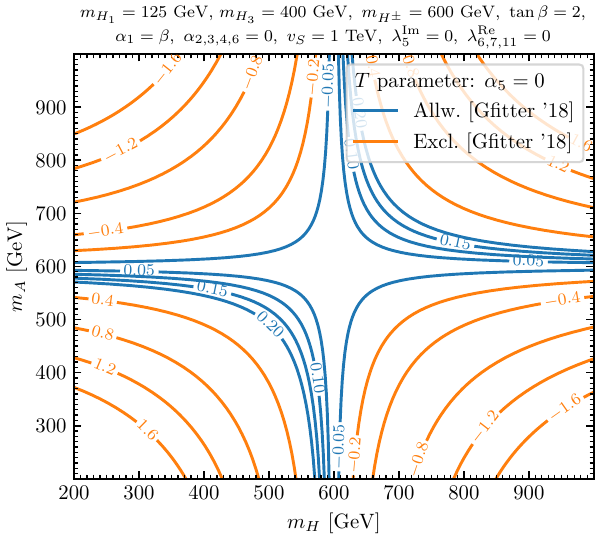}~
\includegraphics[width=0.48\textwidth]{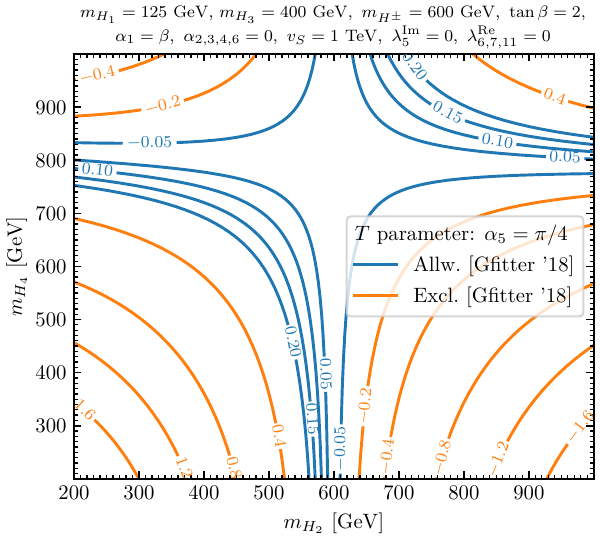}
\caption{Left: predictions for the $T$-parameter in
a CP-conserving scenario in the
$\{ m_H, m_A \}$ plane, where $m_H$ is the mass of the
second CP-even doublet Higgs state, and $m_A$ is the
mass of the CP-odd Higgs state. The mixing angles
$\alpha$ are set according to the alignment limit given
in \cref{eq:alignlim}, such that the singlet field
corresponding to the Higgs boson with a mass
of $m_{H_3} = 400$~GeV
is decoupled, and there is no CP-violating mixing
in the scalar sector. All other input parameters
are set as shown on top of the plot.
The contour lines indicate
fixed values of~$T$. Blue lines correspond to values
of $T$ that are compatible with the experimentally
allowed range from a global fit to EWPOs from
\citere{Haller:2018nnx}, see text for details,
while orange lines correspond to values of the
$T$-parameter that are outside of the experimentally
allowed range.
Right: the same as in the left plot, but
for a scenario with CP-violation via
a non-zero value of $\alpha_5 = \pi / 4$.
In this case, the CP-odd component of the Higgs
doublet fields mixes with the singlet scalar,
giving rise to two CP-mixed mass eigenstates
$H_3$ and $H_4$. Since in this case there is no
CP-odd eigenstate anymore, we change the axes labels
to $m_{H_2}$ and $m_{H_4}$ instead of
$m_H$ and $m_A$ used in the left plot.}
\label{fig:ewpo}
\end{figure}

In \cref{fig:ewpo} we show the predictions for
the $T$ parameter for two representative
parameter planes. In the left plot, we consider
the CP-conserving limit, and the mixing between
the singlet state and the doublet states
is zero. The neutral scalar spectrum consists of
a SM-like Higgs boson $H_1$, a second CP-even
Higgs boson $H$ and a CP-odd Higgs boson $A$
whose masses are varied between 200~GeV and 1~TeV,
and a singlet Higgs boson $H_3$ whose mass
is chosen to be 400~GeV. Finally, the mass of
the charged scalars is set to $m_{H^\pm} = 600$~GeV.
The remaining parameters are fixed as shown
above the plot. One can observe that the values
of $T$ in agreement with the experimental data,
indicated with blue contours, are confined to the
parameter space regions where either $H$ or $A$
is approximately mass degenerate with $H^\pm$.
This is the expected and known result from the
2HDM, which we here reproduce by decoupling
the singlet field.

This can be compared to the right plot
of \cref{fig:ewpo}, where we change the value
of one of the mixing angles from $\alpha_5 = 0$
(left plot) to $\alpha_5 = \pi / 4$ (right plot).
This gives rise to a mixing in the scalar
sector between the pseudoscalar
component and the singlet component. Accordingly,
we change the labels on the axes to $m_{H_2}$ and
$m_{H_4}$ since there are no CP eigenstates anymore.
As a consequence of the CP-violation, the experimentally
allowed region is shifted to larger values of
$m_{H_4}$. We observe that the $T$-parameter goes to zero if
$m_{H_2} = m_{H^\pm}$ (as before) and if
the mean of the two masses of $H_3$ and $H_4$
is approximately equal to the mass of
$m_{H^\pm}$, i.e.~$(m_{H_3} + m_{H_4}) / 2
\approx m_{H^\pm}$ (see also Ref.~\cite{Li:2025zga}).
In the latter case,
the $T$-parameter is predicted to be in
agreement with the experimental value
without the presence of approximately
mass-degenerate scalars, in contrast to
the CP-conserving limit.
This is an important difference between pure
2HDMs and models that in addition to a second
Higgs doublet contain one or more gauge singlet
fields with non-zero vevs.

\begin{figure}[t]
\centering
\includegraphics[width=0.54\textwidth]{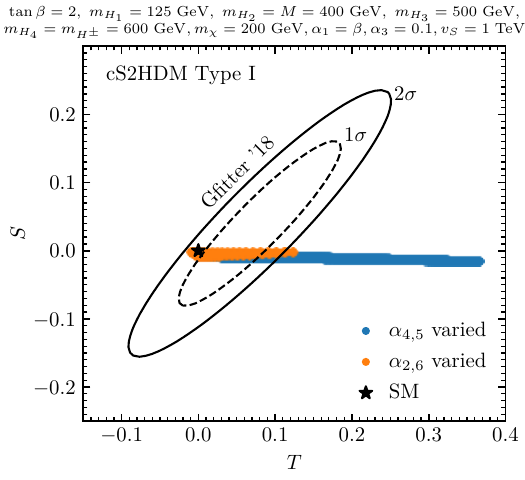}
\caption{\TB{Predictions for the $S$- and the $T$-parameters
for two scenarios with the parameters fixed as shown
on top of the plot. For the blue points, the
mixing angles $\alpha_4$ and $\alpha_5$ where varied
from $-\pi/2$ to $\pi / 2$,
with $\alpha_2 = \alpha_6 = 0$.
For the orange points, instead $\alpha_2$ and $\alpha_6$
were varied, while $\alpha_4 = \alpha_5 = 0$
(see main text for details).
The black dashed and solid contours indicate the
experimentally allowed regions in the $S$-$T$-plane
at $1\sigma$ and $2\sigma$ confidence level,
respectively~\cite{Haller:2018nnx}.
The black star indicates the SM prediction.}}
\label{fig:ewpo-s}
\end{figure}

\TB{
To assess the relevance of constraints from EWPOs
beyond custodial-symmetry breaking effects, we have
additionally evaluated the predictions for the $S$-parameter,
which is defined as~\cite{Peskin:1990zt}
\begin{align}
 \frac{\alpha}{4 s_w^2 c_w^2} S =
  \left[\frac{\Pi_{ZZ}(M_Z^2) - \Pi_{ZZ}(0)}{M_Z^2} -
  \left.\frac{\partial A_{\gamma\gamma}(q^2)}
   {\partial q^2}\right|_{q^2 = 0} +
  \frac{c_w^2 - s_w^2}{c_w s_w}
   \left.\frac{\partial A_{\gamma Z}(q^2)}
    {\partial q^2}\right|_{q^2 = 0}\right]_{\mathrm{BSM}}
  &- \notag \\
  \left[\frac{\Pi_{ZZ}(M_Z^2) - \Pi_{ZZ}(0)}{M_Z^2} -
  \left.\frac{\partial A_{\gamma\gamma}(q^2)}
   {\partial q^2}\right|_{q^2 = 0} +
  \frac{c_w^2 - s_w^2}{c_w s_w}
   \left.\frac{\partial A_{\gamma Z}(q^2)}
    {\partial q^2}\right|_{q^2 = 0}\right]_{\mathrm{SM}}& \, .
\end{align}
Here, $\Pi_{\gamma\gamma}$ and $\Pi_{\gamma Z}$ are the
transverse parts of the photon and mixed photon-$Z$ boson
self-energies, respectively, and $q$ is the external momentum.
The $S$-parameter measures new-physics contributions to the
momentum dependence of neutral-current interactions.
The cS2HDM predictions for the $T$- and the
$S$-parameters at one-loop level
are shown in \cref{fig:ewpo-s} for two
representative scenarios. In both cases, non-zero contributions
arise from mass splittings among the neutral BSM scalars,
if the state $H_4$, which is chosen to be mass degenerate
with $H^\pm$, acquires CP-even admixtures and thus deviates
from a purely CP-odd eigensate (see discussion above).

In the first scenario (blue points), we vary the mixing angles
$\alpha_4$ and $\alpha_5$ in the range
$[-\pi/2,\pi/2]$. Non-zero values of these two
mixing angles induce CP-even components in $H_4$ through
mixing with the doublet-like state $H_2$ and
the singlet-like state $H_3$.
The other free parameters of the cS2HDM are kept fixed to
the values shown on top of the plot. In particular,
the alignment limit is maintained by fixing $\alpha_1 = \beta$
and $\alpha_2 = \alpha_6 = 0$, such that the lightest
scalar $H_1$ behaves as the SM Higgs
boson for all blue points.
In the second scenario (orange points), we
instead vary $\alpha_2$ and $\alpha_6$, while fixing
$\alpha_4 = \alpha_5 = 0$. This generates a singlet
admixture and a CP-odd component in the lightest
state $H_1$. As a consequence, its couplings to fermions
and gauge bosons are modified compared to the ones of
the SM Higgs boson.
We therefore vary $\alpha_2$ and
$\alpha_6$ in the range in which the constraints
from LHC Higgs-boson cross section measurements are
satisfied, as discussed in \cref{sec:h125}.
The experimentally allowed regions at the $1\sigma$ and
$2\sigma$ confidence level are indicated by the dashed
and solid black contours, respectively~\cite{Haller:2018nnx}.
As can be seen, the predictions for $S$ and $T$ are
strongly correlated in extensions of the SM
with a second Higgs doublet such as the cS2HDM.
Values of $S$ that are compatible with the experimental
bounds on $T$ are in general significantly smaller than
the corresponding values of $T$. We stress that, while
showing here only the results for two specific scenarios,
we have verified within our numerical discussion
in \cref{sec:numdis} that this behaviour persists
across the viable parameter space of the model in our scan.
Consequently, we impose only the experimental constraint
on the $T$-parameter in the parameter scan
presented in \cref{sec:numdis}.}

\subsection{Dark-Matter Constraints}

The presence of the DM scalar in our model
imposes significant constraints on the parameter
space due to several factors.
First, the model must reproduce the experimentally
observed relic abundance of DM,
or at the very least, avoid predicting an
overabundance that exceeds current cosmological
limits. While direct detection experiments generally
place stringent bounds on the interaction
cross-section between the DM and nucleons,
these constraints are of minor relevance in our
model due to the cancellation of
DM-nucleon scattering, a consequence of the
pNG nature of the DM particle.
Finally, indirect detection constraints arise
from astronomical observations that are sensitive
to DM annihilation signals.

\subsubsection{Relic Abundance}
\label{sec:relic}

In our model the DM candidate $\chi$ is produced
in the early universe through the
thermal freeze-out mechanism.
To calculate the predicted DM relic abundance,
we created a model file for
\texttt{FeynRules v.12.2}~\cite{Christensen:2008py,
Alloul:2013bka}, which was then used to generate
the corresponding \texttt{CalcHEP}~\cite{Belyaev:2012qa}
files that are required
to implement the model in the public code
\texttt{micrOMEGAS v.5.3.41}~\cite{Belanger:2006is,Belanger:2018ccd}.
This allows us to compute the relic abundance, taking into
account all relevant DM annihilation processes
at leading order.
We confront the theoretically predicted values for the
relic abundance assuming the standard freeze-out
scenario, $(\Omega h^2)_{\mathrm{FO}}$, with
the experimental value 
$(\Omega h^2)_{\mathrm{Planck}} = 0.1200 \pm 0.0012$
obtained by the Planck collaboration
by measuring the power spectrum
of the cosmic microwave
background~\cite{Planck:2018vyg}.
We define the ratio
\begin{equation}
  \xi_{\mathrm{Planck}} =
  \frac
    {(\Omega h^2)_{\mathrm{FO}}}
    {(\Omega h^2)_{\mathrm{Planck}}} \ .
  \label{eq:relicratio}
\end{equation}
Parameter points predicting
$\xi_{\mathrm{Planck}} > 1$ overclose the universe
and are regarded as excluded by the measurement of the
relic abundance of DM (under the assumption of
a standard cosmological history and the freeze-out
mechanism).
Parameter points predicting $\xi_{\mathrm{Planck}}
\approx 1$ are allowed and predict the total amount
of the observed relic abundance of DM. Finally,
parameter points predicting $\xi_{\mathrm{Planck}}
\leq 1$ can be regarded as allowed, but
in order to fully account for the observed abundance
of DM additional components would be required that
are not contained in our model.

The presence of the pNG DM state requires
that the real component of $\Phi_S$ has
a non-zero vev $v_S$. This 
allows for
a mixing of the neutral components of the
Higgs doublets and the real singlet
component, see discussion in \cref{sec:higgssector}.
As a consequence, a phenomenologically viable
pNG DM state is only present if
the singlet field $\varphi_S$ mixes with
at least one of the other neutral fields
$\varphi_1$, $\varphi_2$ or $A_0$ from the
Higgs doublets.
If, on the other hand, the real
singlet component is not mixed with the
doublet fields, the singlet sector decouples
from the visible sector, and the portal couplings
required for the thermal freeze-out mechanism
vanish.

In the CP-conserving limit, the portal
interactions between the DM state $\chi$ and
the visible matter can be confined to the
CP-even part of the Higgs sector.
As a consequence, if the state $H_1$ corresponding
to the 125~GeV Higgs boson acts as a portal,
the state $H_1$ has to have a non-zero singlet admixture,
and the model predicts
modifications of the couplings of the Higgs boson
at~125~GeV with respect to the SM predictions.
This is an important difference to models like
the 2HDM+a in which the CP-even Higgs sector,
and thus the Higgs boson at~125~GeV, does not act
as a portal between the DM and the ordinary matter
in the annihilation processes, such that
no departures from the alignment limit are required.

\begin{figure}[t]
\centering
\includegraphics[width=0.48\textwidth]{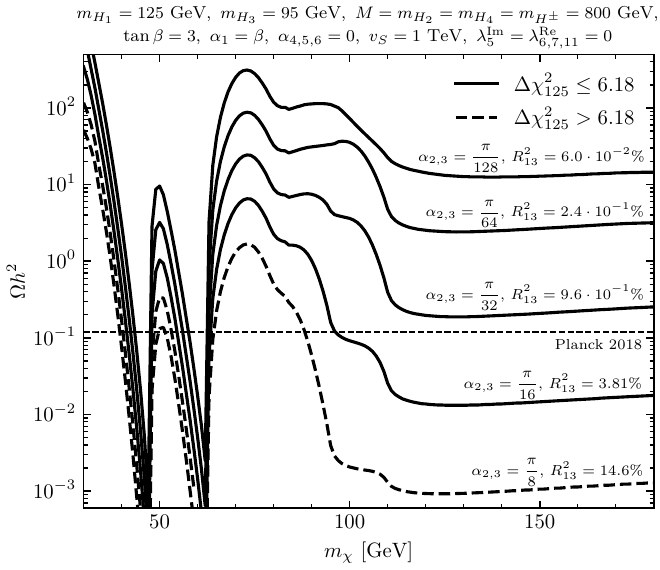}~
\includegraphics[width=0.48\textwidth]{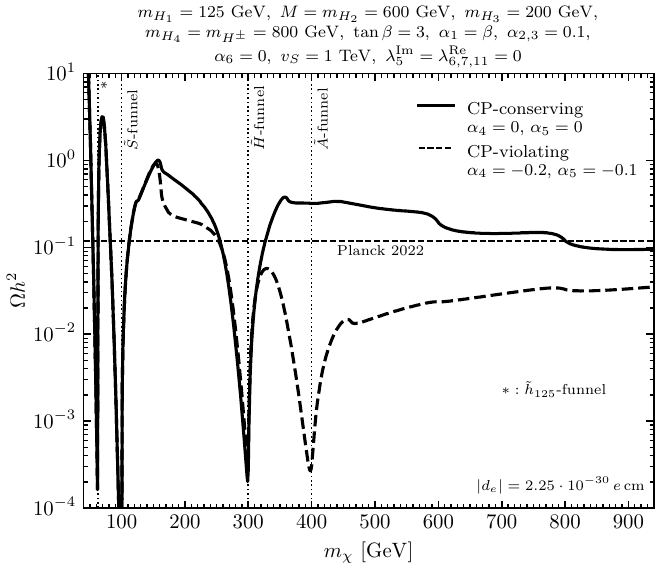}
\caption{Left: Predicted relic abundance
in the CP-conserving limit
with Yukawa type~I with a SM-like
Higs boson at $m_{h_1} = 125.1\gev$,
a singlet-like Higgs boson at
$m_{h_3} = 95\gev$, and the other doublet-like
Higgs-bosons at
$M = m_{h_2} = m_{h_4} = m_{H^\pm} = 800\gev$.
The remaining free parameters are set
as follows: $\tan\beta = 3$, $\alpha_1 = \beta$,
$\alpha_{4,5,6} = 0$, $v_S = 1000\gev$,
$\lambda_5^{\rm Im} = \lambda_6^{\rm Re} =
\lambda_7^{\rm Re} = \lambda_{11}^{\rm Re} = 0$,
and the mixing angles $\alpha_2 = \alpha_3$
are varied as shown in the plot.
The values of $R_{13}^2$ shown in the
plot are the singlet-admixture of the
SM-like Higgs boson~$h_1$. The solid lines
correspond to the regions of the DM mass
$m_\chi$ compatible with the LHC Higgs-boson
measurements for a given value of
$\alpha_2 = \alpha_3$, whereas the dashed
lines indicate regions that are excluded,
see \cref{sec:h125} for details.
The horizontal dashed line shows the experimental
value of the DM relic abundance measured
by Planck~\cite{Planck:2018vyg}.
Right: Predicted relic abundance as a function
of $m_\chi$ in a scenario with CP-violation
($\alpha_4 = -0.2$, $\alpha_5 = \TB{-0.1}$)
and without ($\alpha_4 = \alpha_5 = 0$)
CP-violation indicated with the solid and dashed
lines, respectively. All other parameters are
set as shown on top of the plot. The value of the
electron EDM $|d_e|$ shown in the lower right
corner is the one predicted for the parameter
points corresponding to the dashed line, independently
of the DM mass $m_\chi$, see discussion in
\cref{sec:edms}. The horizontal dashed line
shows the experimental
value of the DM relic abundance measured
by Planck, and the vertical dashed lines indicate
the mass of the particle that is responsible
for the main annihilation channel in the funnel regions.}
\label{fig:relic-align}
\end{figure}

To quantify the coupling modifications of the 125~GeV
Higgs boson that are required for
a prediction of the
DM relic abundance equal or below
the Planck measurement in a CP-conserving scenario
in which the 125~GeV Higgs boson acts as a portal,
we show in the left plot of
\cref{fig:relic-align} the predicted
values of the relic abundance
$\Omega h^2$ as a function of the DM
mass $m_\chi$ for a representative
benchmark point with a singlet-like
Higgs boson lighter than 125~GeV. Each line corresponds to
different values of the mixing angles
$\alpha_2$ and $\alpha_3$, where all other
parameters were fixed to the values shown
on top of the plot. Since $\alpha_{4,5,6} = 0$,
in the limit
$\alpha_{2,3} \to 0$ the singlet decouples
and the portal couplings vanish,
while for increasing values of
$\alpha_2$ and $\alpha_3$
the mixing between
the Higgs boson $H_1$ at 125~GeV and the
singlet-like Higgs boson $H_3$ at~95~GeV
increases. According to the discussion above, we observe that
the predicted relic abundance
becomes overall smaller when the mixing angles are
increased.
For the upper three lines, where the singlet
admixture of $H_1$, given by $R_{13}^2  =
\sin^2\alpha_2$ here (see \cref{eq:R13}), is below
one percent, the predicted abundance of DM
is larger than the value observed by
the Planck collaboration. As a result,
the corresponding parameter points are excluded
except for the \textit{funnel} regions
$m_\chi \approx m_{H_1} / 2$ and $m_\chi \approx m_{H_3} / 2$ 
where the annihilation is resonantly enhanced.
On the other hand, for larger singlet admixtures
in the SM-like Higgs boson of $R_{13}^2 = 3.81\%$ and
$R_{13}^2 = 14.6\%$ (the lowest two lines), the predicted
abundance is below the Planck value across most
of the DM mass interval depicted in
\cref{fig:relic-align}.
However, if the singlet admixture becomes too
large, the coupling modifications of the 125~GeV Higgs boson
are not in agreement with the signal-rate
measurements from the LHC. In the plot, we indicate the mass
intervals of $\chi$ which are incompatible with the
Higgs data at more than $2\sigma$ confidence level
(see discussion in \cref{sec:h125}) with dashed lines.
The lowest line with a singlet component of about
15\% in 125~GeV Higgs boson is shown entirely as a dashed line,
indicating that all parameter points  with
$\alpha_{2,3} = \pi / 8$ are excluded. In addition,
the line for $\alpha_{2,3} = \pi / 16$, corresponding
to a singlet component of about 4\%, is excluded for
DM masses of $m_\chi \lesssim 62$~GeV, where the invisible
decay mode of the 125~GeV Higgs boson gives rise to
an additional modification to the signal rates, rendering
them incompatible with the LHC measurements.
Assuming the absence of CP-violation in the Higgs sector,
we conclude that large parts of the parameter space
of the model require coupling modifications of the
detected Higgs boson of a few percent in order to be
compatible simultaneously with LHC Higgs-boson measurements
and the Planck measurement of the DM relic abundance
(taken as upper limit), apart from corners of the parameter
space with tuned DM mass values
in which the DM annihilation is resonantly enhanced.

This picture changes if CP-violation in the Higgs sector
is present, which is demonstrated
for an example scenario
in the right plot of \cref{fig:relic-align}.
Here, we set $\alpha_2 = 0.1$ and $\alpha_6 = 0$, such that
the singlet admixture of the 125~GeV Higgs boson is
$R_{13}^2 = 1\%$, which is substantially below the current
and near future
experimental sensitivity of the LHC to modifications of
the Higgs-boson signal rates, rendering the scenario well
compatible with the LHC data.
CP-violation in the Higgs sector is introduced with
non-zero values of the two mixing angles $\alpha_4$
and $\alpha_5$ (dashed line), whereas there is no
CP-violation if these angles are zero (solid line).
The specific values $\alpha_4 = -0.2$ and
$\alpha_5 = \TB{-0.1}$ 
were chosen
in order to obtain a prediction for the
eEDM of $|d_e| = \TB{2.25 \cdot 10^{-30}}~e\,$cm
that is
well below the current experimental upper bounds,
see the discussion in \cref{sec:edms}.
All other free parameters are set to the values shown
above the plot.
We depict the predicted DM relic abundance
for both cases
($\alpha_{4,5} = 0$ and $\alpha_{4,5} \neq 0$)
as a function of the DM mass $m_\chi$.
One can see that in \TB{large} 
parts of the depicted
range of $m_\chi$ both lines approximately lie
on top of each other.
However, at $m_\chi \approx m_{H_4} / 2 \approx
400$~GeV, the dashed line drops by 
\TB{several orders} of
magnitude in $\Omega h^2$ below the solid line.
Here, the non-zero values of $\alpha_4$ and
$\alpha_5$ give rise to a non-zero coupling
for $\chi\chi H_4$ interactions, and the DM
annihilation is resonantly enhanced
via $s$-channel annihilation with a dominantly
CP-odd state $H_4$ in the $s$-channel.
On the other hand, without CP-violation
the coupling $\chi \chi H_4$, with $H_4$ being
a pure CP-odd state, is absent, and the solid line
remains flat in the ``$\tilde{A}$-funnel'' region
indicated in the plot, where the tilde in
the notation $\tilde A$ indicates that for
$\alpha_4 \neq 0$ and $\alpha_5 \neq 0$
the state is only
an approximate CP-odd state.
This demonstrates a generic feature of the considered
model: under the presence of CP-violation, the
DM-portal interactions can be governed by the
coupling of the DM state to a mostly CP-odd state,
in which case the DM relic abundance can be
accommodated in agreement with the Planck measurement
without sizable modifications to the properties
of the 125~GeV Higgs boson with respect to the
SM predictions. Thus, allowing for CP-violating
phases in the scalar sector significantly widens
the allowed parameter space of the model compared
to the case in which no CP-violating is considered.

We end this section by mentioning
that in regions of the parameter space where the cS2HDM
predicts a relic density below the value measured
by Planck, there is room for an extended
dark sector that could provide the correct abundance.
For instance, as discussed in Ref.~\cite{Capucha:2024oaa},
one possibility is to add a second singlet field
stabilized by an independent $\mathbb{Z}_2$
symmetry, which couples only feebly to the
visible sector. In this case the additional singlet
can act as a freeze-in DM candidate, while leaving
the collider and Higgs-sector phenomenology of
the cS2HDM essentially unaffected.

\subsubsection{Direct Detection}
\label{sec:directdetection}

\begin{figure}
\centering
\includegraphics[height=5cm]{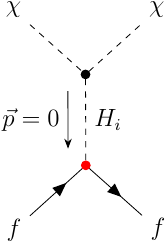}
\hspace*{2cm}
\includegraphics[height=5cm]{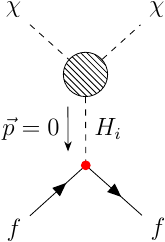}
\hspace*{2cm}
\includegraphics[height=5cm]{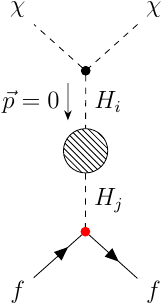}
\caption{Feynman diagrams for the scattering
process $\chi f \to \chi f$ at the classical
level (left) and at the quantum level, with
vertex corrections (center) and propagator
corrections (right). The red vertex is in
general CP violating. The hashed blobs
represent loop corrections that give
rise to non-vanishing amplitudes
in the limit of zero momentum transfer
$\vec p = 0$.}
\label{fig:dddiags}
\end{figure}

Due to the pNG-nature of the DM
state~$\chi$, the scattering process
$\chi f \to \chi f$ is absent at classical
level in the limit of zero momentum
transfer.
This is a consequence of a cancellation between
the $t$-channel scattering amplitudes
with neutral-scalar exchange corresponding
to the Feynman diagrams shown on the
left-hand side of \cref{fig:dddiags}.
\TB{This cancellation becomes manifest
by expressing
the $H_i \chi \chi$ couplings as shown on the
right-hand side of \cref{eq:chichihcpl}
and exploiting the
orthogonality of the scalar mixing matrix~$R$.}
As a result, the scattering rate at DM
direct-detection experiments is strongly
suppressed.
However,
the cancellation mechanism does not
hold at the quantum level
since the global $U(1)$
symmetry is softly broken in order
to give a finite mass to~$\chi$,
see the term proportional to
$m_\chi^2$ in the scalar potential
shown in \cref{eq:scapot}.
The potentially numerically important
contributions arise at one-loop level
from diagrams in which $\chi$ appears
as virtual particle
either in the scalar trilinear vertex
or in the $t$-channel propagator,
see the two diagrams on the right-and
side of \cref{fig:dddiags}.
Box contributions where two Higgs bosons
are exchanged between the DM state $\chi$
and the fermion line are highly suppressed
by additional factors of first- and
second-generation quark Yukawa couplings
and are therefore neglected here,
see also the discussions in \citeres{Azevedo:2018exj,
Glaus:2020ihj,Biekotter:2022bxp}. 

In order to confront the model with
experimental limits from DM direct-detection
experiments, we computed the scattering
amplitude for the process $\chi f \to \chi f$,
where the fermion $f$ can be either an up-type
quark $u$ or a down-type quark $d$,
at the one-loop level in the limit of
vanishing momentum-transfer.\footnote{The contributions
from finite-velocity effects are several orders
smaller than the quantum corrections assuming
cold, i.e.~non-relativistic, DM~\cite{Azevedo:2018exj}.
Thus, they are not taken into account here.}
Under the presence of CP-violation, we have to consider
also spin- and momentum-dependent operators in
the non-relativistic effective theory that
describes the DM-nucleon scattering.
In the UV theory, the loop diagrams shown in
\cref{fig:dddiags} generate four
dimension-6 operators
\begin{equation}
\mathcal{L}_{\rm UV} = \sum_q \left(
\hat{\mathcal{C}}^{(6)}_{3,q} \mathcal{O}_{3,q}^{(6)} +
\hat{\mathcal{C}}^{(6)}_{4,q} \mathcal{O}_{4,q}^{(6)}
\right) + 
\hat{\mathcal{C}}^{(6)}_{5,g} \mathcal{O}_{5}^{(6)} +
\hat{\mathcal{C}}^{(6)}_{6,g} \mathcal{O}_{6}^{(6)}
\, ,
\label{eq:lagdduv}
\end{equation}
with the CP-even operators
\begin{equation}
\mathcal{O}_{3,q}^{(6)} =
m_q (\chi \chi) (\bar q q) \qquad \textrm{and} \qquad
\mathcal{O}_{5,g}^{(6)} =
\frac{\alpha_s}{12 \pi} (\chi \chi) G^{a \mu\nu} G_{\mu\nu}^a
\, ,
\label{eq:relopseven}
\end{equation}
and the CP-odd operators
\begin{equation}
\mathcal{O}_{4,q}^{(6)} =
m_q (\chi \chi) \ii (\bar q \gamma_5 q) \qquad \textrm{and} \qquad
\mathcal{O}_{6,g}^{(6)} =
\frac{\alpha_s}{8 \pi} (\chi \chi) G^{a \mu\nu} \widetilde G_{\mu\nu}^a
\, ,
\end{equation}
and where we follow the notation and conventions
of \citere{Bishara:2017pfq}.
We make use of the three-flavour scheme
in which the sum in \cref{eq:lagdduv} runs over
the light quark flavours $q=u,d,s$. $m_q$ is the
mass of the light quark $q$, $\alpha_s$ is the strong
coupling constant at hadronic scales $\mu \approx 1$~GeV,
and $G^{a \mu\nu}$ and
$\widetilde G^{a \mu\nu}$ are the gluon field strength
tensor and its dual, respectively.

We compute the Wilson coefficients 
$\hat{\mathcal{C}}_{3,q}^{(6)}$
and $\hat{\mathcal{C}}_{4,q}^{(6)}$
at one-loop level taking into account
the set of diagrams depicted generically
in the middle and right diagrams shown
in \cref{fig:dddiags} with $f = q$ and factoring out a factor
of $m_q$. The resulting expressions are
UV-finite without renormalization
and thus also independent of the renormalization scale.
We will present a detailed discussion
about this loop computation in an
accompanying paper.\footnote{The
computations for the scattering rates of pNG DM
on nucleons at the one-loop level can be found
for the SM+singlet model in
\citeres{Ishiwata:2018sdi,Azevedo:2018exj,Glaus:2020ihj} and
for the CP-conserving 2HDM+singlet model
in \citere{Biekotter:2022bxp}, respectively.}

The 
effective operators $\mathcal{O}_{5,g}^{(6)}$
and $\mathcal{O}_{6,g}^{(6)}$
account for the
interaction of $\chi$ with gluons
which is generated
via the exchange of the neutral scalars $H_i$
between $\chi$ and the heavy quarks $Q$ contained
as virtual particles in the nucleons.
In the so-called heavy quark
expansion~\cite{Witten:1975bh} one
can integrate
out the heavy quarks, giving rise to an effective
gluon interactions that can be determined via the
matching relations~\cite{Fan:2010gt}
\begin{equation}
m_Q (\bar Q Q) \rightarrow - \frac{2}{3} \frac{\alpha_s}{8 \pi}
  G^{a \mu\nu} G_{\mu\nu}^a \qquad \textrm{and} \qquad
 m_Q \ii (\bar Q \gamma_5 Q) \rightarrow - \frac{\alpha_s}{16 \pi}
   G^{a \mu\nu} \widetilde G_{\mu\nu}^a \, ,
\end{equation}
where $m_Q$ denotes the
mass of the heavy quark. Comparing to \cref{eq:lagdduv},
we therefore obtain the Wilson
coefficients for the effective gluon interaction via
\begin{equation}
\hat{\mathcal{C}}_{5,g}^{(6)} =
  \sum_Q \hat{\mathcal{C}}_{3,Q}^{(6)}
\qquad \textrm{and} \qquad
\hat{\mathcal{C}}_{6,g}^{(6)} =
  \frac{1}{2} \sum_Q \hat{\mathcal{C}}_{4,Q}^{(6)}
\, .
\end{equation}
where the sum runs over the heavy quark
flavours $Q=t,b,c$. 
Here $\hat{\mathcal{C}}_{3,Q}^{(6)}$ and
$\hat{\mathcal{C}}_{4,Q}^{(6)}$ are the Wilson coefficients
of the operators $\mathcal{O}_{3,Q}^{(6)}$
and $\mathcal{O}_{4,Q}^{(6)}$, respectively.
Similarly to $\hat{\mathcal{C}}_{3,q}^{(6)}$
and $\hat{\mathcal{C}}_{4,q}^{(6)}$,
these are given by the CP-even
and CP-odd components of the amplitude
resulting from the loop diagrams shown in
\cref{fig:dddiags} with $f = Q$, respectively.

With the Wilson coefficients in the UV theory
at hand, one can compute the scattering rates
for different DM direct detection experiments.
In the absence of CP-violation, one finds
$\hat{\mathcal{C}}_{4,q}^{(6)} = \hat{\mathcal{C}}_{6,g}^{(6)} = 0$,
and only spin-independent scattering cross sections
are generated which for a given nucleon $N = p,n$,
with $p$ and $n$ being the proton and the
neutron, respectively,
can be written as~\cite{Cerdeno:2010jj}
\begin{equation}
\sigma_{\rm SI}^N = \frac{1}{\pi}
  \frac{m_N^4}{(m_N + m_\chi)^2}
  \left|
    \sum_q F_S^{q/N} \hat{\mathcal{C}}_{3,Q}^{(6)} 
    + \frac{2}{27} F_G^N \hat{\mathcal{C}}_{5,g}^{(6)}
  \right|^2 \, ,
  \label{eq:ddcseven}
\end{equation}
where $m_N$ is the nucleon mass, and where the
hadronic form factors $F_S^{q/N}$ for scalar
currents have
to be extracted from experimental data or derived
using lattice simulations in QCD, see discussion
below, and $F_G^N \approx 1 - \sum_q F_S^{q/N}$.
The predicted spin-independent nucleon scattering cross
section $\sigma_{\rm SI}^N$ can then
readily be confronted with
the upper limits reported by the various
DM direct-detection experiments.

\begin{figure}
\centering
\includegraphics[width=0.48\textwidth]{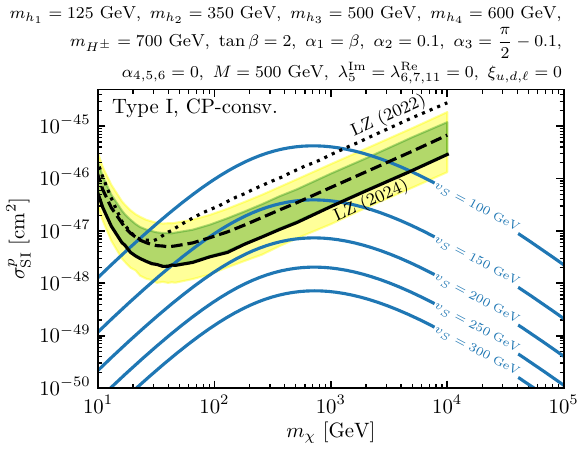}~
\includegraphics[width=0.48\textwidth]{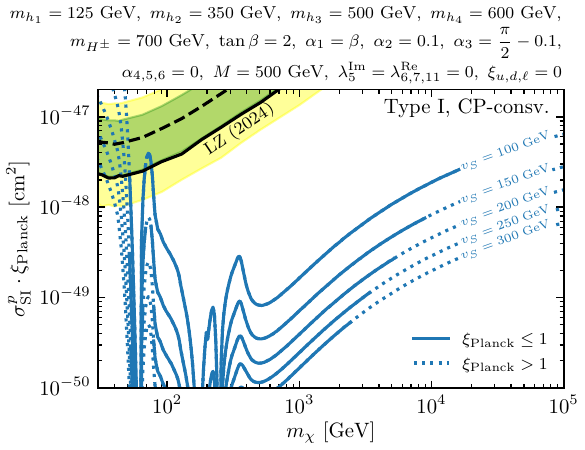}
\caption{Prediction for the spin-independent DM-proton
scattering cross section in the absence
of CP-violation in a benchmark scenario
taken from \citere{Biekotter:2022bxp}.
In the right plot, the predicted values
of the scattering cross section are rescaled
with the ratio $\xi_{\mathrm{Planck}} =
(\Omega h^2)_{\mathrm{FO}} /
(\Omega h^2)_{\mathrm{Planck}}$,
where $(\Omega h^2)_{\mathrm{FO}}$ is the
predicted DM relic abundance
assuming standard freeze-out mechanism and
$(\Omega h^2)_{\mathrm{Planck}} $
is the experimental value
measured by Planck~\cite{Planck:2018vyg}.
If the blue lines are displayed as dashed lines,
then the parameter points predict
$\xi_{\mathrm{Planck}} > 1$ and are excluded
based on the measured value of the DM relic abundance.
Also shown are the currently strongest
experimental expected and observed
90\% confidence-level limits from the
LZ experiment with the black dashed and
solid line, respectively~\cite{LZCollaboration:2024lux}.
The green and
yellow shaded regions indicate the
$1\sigma$ and $2\sigma$ uncertainty bands,
respectively. The left plot also shows
with the dotted black line the previous LZ limit
published in 2022~\cite{LZ:2022lsv}.}
\label{fig:ddcpcons}
\end{figure}

In \cref{fig:ddcpcons} we show a comparison between
the predicted values for the spin-independent
DM-proton scattering
cross section $\sigma_{\rm SI}^p$ as a function
of the DM mass $m_\chi$ and for different
values of $v_S$ in a CP-conserving
benchmark scenario taken from \citere{Biekotter:2022bxp}
for which the expression given in
\cref{eq:ddcseven} can be applied.
We verified that this scenario satisfies the
theoretical constraints from the boundedness of
the scalar potential and from perturbative unitarity.
In the left plot of \cref{fig:ddcpcons} we validate
the result from \citere{Biekotter:2022bxp}, demonstrating
that the scattering rates generated at one-loop level
can be of the order of the currently strongest
experimental limits from the LZ
collaboration~\cite{LZ:2023lvz}
for values of $v_S$ substantially below the
masses of the BSM neutral Higgs bosons, where
the scalar couplings related to the
singlet field $\lambda_8$, $\lambda_9$,
$\lambda_{10}$ and $\lambda_{11}$ become sizable,
see \cref{eq:lam8,eq:lam9,eq:lam10,eq:lam11}.
On the other hand, for values of $v_S \gtrsim m_{H_2},
m_{H_3},m_{H_4}$ the predicted scattering cross sections
rapidly fall below the current and near future
experimental sensitivity of DM direct detection
experiments.

The same scalar portal couplings governing
the scattering amplitudes are also responsible
for the DM annihilation process in the early
universe that determines the predicted relic
abundance of DM.
As a result,
for pNG DM 
the parameter space regions that can potentially
be probed by direct detection experiments typically
predict a significant under-abundance of
DM, i.e.~$\xi_{\rm Planck} \ll 1$ as defined
in \cref{eq:relicratio}.
In this case, the
reduced density of DM particles that reach the
detector, and thus the expected number of
scattering events, are suppressed by factors
of $\xi_{\rm Planck}$. Since the experimental
collaboration assume the measured value of
the DM relic abundance when setting limits,
one can account for the
impact of a smaller DM density
by comparing the product
$\sigma_{\rm SI}^p \cdot \xi_{\rm Planck}$ to
the experimental limit on $\sigma_{\rm SI}^p$.
This is what we show in the right plot
of \cref{fig:ddcpcons} with $\xi_{\rm Planck}$
computed using
\texttt{micrOMEGAS}~\cite{Belanger:2018ccd}
(see the discussion in \cref{sec:relic})
and assuming the standard freeze-out mechanism.
Here the lines indicate the value of
$\sigma_{\rm SI}^p \cdot \xi_{\rm Planck}$ for
the same parameter point shown in the left plot.
The intervals in which solid lines are displayed
correspond to values of $\xi_{\rm Planck} \leq 1$,
and dashed lines are used for
$\xi_{\rm Planck} > 1$.
In the latter case, the corresponding
parameter points are excluded because the
DM relic abundance is predicted to be too large.
One can see that now the prospects for
the direct detection of~$\chi$ are severely
reduced. Conversely, a hypothetical detection
of DM at direct detection experiments currently
in operation would be incompatible with the depicted
benchmark scenario for all values of~$v_S$ except
for $v_S = 100$~GeV and a DM mass of
$m_\chi \approx 80$~GeV.

So far we have focused on a CP-conserving scenario.
However, the cancellation mechanism suppresses
the tree-level contribution to the DM-nucleon scattering
irrespective if explicit sources of CP-violation
are present in the scalar potential or the Yukawa sector,
in contrast to the well studied 2HDM+a DM model.
Under the presence of
CP-violating parameters,
the comparison to the experimental data from
DM direct detection experiments becomes
significantly more involved because also
momentum-dependent and spin-dependent
interactions between the DM state $\chi$ and
the nucleons are generated.
In order to take into account these effects,
it is required to consider additional operators
in the non-relativistic effective description
of the DM-nucleon scattering.
To this end, we perform a
matching of the Wilson coefficients in the
UV theory shown in \cref{eq:lagdduv} to the relevant
set of Galilean-invariant operators in the
non-relativistic effective field theory describing
the interaction between $\chi$ and
nucleons~\cite{Bishara:2017pfq},
\begin{equation}
\mathcal{L}_{\rm NR} = c_1^N \mathcal{O}_1^N +
  c_{10}^N \mathcal{O}_{10}^N \qquad \textrm{with} \qquad
\mathcal{O}_1^N = \mathbbm{1}_N \qquad \textrm{and} \qquad
\mathcal{O}_{10}^N = - \ii \left( \vec{S}_N \cdot
  \frac{\vec{q}}{m_N} \right)
\, ,
\end{equation}
where $N=p$ for protons and $N=n$ for neutrons.
Here, $c_1^N$ and $c_{10}^N$ are the Wilson coefficients
of spin-independent and spin-dependent interactions
contained in the operators $\mathcal{O}_1^N$
and $\mathcal{O}_{10}^N$, respectively.\footnote{The lower
index of the non-relativistic
operators $\mathcal{O}_{i}^N$ is a naming convention.
In total there are ten different operators
that can be obtained in a
non-relativistic reduction of
a manifestly relativistic operator basis
up to second order in $\vec p$
if one considers only
spin-0 or spin-1 mediators~\cite{Anand:2013yka}.}
$\vec{S}_N$ denotes the spin of the nucleon $N$,
and $\vec{q}$ is the three-momentum vector of
the exchanged momentum in the $\chi$-$N$ scattering.
No other operators are generated in the
non-relativistic effective field theory at leading order
in the chiral expansion from the operators present
in our model in the UV shown in
\cref{eq:lagdduv}.
The Wilson coefficients are in general momentum dependent,
but working in the limit of vanishing momentum exchange
we can perform the matching between the UV theory
and the non-relativistic EFT to leading order in chiral
counting. One finds the relations~\cite{Bishara:2017nnn}
\begin{align}
c_1^N = \frac{1}{2 m_\chi}
  \left( \sum_q F_S^{q/N} \hat{\mathcal{C}}_{3,q}^{(6)}
  + F_G \, \hat{\mathcal{C}}_{5,g}^{(6)}
  \right) 
  \, , \\
c_{10}^N =  \frac{1}{2 m_\chi}
  \left( \sum_q F_P^{q/N} \hat{\mathcal{C}}_{4,q}^{(6)}
  + F_{\widetilde{G}} \, \hat{\mathcal{C}}_{6,g}^{(6)}
  \right)
  \, ,
\end{align}
where $F_S^{q/N}$, $F_P^{q/N}$, $F_G$ and
$F_{\widetilde{G}}$ are hadronic form factors
whose values can be computed in lattice QCD.
We use the numerical values summarized in
\citere{Bishara:2017nnn} using lattice QCD
results from \citeres{Junnarkar:2013ac,Yang:2015uis,
Durr:2015dna}.
The Wilson coefficients for the scattering on
protons ($N = p$) and neutrons ($N = n$) can be decomposed into
isoscalar ($s$) and isovector ($v$) components,
\begin{equation}
 c_i^s = \frac{1}{2} (c_i^p + c_i^n) \, , \qquad
 c_i^v = \frac{1}{2} (c_i^p - c_i^n) \, ,
 \label{eq:wilson_cs_cv}
\end{equation}
which are often used as independent interactions in
experimental analyses.

\begin{figure}[t]
\centering
\includegraphics[width=0.9\textwidth]{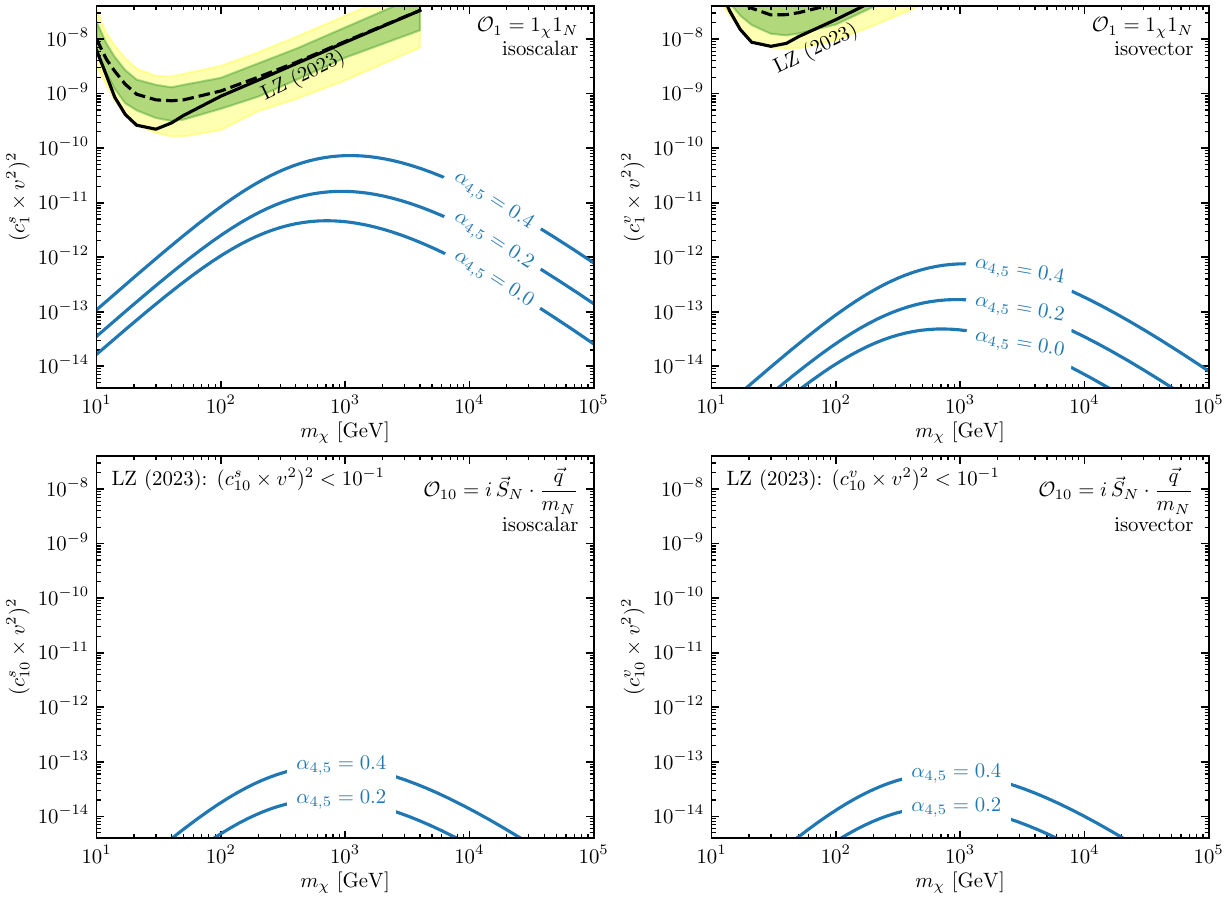}
\caption{Predicted values of the squared
Wilson coefficients $c_1^s$, $c_1^v$, $c_{10}^s$
and $c_{10}^v$ as defined in
\cref{eq:wilson_cs_cv} as a function of the DM mass
$m_\chi$ in the top left, top right, bottom left
and bottom right plot, respectively, normalized
to $1 / v^2$, and for different values of the
mixing angles $\alpha_4$ and $\alpha_5$.
The other free parameters are fixed as shown
in \cref{fig:ddcpcons}.
The plots in the top row also show the current
experimental upper limits on $c_1^s$, $c_1^v$
from the LZ collaboration~\cite{LZ:2023lvz}.}
\label{fig:ddeftcoefs}
\end{figure}

Once $c_1^N$ and $c_{10}^N$ are known, the scattering
rates can be computed for different target materials
and exposure time of the different experiments.
We can circumvent this step because recently the
LZ collaborations has presented exclusion
limits directly on the Wilson coefficients $c_i^{s,v}$
of the non-relativistic EFT based on
the data collected until 2022~\cite{LZ:2023lvz}. 
In \cref{fig:ddeftcoefs} we show a comparison between
the LZ limits on the EFT Wilson coefficients
and the predicted values
$c_{1,10}^{s,v}$
(normalized to $1 / v^2$)
for the benchmark point used also
in \cref{fig:ddcpcons},
but with a fixed value of
$v_S = 200$~GeV, and we use non-zero values
for $\alpha_4 = \alpha_5$
in order to study the phenomenological
impact of the presence of CP-violation.
In the two plots in the bottom row one can
see that the coefficients $c_{10}^s$ and
$c_{10}^v$, corresponding to the CP-odd operator
$\mathcal{O}_{10}$, are several orders of magnitudes
below the experimental sensitivity of the
LZ experiment (or other comparable DM direct detection
experiments).
We therefore conclude that for the pNG DM state
contained in our model the experimental constraints
on the spin-dependent scattering of~$\chi$ on
nucleons do not play a role.
In the top row, in which we show
the predictions for the coefficients $c_1^s$ and
$c_1^v$, corresponding to the CP-even
operator $\mathcal{O}_1$,
one can see that the isoscalar interactions
give rise to the dominant contributions to the
spin-independent scattering.
For fixed values of the DM mass, the values of
$(c_1^s \times v^2)^2$ vary by almost a factor
of $100$ by changing the mixing angles
$\alpha_4$ and $\alpha_5$ from 0 to 0.4.
As a consequence, while the spin-dependent interactions
directly induced by the CP-violation is
not observable, the
indirect effects of the CP-violation on the
magnitude of the spin-independent interactions
are significant and should be taken into account
in a prediction for scattering rates at
DM direct-detection experiments.
In our presentation of
benchmark scenarios in \cref{sec:numdis}, we will
apply as a constraint the most recent LZ limits on
the spin-independent DM-nucleon scattering cross
sections (see \cref{fig:ddcpcons}), including the indirect effects
of the CP-violating effects via their
impact on the CP-even operators
$\mathcal{O}_{3,q}$ and $\mathcal{O}_{5,g}$
(see \cref{eq:relopseven}), and not considering any further
the experimental limits on the spin-dependent
interaction rates.

\subsubsection{Indirect Detection}
\label{sec:indirectdetect}

DM indirect detection experiments search for
astrophysical signals that arise from the
annihilation or decay of DM particles, which
would result in gamma rays, neutrinos or
cosmic rays consisting of high-energetic
electrons, positrons and antiprotons.
The primary constraints from DM indirect detection
come from the observations of astronomical objects
with high DM density, such as the galactic
center of the milky way, dwarf spheroidal galaxies
or galaxy clusters.
In the present model, the scalar DM particle
$\chi$ interacts
with the visible sector via the Higgs sector,
such that it mostly annihilates into
third-generation fermions or gauge bosons, which
subsequently decay into stable particles.
The dominant annihilation processes depend
on the DM mass $m_\chi$, the mass spectrum
of the neutral Higgs bosons, $m_{H_i}$,
and their coupling to $\chi$.

Experimental constraints on today's DM
annihilation cross sections have been set by a
variety of indirect detections experiments,
including observations from the
Fermi-LAT telescope~\cite{Fermi-LAT:2015att},
HESS~\cite{HESS:2018cbt},
MAGIC~\cite{MAGIC:2022acl}
and AMS-02~\cite{AMS:2016oqu}.
In the context of our model, however,
the potential impact of these constraints
is very limited, in particular 
on rather
light DM masses 
with
$m_\chi \lesssim 70$~GeV~\cite{Biekotter:2021ovi}
where the parameter space is much more
stringently constrained by LHC limits on the
invisible branching ratio of the 125~GeV
Higgs boson and the Planck measurement
of the DM relic abundance.\footnote{See
\citere{Biekotter:2021ovi} for a discussion
of the parameter space regions favoured by
the galactic-center excess~\cite{Fermi-LAT:2017opo}
in the CP-conserving limit of our model.}
For larger DM masses
the particle density of DM particles is
substantially smaller, suppressing the expected
rate of cosmic rays generated via DM
annihilation, and the measured value of the
DM relic abundance (taken as an upper bound)
typically
results in stronger constraints on the
parameter space~\cite{Jiang:2019soj,
Biekotter:2021ovi,
Darvishi:2022wnd}.\footnote{It has recently been pointed
out in Ref.~\cite{Fairbairn:2025cae}
that primordial magnetic fields may seed the formation
of dense early minihalos, potentially enhancing the
DM annihilation rate and thereby, depending on the main
annihilation channel, leading to significantly
stronger indirect detection constraints on WIMP scenarios.
We do not take these effects into account in the present analysis.}
Additionally, the constraints from DM
indirect detection experiments suffer from
sizable systematic uncertainties which arise from
uncertainties in the DM density profile
within the observed astronomical objects and along
the propagation path of the cosmic rays,
the propagation models of cosmic-ray
particles, and the modeling of different background
signals from other astrophysical sources.
As a consequence, we do not consider constraints
from DM indirect detection in this work,
leaving a comprehensive analysis for this model,
including the impact of CP-violating interactions,
for future work.

\subsection{Electric dipole moments}
\label{sec:edms}

\begin{figure}[t]
    \centering
    \includegraphics[width=0.45\linewidth]{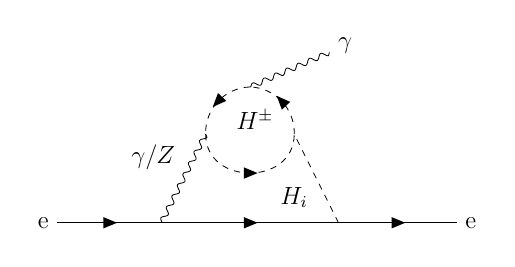}
    \caption{Example diagram with a charged Higgs boson
    loop contributing to the electron EDM at
    the two-loop level.}
    \label{fig:edmdiag}
\end{figure}

The most stringent constraints on CP-violation
in the Higgs potential
typically arise from the non-observation of
electron electric dipole moments~(eEDMs).
This holds true in the model considered here.
The eEDM $d_e$ is defined in the $q^2  = 0$ limit
of the CP-odd Pauli form factor in the
$e^+e^- \gamma$ vertex function~\cite{Altmannshofer:2020shb}
\begin{equation}
  i d_e \bar{u}(p') \sigma^{\mu\nu} q_\nu \gamma_5 u(p) \, ,
  \qquad 
  \sigma^{\mu\nu} = \frac{i}{2} [ \gamma^\mu, \gamma^\nu ] \, ,
\end{equation}
where $u(p)$ is the electron Dirac spinor,
$q$ is the incoming momentum of the photon,
and $\gamma^\mu$ are the Dirac matrices.
The theoretical predictions for the eEDM in this
model have been investigated 
in \citere{Darvishi:2022wnd}.
There it was shown that (as in the CP-violating
2HDM) the numerically dominant contributions
in most parts of the parameter space arise
first at the two-loop level, whereas the one-loop
contributions to the eEDM are subleading barring
sizable destructive interferences at the two-loop level.
Taking this into account, in our analysis we
consider the two-loop Barr-Zee type
contributions~\cite{Barr:1990vd}.
An example diagram is shown in \cref{fig:edmdiag},
where CP-violation can be present in the $H^\pm H^\mp H_i$
couplings (see \cref{sec:cpviolationscalar})
and in the $H_i e^+ e^-$ couplings
(see \cref{sec:cpviolationferm}).
This diagram is special since,
in the absence of CP-mixing in the scalar sector
and without CP-violating $H_i f \bar f$ couplings, this
diagram can still give rise to non-vanishing EDMs
for $\lambda_5^{\rm Im} \neq 0$ from the diagram
with the state $A_0$ on the internal line connecting
the electron propagator with the $H^\pm$-loop,
via the CP-violating $A_0 H^\pm H^\mp$ coupling
given for the alignment limit in \cref{eq:axxalignlim}.
Another set of diagrams which contributes only under
the presence of CP-violating $H_i f \bar f$ couplings
is obtained, for instance,
by replacing the $H^\pm$-loop with a fermion-loop,
in which case also CP-violating couplings of the neutral
scalars $H_i$ to top and bottom quarks become relevant,
which has great implications for the interplay between
the eEDM and the phenomenology at the LHC.

Our computation of the eEDM is based on the complete
set of gauge-invariant Barr-Zee type contributions
published for the complex 2HDM in
\citere{Abe:2013qla}.\footnote{We did not include the so-called kite diagrams  (cf.~Ref.~\cite{Altmannshofer:2020shb} and Ref.~\cite{Ning:2025zfh} for recent discussions in the context of the CP-violating 2HDM and the next-to-minimal supersymmetric extension of the SM (NMSSM), respectively). Their contribution will be discussed in a future publication.}
In order to make these expressions applicable to
our model, we replaced the effective couplings of
the Higgs bosons contained in the complex 2HDM
by the corresponding coupling factors of the
Higgs bosons contained in our model, taking into
account the additional neutral scalar state
originating from the real component of the
singlet field $\Phi_S$.
To approximately account for sizable QCD
corrections~\cite{Brod:2023wsh} in the
contributions with bottom- and charm-quarks,
we use running quark masses at the electroweak
scale $\mu = M_Z$ as put forward
in \citere{Altmannshofer:2024mbj}.
The DM state (originating from the imaginary
component of $\Phi_S$) does not enter in the
eEDM predictions at two-loop level.
As a consequence, at the considered loop order
the DM mass $m_\chi$ can be
varied in our model without having an impact
on the eEDM.

\begin{figure}
\centering
\includegraphics[width=0.48\textwidth]{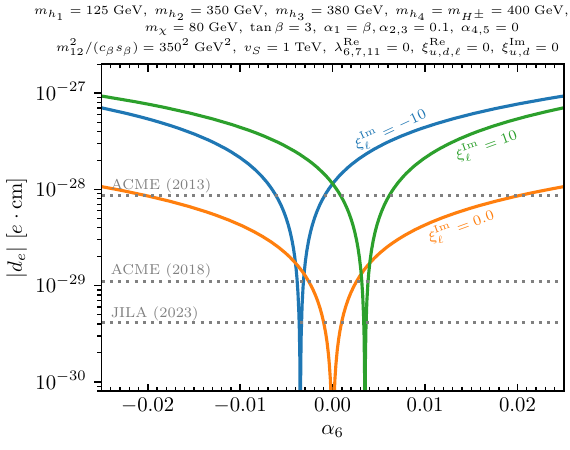}~
\includegraphics[width=0.48\textwidth]{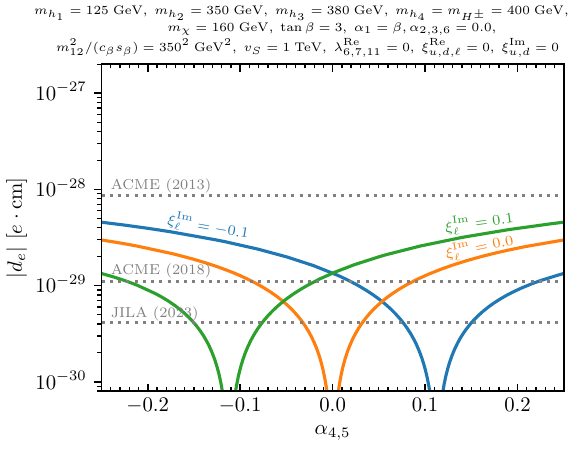}
\caption{Predicted values of the eEDM
$|d_e|$ (see text for details)
as a function of $\alpha_6$ (left) and $\alpha_4=\alpha_5$ (right) for $\xi_\ell^{\rm Im} = - \TB{10}$, 
$\xi_\ell^{\rm Im} = 0$ and $\xi_\ell^{\rm Im} = \TB{10}$ 
(left) and $\xi_\ell^{\rm Im} = -0.1$,
$\xi_\ell^{\rm Im} = 0$ and $\xi_\ell^{\rm Im} = 0.1$ (right)
in blue, orange and green, respectively,
with the other free parameters set as shown on top
of the plot. The horizontal gray dashed lines indicate
the experimental 90\% confidence-level upper limits
from the ACME~\cite{ACME:2013pal,ACME:2018yjb}
and JILA~\cite{Roussy:2022cmp} collaborations.}
\label{fig:edmexamples}
\end{figure}

To get an idea of the typical size of the eEDM,
we show the predictions for $|d_e|$ in an example
scenario as a function of the mixing angle
$\alpha_6$ and for three different values
of $\xi_\ell^{\rm Im}$ in the left plot
of \cref{fig:edmexamples}.
All other free parameters are set as shown
on top of the plot.
Also indicated with horizontal dotted lines
are the experimental upper limits on
$|d_e|$ at 90\% confidence level obtained
by the ACME collaboration in 2013~\cite{ACME:2013pal}
and 2018~\cite{ACME:2018yjb}, and by JILA in
2023~\cite{Roussy:2022cmp}.
Non-zero values of the mixing angle $\alpha_6$
in general give rise to a non-zero CP-odd admixture
in the state $H_1$, which here corresponds to the
Higgs boson at 125~GeV.
Moreover, non-zero values of $\xi_\ell^{\rm Im}$ give
rise to CP-violating couplings between the Higgs bosons
$H_i$ and charged leptons, also inducing non-zero values
for the eEDM.
The orange line showing the case $\xi_\ell^{\rm Im} = 0$
goes to zero for $\alpha_6 = 0$ where the model is
CP-conserving. For $\alpha_6 \neq 0$ the orange line
grows to values of $|d_e| \gtrsim \TB{10^{-28}}~e \cdot$~cm,
which is more than \TB{an order} 
of magnitudes larger
than the current experiment limits.
The blue and the green lines show the predictions
for the eEDM for non-zero values of
$\xi_\ell^{\rm Im} = - \TB{10}$ 
and $\xi_\ell^{\rm Im} = \TB{10}$, 
respectively.
This leads to a shift to non-zero values
of the mixing angle $\alpha_6$
where the considered example parameter point
is in agreement with the experimental upper limits
on the eEDM. This demonstrates that it is possible
to have a cancellation in the eEDM between the CP-violation
from the Higgs potential (for $\alpha_6 \neq 0$ here)
and the CP-violation from the Yukawa sector
(for $\xi_\ell^{\rm Im} \neq 0$ here), such that
the predicted value for the eEDM is below the current bounds.
However, if the CP-violating effects are present in the
couplings of the 125~GeV Higgs boson
as is the case here, the cancellation
requires a substantial amount of fine-tuning of parameters,
as is visible in the left plot of \cref{fig:edmexamples}.
It should be noted here that if large cancellations between
different amplitudes at the considered two-loop level
are present, the numerically most important contributions
to the eEDM are at some point expected to arise at the
one-loop level or the three-loop level
which are not taken into account here.

In the right plot of
\cref{fig:edmexamples} we show $|d_e|$ as
a function of $\alpha_4 = \alpha_5$, and for three
different values of $\xi_\ell^{\rm Im} = -0.1$,
$\xi_\ell^{\rm Im} = 0.0$ and
$\xi_\ell^{\rm Im} = 0.1$ in blue, orange and
green, respectively.
Non-zero values of the mixing angles $\alpha_4$
and $\alpha_5$ result in CP-violating mixing
between the BSM Higgs bosons
$H_{2,3,4}$, whose masses are set to $m_{H_2} = 350$~GeV,
$m_{H_3} = 380$~GeV and $m_{H_4} = 400$~GeV.
For the SM-like Higgs boson $H_1$ at
125~GeV, on the other hand,
there is no CP-odd admixture since here $\alpha_6 = 0$.
By comparing both plots in
\cref{fig:edmexamples} one can see that the predicted
eEDM induced by non-zero values of $\alpha_4$ and
$\alpha_5$ (right plot) is substantially smaller compared to
the predictions for non-zero values of $\alpha_6$
\TB{(left plot).}
This demonstrates that
the cancellation that is required to predict a sufficiently
small eEDM under the presence of CP-violation is
significantly less severe if the CP-violating effects
are mostly confined to the properties of the BSM Higgs bosons
with masses that are greater than 125~GeV, while
the 125~GeV Higgs boson acts mostly as a CP-even state.
This is in line with
the fact that (in the absence of accidental
cancellations) the generated eEDM scale via~\cite{Morrissey:2012db}
\begin{equation}
  |d_e| \sim \frac{m_e}{1~\mathrm{MeV}}
    \left( \frac{1~\mathrm{TeV}}{M} \right)^2
    \cdot 10^{-26}~e~\mathrm{cm} \, ,
    \label{eq:edmscale}
\end{equation}
where $M$ represents the mass scale of the particles
whose interactions are CP-violating. Hence, the
heavier the Higgs bosons $H_i$ with CP-violating couplings,
the smaller is the resulting eEDM.

\section{Benchmark Planes and Scans}
\label{sec:numdis}

In this section, we discuss the results of our numerical analysis regarding 
the impact of the theoretical and experimental constraints
on the parameter space of the cS2HDM, 
with a special focus on the eEDM experimental upper bound,
the DM direct detection limits,
and the LHC Higgs data. 
In \cref{sec:num_edm} and \cref{sec:num_dd},
we discuss the results from
a scan over the parameter space of the model.
We generated random
parameter points and tested them against theoretical and
experimental constraints as described in Section \ref{sec:constraints}.
The values for each free parameter (see \cref{eq:freeparas})
of the parameter points were selected within the intervals
summarised 
in \cref{tab:scan_ranges}.
\TB{In this parameter scan, all points were
required to satisfy the theoretical, collider,
and EWPO constraints, except for DM direct detection
and eEDM bounds. These two constraints were instead
studied in detail using the scan results in order
to investigate their impact on the parameter space.
In practice, the direct detection limits turned out to not
exclude any parameter point due to the cancellation mechanism (see \cref{sec:directdetection}),
while the parameter points excluded or allowed by
the eEDM are explicitly indicated in
\cref{fig:edmexamples} and \cref{fig:DD_EDM}.
After discussing the results of the parameter scan},
in \cref{sec:num_lhc}
we compare the sensitivity of LHC cross section
measurements of the 125~GeV Higgs boson to CP-violating couplings,
and compare it to the sensitivity of eEDM experiments.

\TB{
Throughout our numerical analysis,
we consider parameter space regions
where the masses of the BSM Higgs bosons
are restricted
to values below 1~TeV. We stress that this
is a choice rather than a limitation
of the model. The cS2HDM features a well-defined decoupling
limit in which all BSM scalars can become heavy:
$M \approx v_S \approx m_{H_{2,3,4}} \approx m_{H^\pm}
\gg v$, where $v$ denotes the Higgs vacuum expectation value
and represents the electroweak scale.
We focus on scalar masses
below the TeV scale for several reasons.
First, this mass range allows for direct production of the
additional Higgs bosons at the LHC, making the model
testable at current and future colliders.
Second, DM direct detection experiments are most sensitive
to scenarios in which the DM mass (which is naturally bound
to be close to the Higgs boson masses for efficient
annihilation in the early universe) are below the
multi TeV-range. Third, Higgs bosons with masses substantially
larger than the electroweak scale effectively decouple,
rendering a strong first-order electroweak phase transition
unlikely, such that electroweak baryogenesis cannot be realized.}

\begin{table}[t]
    \centering
    \begin{tabular}{ c | c || c | c}
        \textbf{Parameter} & \textbf{Range} & \textbf{Parameter} & \textbf{Range} \\
        \hline
        \hline
        $m_{H_1}$     & $125.09$                                    & $M$                 & $\left[200,1000\right]$ \\
        $m_{H_2}$     & $\left[20,1000\right]$                      & $v_S$               & $\left[200,1000\right]$ \\
        $m_{H_3}$     & $\left[20,1000\right]$                      & $\lambda_{5}^{\rm Im}$  & $0$ \\
        $m_{H_4}$     & $\left[20,1000\right]$                      & $\lambda_{6}^{\rm Re}$  & $0$ \\
        $m_{H^{\pm}}$ & $\left[200,1000\right]$                     & $\lambda_{7}^{\rm Re}$  & $0$ \\
        $m_{\chi}$    & $\left[20,1000\right]$                      & $\lambda_{11}^{\rm Re}$ & $0$ \\
        $\tan{\beta}$ & $\left[1,10\right]$                                      & $\xi_{u}^{\rm Re}$      & $\left[-15,15\right]$ \\
        $\alpha_{1}$  & $\beta$                                     & $\xi_{d}^{\rm Re}$      & $\left[-15,15\right]$ \\
        $\alpha_{2}$  & 0                                           & $\xi_{\ell}^{\rm Re}$      & $\left[-15,15\right]$ \\
        $\alpha_{3}$  & $\left[-\frac{\pi}{2},\frac{\pi}{2}\right]$ & $\xi_{u}^{\rm Im}$      & $\left[-15,15\right]$ \\
        $\alpha_{4}$  & $\left[-\frac{\pi}{2},\frac{\pi}{2}\right]$ & $\xi_{d}^{\rm Im}$      & $\left[-15,15\right]$ \\
        $\alpha_{5}$  & $\left[-\frac{\pi}{2},\frac{\pi}{2}\right]$ & $\xi_{\ell}^{\rm Im}$      & $\left[-15,15\right]$ \\
        $\alpha_{6}$  & 0 
    \end{tabular}
    \caption{List of parameter ranges for the random scan of the parameter space of the cS2HDM. The values for the masses and $v_S$ are expressed in GeV and single values represent constant values for the
    corresponding parameter.}
    \label{tab:scan_ranges}
\end{table}

\subsection{The Electron EDM Constraint}
\label{sec:num_edm}
\begin{figure}
    \centering
    \includegraphics[width=0.6\linewidth]{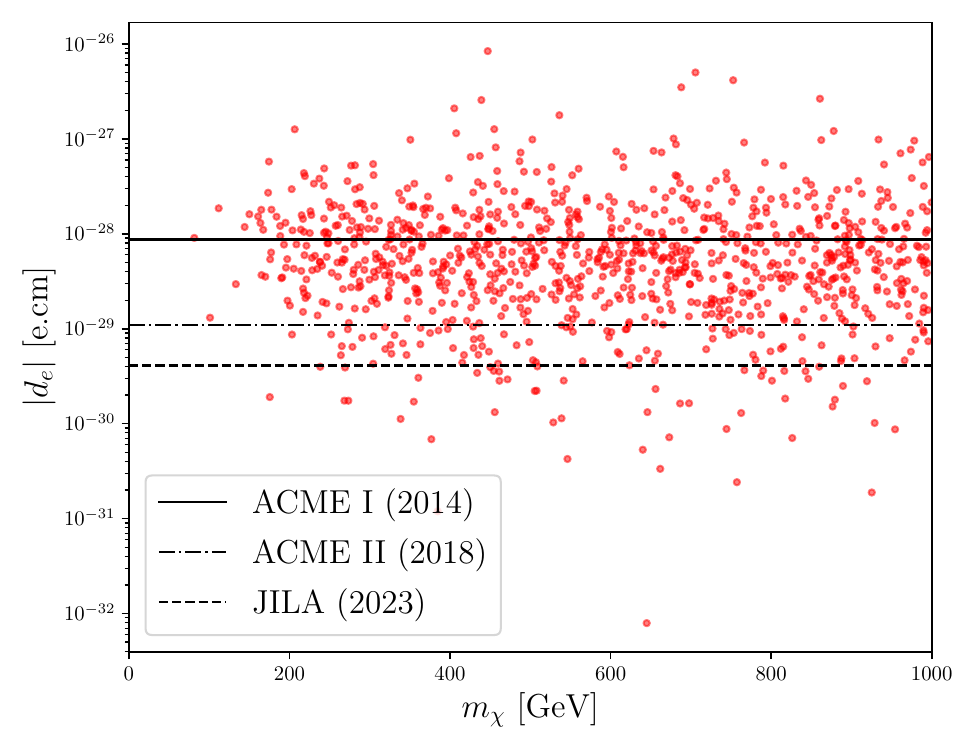}
    \caption{Predicted values of the electron EDM as a function of the DM candidate mass for a parameter space scan. The horizontal black lines represent the experimental upper limits from the ACME~\cite{ACME:2013pal,ACME:2018yjb} and JILA~\cite{Roussy:2022cmp} collaborations.}
    \label{fig:plot_EDM_MDM}
\end{figure}

As explained in Section \ref{sec:edms}, the eEDM is one of the most
stringent constraints on CP-violation in the Higgs potential.
In light of the recent improvements of the experimental upper
limit on the eEDM, see the discussion in \cref{sec:edms},
we performed a random parameter scan in order to
confirm that it is possible to find points in the parameter space
of the cS2HDM that are allowed by this constraint.
In \cref{fig:plot_EDM_MDM}, we show the predicted values for the
eEDM $|d_e|$ for each parameter point
plotted against
the mass of the DM candidate $m_{\chi}$.
All displayed points we generated by
demanding that the whole set of theoretical
and experimental constraints discussed in \cref{sec:constraints}
is satisfied,
with the exception of the upper limit on the eEDM.
The black lines represent the latest upper bounds on the eEDM as
presented by the ACME Collaboration in 2014~\cite{ACME:2013pal}
(solid) and 2018~\cite{ACME:2018yjb} (dash-dotted), and more
recently by the JILA 
in 2023~\cite{Roussy:2022cmp} (dashed). One can see that
while most parameter points are excluded by the 
most stringent experimental limit from JILA,
in line with the expectation from \cref{eq:edmscale},
there are also parameter points over the whole
scan range of the DM mass that still evade this constraint.
This shows that in the cS2HDM
the 
possibility of having CP-violation in both the
scalar and Yukawa sectors 
provides the necessary cancellation to sufficiently suppress
the eEDM down to the allowed range.
In addition, 
the allowed parameter points are distributed
down to rather low DM masses of $m_\chi \lesssim 200\gev$.
This shows that the eEDM constraint does not exclude
the part of the parameter space that could potentially be
probed
with DM searches at the LHC.
This enables an interesting interplay between the eEDM measurements
and LHC DM searches that can be exploited to constrain the model.

\subsection{DM Direct Detection}
\label{sec:num_dd}
As mentioned before, one of the most attractive features of pNG DM models is the fact that the DM-nucleon scattering cross section is naturally suppressed, making it easier to evade the DM direct detection constraint.
To demonstrate this, and to estimate to what extend current
direct detection limits are able to probe the parameter space
of the cS2HDM, we analyze here the predicted DM-nucleon scattering
cross sections focusing on the dominant spin-independent interactions
(see the discussion in \cref{sec:directdetection}.
We compare the theoretical predictions against the currently
most stringent cross-section limits that were published
by the LZ collaboration.

\begin{figure}[t]
    \centering
    \includegraphics[width=1\linewidth]{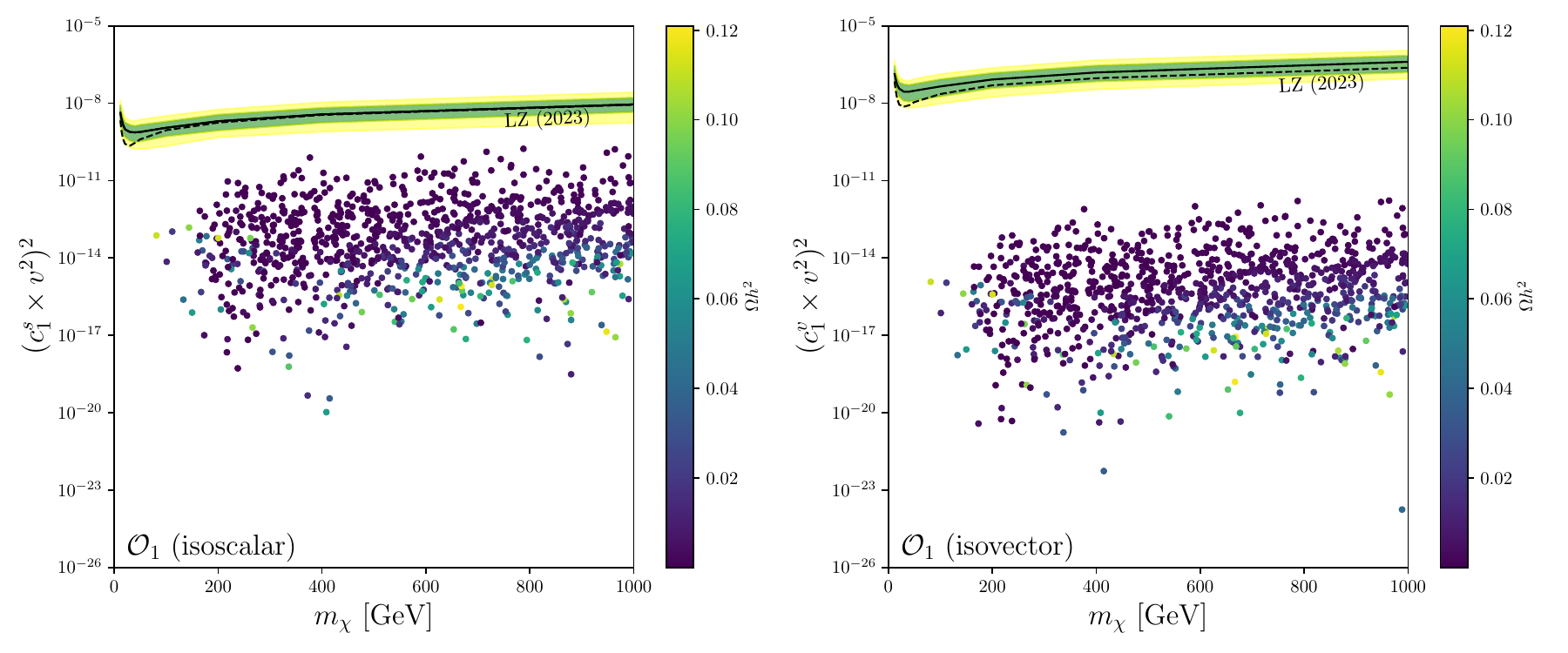}
    \includegraphics[width=1\linewidth]{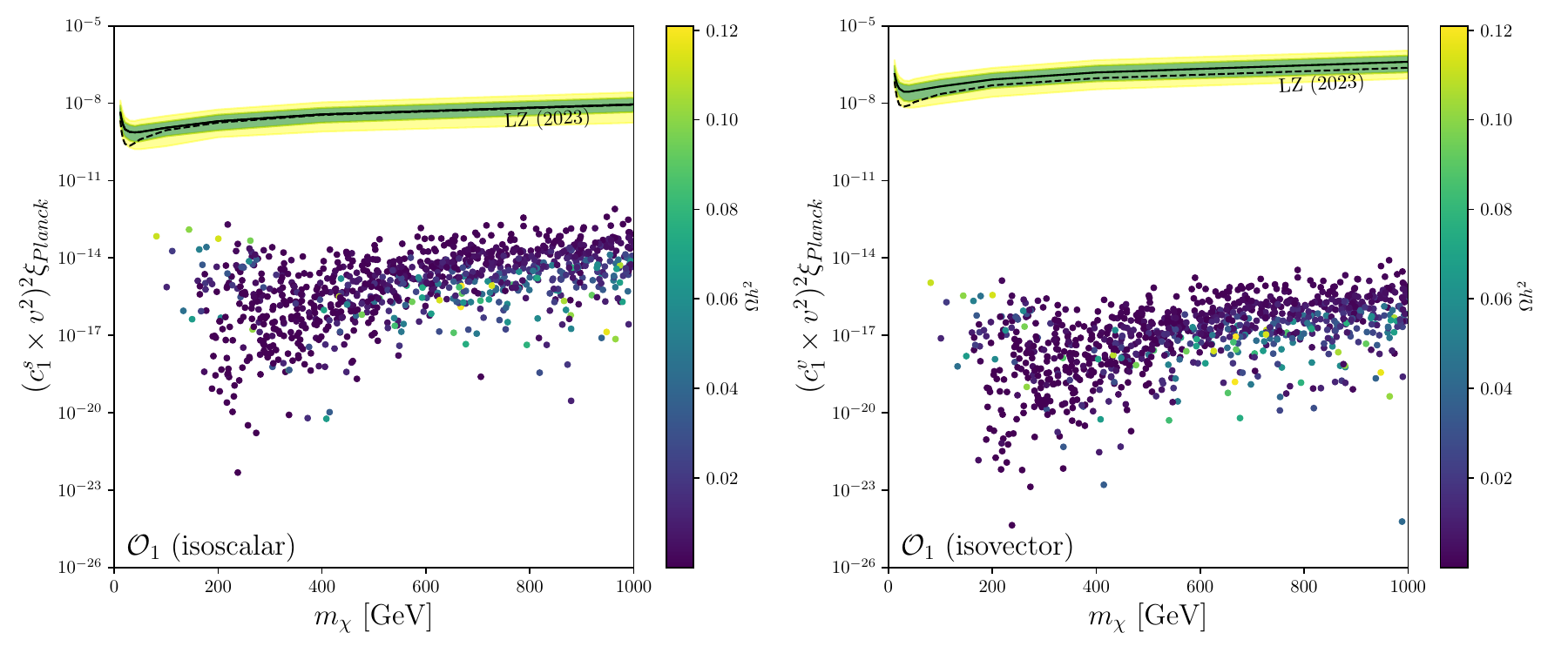}
    \caption{Predicted values of the squared non-relativistic Wilson coefficients $c_{1}^{s}$ (left) and $c_{1}^{v}$ (right)
        plotted against
    the DM mass $m_{\chi}$ for a scan of the parameter space.
        Also shown with the solid and dashed
    black lines are the observed and
    expected upper limits set by the LZ experiment, respectively~\cite{LZ:2023lvz}.
    The green and orange bands show the
    $1\sigma$ and $2\sigma$ uncertainty bands,
    respectively.
    The color of each point indicates the predicted DM relic abundance. In the bottom row, the values for each point were scaled by $\xi_{Planck}$ to adjust for the respective fraction of the DM relic density.}
    \label{fig:direct_detection}
\end{figure}

In the top row of Fig.~\ref{fig:direct_detection} we show the predicted values for the squared Wilson coefficients $c_1^s$ (left) and $c_1^v$ (right) respectively referring to
the isoscalar and isovector components of
the non-relativistic spin-independent operator $\mathcal{O}_1$,
as defined in \cref{eq:wilson_cs_cv},
plotted against the mass of the DM candidate $m_{\chi}$.
The color code represents the predicted relic abundance $\Omega h^2$
for each point. Finally, the black line represents the
observed experimental upper bound as presented by the
LZ Collaboration in 2023~\cite{LZ:2023lvz}.
We observe that 
most 
points are several orders of magnitude below the experimental limit
for both the isoscalar and isovector Wilson coefficients,
confirming that the pNG nature of the DM candidate makes
it possible to evade the DM direct detection constraint.
Nevertheless, we also find parameter points which are
only about two orders of magnitude below the LZ limit,
and which therefore might be in reach of future
experiments~\cite{DARWIN:2016hyl}.
For most points, the DM candidate accounts only for a small fraction of the measured DM relic abundance of the Universe, meaning that in these cases some additional DM mechanism not present in our model would be needed to explain the remaining relic abundance.
Regarding the DM mass, we observe that
the allowed points cover the range from around 100~GeV to 1~TeV.
The lower limit of the range of $m_\chi$ arises mainly
from LHC Higgs data because for $m_{\chi}\leq 62$~GeV
the decay channel $H_1\rightarrow\chi\chi$ becomes kinematically
allowed, which is highly constrained by 
direct searches for invisible Higgs boson decays and
from global fits to the LHC cross section
measurements~\cite{Biekotter:2022ckj}.
However, it should be noted that in a parameter
scan that is more targeted to the low-mass region it is
possible to find valid parameter points also for
$m_{\chi}\leq 62$~GeV~\cite{Biekotter:2021ovi}.

We included CP-violating phases in the parameter scan
via non-zero values of the flavour alignment parameters
$\xi_{u,d,\ell}$ and the mixing angles $\alpha_{4,5}$.
Thus, our parameter scan demonstrates that
the introduction of CP-violation in the cS2HDM does not
spoil the suppression of the DM-nucleon scattering rates.
For models like the 2HDM+a, where the presence of CP-violation
in the Higgs sector is expected to weaken the suppression
of the scattering rates, it would be worthwhile to investigate
how significant this effect is. We leave an analysis of the
size of scattering cross sections in the 2HDM+a in the presence
of CP-violation in the Higgs sector,
and a comparison to the cS2HDM studied here, for future work.

The LZ experimental limit, as shown in Fig.~\ref{fig:direct_detection},
is determined 
assuming that the DM candidate represents the full measured DM relic abundance. For our scan we only required that the predicted DM relic abundance is below the measured limit. 
To 
compare our theoretical predictions 
with the experimental measurements taking into account
this effect of a reduced particle density of $\chi$
according to the predicted relic abundance, 
one can in an approximate approach re-scale the predicted value
for the DM-nucleon scattering cross section for each point
by the fraction of the predicted over the
measured DM relic abundance. 
In the bottom row of Fig.~\ref{fig:direct_detection}, we show the squared Wilson coefficients $c_1^s$ (left) and $c_1^v$ (right) re-scaled by
the ratio $\xi_{\rm Planck}$ as defined in Eq.~(\ref{eq:relicratio}),
plotted against
the mass of the DM candidate $m_{\chi}$. Again, the color code represents the predicted DM relic abundance $\Omega h^2$ for each point. We see that, in comparison with the unscaled predictions shown in
the plots in the top row,
most points move substantially further away from the
experimental upper bound.
We carried out the same analysis 
for the non-relativistic spin-dependent operator $\mathcal{O}_{10}$. In this case, the values for the correspondent Wilson coefficients $c_{10}^s$ and $c_{10}^v$, were many orders of magnitude lower than the spin-independent counterparts, therefore reinforcing the conclusions drawn at the end of
\cref{sec:directdetection} that for the pNG DM state the experimental constraints on the spin-dependent DM-nucleon scattering do not play a role
even if explicit CP-violation is considered.

\begin{figure}[t]
    \centering    
    \includegraphics[width=1\linewidth]{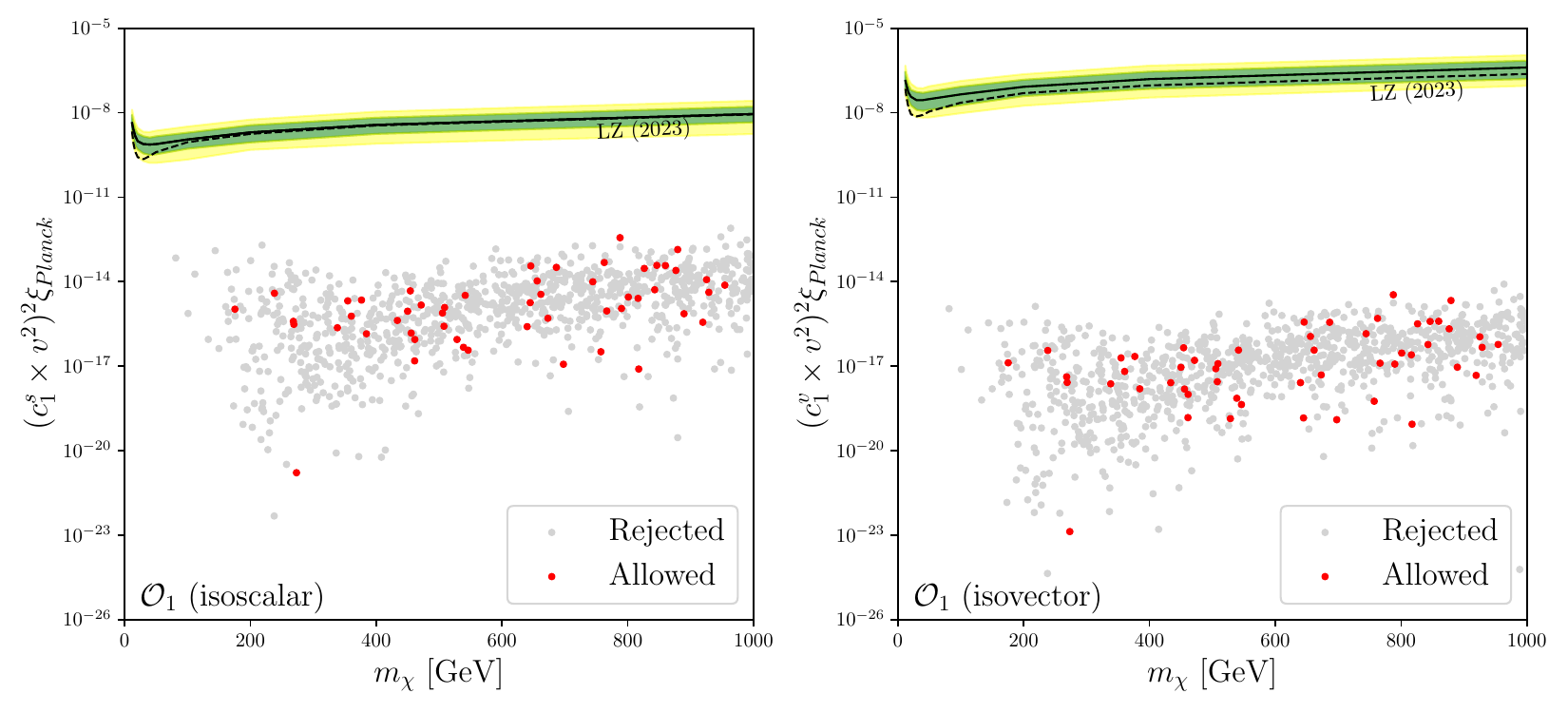}
    \caption{Predicted 
    values of the squared non-relativistic Wilson coefficients $c_{1}^{s}$ (left) and $c_{1}^{v}$ (right) re-scaled with the ratio
    of predicted over observed relic abundance
    $\xi_{\rm Planck}$ plotted against
        the DM mass $m_{\chi}$ for a scan of the parameter space. The red points are allowed by the JILA 2023 electron EDM constraint while the gray points are excluded.}
    \label{fig:DD_EDM}
\end{figure}

As mentioned already above, the most sensitive
observable to new sources of CP-violation is the
eEDM. It is therefore interesting to investigate
possible correlations between the predicted
values for the eEDM and the DM-nucleon scattering
cross sections.
In Fig.~\ref{fig:DD_EDM} we show
as in the bottom row of \cref{fig:direct_detection}
the predicted re-scaled values of the squared non-relativistic Wilson coefficients $c_1^s$ (left) and $c_1^v$ (right)
plotted against
the mass of the DM candidate $m_{\chi}$. The gray points represent the points that are excluded by the eEDM experimental upper bound given by the JILA Collaboration, while the red points represent the points that evade this constraint. The allowed points have values between $10^{-17}$ and $10^{-12}$ for $c_1^s$ and between $10^{-19}$ and $10^{-14}$ for $c_1^v$. The allowed points are 
uniformly distributed
across the displayed DM mass interval.
Consequently, the constraints from DM direct
detection experiments (once reaching more sensitivity
to pNG DM) and the ones from the eEDM measurements
are largely independent and complementary.

\subsection{LHC Sensitivity to CP-Violating
Higgs Boson Couplings}
\label{sec:num_lhc}

An important question is to what extent the LHC can probe CP-violation
in the
observed 125~GeV Higgs boson,
or if sizable CP-odd components that could give rise to detectable effects
are already excluded by the stringent upper limits on the
eEDM. In addition, searches for additional Higgs bosons at
the LHC provide complementary sensitivity to CP-violating
phases, since CP-violating coupling components
can enhance or suppress certain production and decay channels.
We stress that in this discussion
we will focus on LHC measurements or experimental limits
on total cross sections that are not themselves direct indications
of CP-violation, but which are modified in the presence of CP-mixed
interactions.\footnote{Truly CP-sensitive Higgs boson observables can be
obtained by utilizing differential information and angular
distributions, which have been employed in measurements
of Higgs boson decays into $\tau$-lepton pairs~\cite{CMS:2021sdq,ATLAS:2022akr},
Higgs boson production
in association with top quarks~\cite{CMS:2022dbt,ATLAS:2023cbt},
and Higgs boson production via vector-boson
fusion~\cite{ATLAS:2022tan,ATLAS:2023mqy}.
The resulting constraints on CP-violating Higgs boson couplings
to $\tau$-leptons, gauge bosons and top quarks are still relatively
weak. We leave an investigation of whether these measurements
can probe parameter space that is not already excluded by
the inclusive LHC Higgs data and BSM Higgs boson searches,
as well as the upper limit on the eEDM, for future studies
(see Ref.~\cite{Biekotter:2024ykp} for an analysis in the complex 2HDM).}
This is in contrast to the eEDM bounds, which provide
an unambiguous probe of CP-violation.

In order to study the interplay between LHC and EDM constraints,
we analyze three representative parameter planes in which we
vary $\sin(\beta-\alpha_1)$ together with the imaginary part
of one of the three flavor-alignment parameters $\xi_f$,
with $f=u,d,\ell$ for up-type quarks, down-type quarks
and leptons, respectively.
For $\sin(\beta-\alpha_1) = 0$ the model is in the alignment
limit, and the light Higgs boson $H_1$ playing the role of
the SM-like Higgs boson has purely CP-even interactions,
regardless of the values of $\xi_f$.
Away from the alignment limit, $\sin(\beta-\alpha_1) \neq 0$,
the CP-odd admixtures in the Yukawa couplings of $H_1$
increase with increasing values of $|\xi_f^{\rm Im}|$.
We therefore consider the cases where (i) only
$\xi_u^{\rm Im}$ is varied, (ii) only $\xi_d^{\rm Im}$ is varied,
and (iii) only $\xi_\ell^{\rm Im}$ is varied, while the
other two imaginary parts are set to zero.  
For this analysis, we fix all remaining parameters
to the following values,
\begin{align}
&m_{H_1} = 125.1\gev \, , \quad
m_{H_{2,3,4}} = m_{H^\pm} = 200\gev \, , \quad
m_\chi = 90.0\gev \, , \quad
\tan\beta = 5 \, , \quad
\xi_{u,d,\ell}^{\rm Re} = 0 \, , \notag \\
&\alpha_{2,4,5,6} = 0 \, , \quad
\alpha_3 = \pi / 4 \, , \quad
M = 150\gev \, , \quad
v_S = 100\gev \, , \quad
\lambda_5^{\rm Im} = 0 \, , \quad
\lambda_{6,7,11}^{\rm Re} = 0 \, .
\end{align}
In this setup the Yukawa sector reduces to type-I in the
limit $\xi_f^{\rm Im}=0$. Moreover, since
$\alpha_{4,5,6}$ and $\lambda_5^{\rm Im}$
are set to zero, the imaginary parts of
the parameters $\xi_f$ are the only sources of CP-violation.

Even though not central to the following discussion,
we have verified that the benchmark planes
satisfy all relevant theoretical and experimental constraints.
In particular, vacuum stability (bounded-from-below conditions)
and perturbative unitarity are ensured by the choice of relatively
low masses for the BSM Higgs bosons, in combination with slightly
smaller values of $M$ and $v_S$. Constraints from the EWPOs
are satisfied since we remain close to the alignment limit and
the BSM Higgs states are taken to be 
mass-degenerate,
thereby suppressing contributions to the $\rho$-parameter.
The choice of $\tan\beta$ is motivated by compatibility with
existing LHC searches in the CP-conserving limit.
For the DM sector, we fix $m_\chi < 100\,\text{GeV}$, which opens
invisible decay modes of the BSM scalars and
thereby reduces their branching ratios for decays into
visible final states such as
$H_{2,3,4} \to \gamma\gamma, \tau^+ \tau^-, ZZ$.
This in turn weakens the corresponding constraints
from LHC searches and provides additional
freedom to vary $\sin(\beta-\alpha_1)$
and the alignment parameters $\xi_f$.
Finally, we set $\alpha_3 = \pi/4$, so that $H_2$ and $H_3$
share an equal singlet component. If $\alpha_3$ were also
set to zero, the singlet field would completely decouple.
\TB{For the parameter planes discussed here,
we did not perform a comprehensive evaluation of the
DM relic abundance. For the chosen value of
$m_\chi = 90$~GeV, however, the pNG DM candidate is expected
to be largely underabundant, as efficient annihilation
in the early universe proceeds via $s$-channel exchange
of the SM-like Higgs boson $H_1$ into gauge-boson pairs.
In addition, the proximity of $m_\chi$ to half of the BSM scalar
masses further enhances annihilation through
exchange of the heavier neutral scalars. Given this
underproduction, and in combination with the
cancellation mechanism suppressing direct-detection
scattering amplitudes, current limits from DM direct
detection experiments are not constraining the shown
parameter planes.}

\begin{figure}
  \centering
  \includegraphics[width=0.32\textwidth]{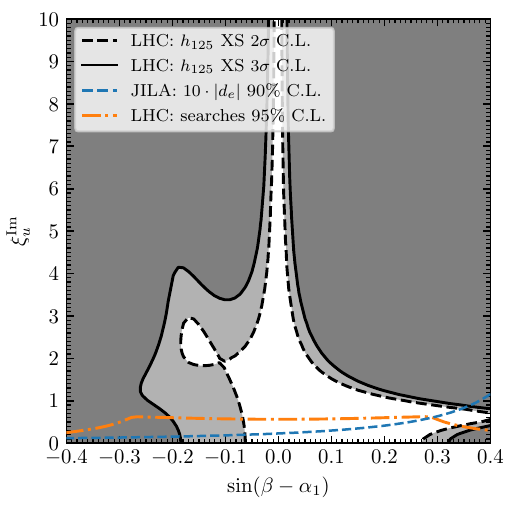}
  \includegraphics[width=0.32\textwidth]{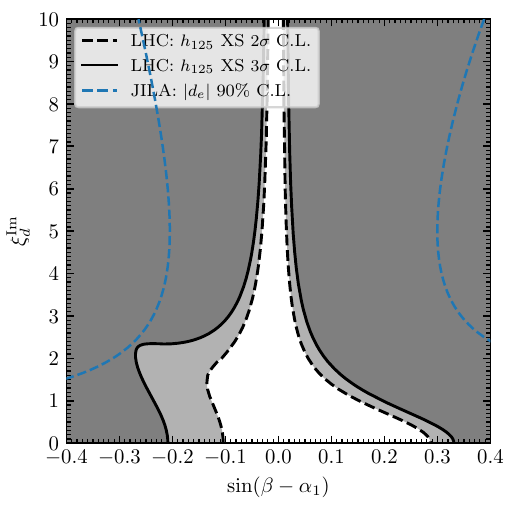}
  \includegraphics[width=0.32\textwidth]{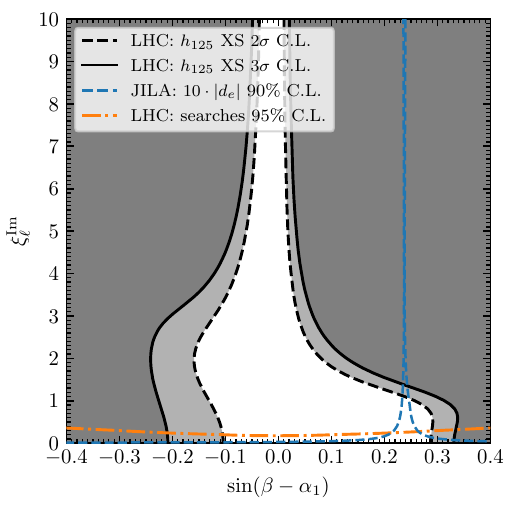}
  \caption{Parameter space excluded by LHC Higgs data
  in planes with $\sin(\beta - \alpha_1)$ on the horizontal
  axes and the flavour alignment parameters on the
  vertical axes: $\xi_u^{\rm Im}$ (left), $\xi_d^{\rm Im}$ (centre), 
  and $\xi_\ell^{\rm Im}$ (right).
  The dashed and solid black lines indicate the exclusion
  region at $2\sigma$ and $3\sigma$ confidence level, respectively,
  obtained with \texttt{HiggsTools}.
  The region above the orange dot-dashed lines are excluded
  by cross section limits from LHC searches for additional
  Higgs bosons. The region above the dashed blue lines
  predict a value for the eEDM that is
  equal to the upper limit from JILA in the centre plot
  and equal to ten times the upper limit from JILA
  in the left and the right plot (the limit was increased
  by a factor of ten for better visibility).}
  \label{fig:BPalphat}
\end{figure}

In \cref{fig:BPalphat} we show the constraints on the
parameter space in three parameter planes. 
Each plot shows the dependence on $\sin(\beta-\alpha_1)$ along
the horizontal axis, while the vertical axis corresponds to one
of the imaginary parts of the flavour-alignment parameters, 
$\xi_u^{\rm Im}$ (left), $\xi_d^{\rm Im}$ (center), and
$\xi_\ell^{\rm Im}$ (right). 
The black dashed and solid curves indicate the regions excluded at
$2\sigma$ ($\sim 95\%$) and $3\sigma$ ($\sim 99\%$)
confidence level, respectively, 
based on Higgs boson cross-section measurements as obtained
with \texttt{HiggsTools}. 
The orange dot-dashed lines represent exclusions from
cross-section limits from LHC
searches for BSM Higgs bosons, with the parameter space above the
line being excluded. 
Finally, the blue dashed lines show where the predicted eEDM
is equal to the current JILA upper limit in the central plot
and to ten times this limit in the left and right plots
(the scaling being introduced for visibility). 

In the left plot we find that, in the CP-conserving limit, 
departures from alignment are allowed in the range
$-0.06 \lesssim \sin(\beta-\alpha_1) \lesssim 0.27$ at 95\% confidence level.
This range is significantly modified when $\xi_u^{\rm Im}$ is non-zero.
For moderate values around $\xi_u^{\rm Im} \sim 0.5$, 
deviations up to $\sin(\beta-\alpha_1) \sim 0.4$ become consistent with
the LHC Higgs data. 
For larger values of $\xi_u^{\rm Im} \gtrsim 4$, the allowed region
in $\sin(\beta-\alpha_1)$ shrinks significantly. 
Additional constraints arise from the eEDM, which in this plane excludes 
values $\xi_u^{\rm Im} \gtrsim 0.1$. This stringent exclusion
is due to CP-violation in the $H_i t\bar{t}$ interactions, 
where the large top-quark Yukawa coupling induces sizable
eEDM contributions at the two-loop level. 
We also find important restrictions from LHC
searches for additional Higgs bosons. Values  of
$\xi_u^{\rm Im} \gtrsim 0.7$ are excluded, driven by cross section
limits from searches for new spin-0 resonances decaying
into photon pairs~\cite{ATLAS:2021uiz},
since the loop-induced $H_i \to \gamma\gamma$ decay receives enhanced
contributions from the top-quark loop when $\xi_u^{\rm Im}$ increases. 
Overall, while a CP-violating coupling of the SM-like Higgs boson
to top quarks can enlarge the allowed departures from the alignment limit, 
compatibility with eEDM and BSM Higgs-boson searches would
require to adjust the additional parameters to make the parameter
space with $\xi_u^{\rm Im} \gtrsim 0.1$ be phenomenologically viable.
For instance, one can modify the mixing angles $\alpha_i$ or
one of the other alignment parameters $\xi_f^{\rm Im}$
to supress the eEDM, or
raise the BSM scalar masses to alleviate the tensions
with LHC searches.

Turning to the second plot in the middle of \cref{fig:BPalphat},
where $\xi_d^{\rm Im}$ is varied, we observe that increasing the 
CP-violating component in the down-type Yukawa sector leads
overall to a reduction of
the allowed range of 
$\sin(\beta-\alpha_1)$ and the 
parameter space that is consistent with LHC Higgs data.
This is a consequence of the fact that
a CP-odd component in the $H_1 b\bar b$ coupling enhances the
partial width for decays of $H_1$ into 
bottom quarks, thereby increasing its total width. As a result, 
the branching ratios for decays into 
detected 
final states 
(most notably $H_1 \to \gamma\gamma$ and $H_1 \to Z Z^*$)
are suppressed, 
which tightens the constraints from Higgs-boson signal-strength measurements.
In this plane the dominant contributions to the eEDM
are proportional to the bottom-quark Yukawa coupling and therefore
strongly suppressed compared 
to the case discussed above in which the top-quark Yukawa coupling
is CP-violating. Accordingly, the blue dashed curve,
which indicates where the JILA bound would 
be saturated, lies entirely within the dark-shaded region already
excluded at the $3\sigma$  level by LHC Higgs data. 
Finally, cross-section limits from searches for additional Higgs
bosons do not impose 
constraints in this parameter plane, and therefore no orange
line is visible in the plot.
Increasing $\xi_d^{\rm Im}$ enhances the 
decays of the heavy scalars into $b\bar b$, but the corresponding
LHC searches are  significantly less sensitive due to
the overwhelming QCD background. At the same time, the 
branching ratios into cleaner final states such as
$\gamma\gamma$ and $\tau^+\tau^-$ are 
suppressed, weakening the impact of these searches.
In summary, in the considered parameter plane the LHC Higgs-boson
cross-section measurements provide the dominant sensitivity
to CP-violation in the down-type quark sector.

The right plot of \cref{fig:BPalphat}
shows the case where $\xi_\ell^{\rm Im}$ is varied,
corresponding to CP-violating couplings of the Higgs bosons
to charged leptons. Here, the
range of $\sin(\beta-\alpha_1)$ allowed by the LHC Higgs data
is less strongly affected by varying the imaginary part
of the flavour alignment parameter
compared to the two parameter planes
discussed above.
For moderate values of $\xi_\ell^{\rm Im} \sim 1$
the upper limit of the allowed region of $\sin(\beta-\alpha_1)$
increases slightly, while for values around
$\xi_\ell^{\rm Im}\sim 2$ the lower limit moves to somewhat 
smaller values. This behaviour is mainly driven by
modifications of the $H_1 \to \tau^+ \tau^-$ 
decay mode, while the production cross sections are essentially
unchanged because quark couplings
remain unaffected. As a result, the overall dependence
of the fit to the LHC Higgs data on
$\xi_\ell^{\rm Im}$ is less 
pronounced than in the cases of $\xi_u^{\rm Im}$ and $\xi_d^{\rm Im}$.
The eEDM provides an extremely strong constraint in this
parameter plane. 
Since $\xi_\ell^{\rm Im}$ directly induces CP-violating
interactions of $H_i$ with electrons, 
the predicted eEDM exceeds the experimental bound across
almost the entire parameter space. 
The only exception is a finely tuned region around
$\sin(\beta-\alpha_1)\sim 0.25$, where 
accidental cancellations between different two-loop contributions
lead to a vanishing eEDM. 
Whether this narrow window is physically viable, however,
would require taking into account 
additional contributions, such as the so-called ``kite''
diagrams at two-loop level, as well as potential higher-loop effects. 
Furthermore, searches for additional Higgs bosons decaying into
$\tau^+\tau^-$ final states~\cite{CMS:2022goy}
provide strong constraints in this plane, as shown with the
orange line. With increasing $\xi_\ell^{\rm Im}$ 
the branching ratios for the decays of the heavy Higgs states into
$\tau^+\tau^-$ are enhanced, and current 
limits exclude values of $\xi_\ell^{\rm Im}\gtrsim 0.4$. 
In summary, the parameter space with sizeable CP-violating
lepton Yukawa couplings is effectively 
excluded in this parameter plane by a combination of eEDM
limits and LHC searches for additional Higgs bosons.
As in  the case of $\xi_u^{\rm Im}$ (see left plot
of \cref{fig:BPalphat}), making the
parameter regions shown in white, which are
in agreement with the LHC Higgs data,
physically viable
would require adjusting other parameters of the model
to suppress the eEDM and 
avoid the strong bounds from direct searches at the LHC.

\subsection{\TB{Benchmark Point with a Strong
First-Order Electroweak Phase Transition}}

\TB{
In order to demonstrate explicitly that the
cS2HDM can accommodate all the ingredients required
for electroweak baryogenesis, while remaining consistent
with current experimental constraints
and predicting a valid DM candidate,
we now discuss a representative benchmark scenario.
Rather than performing a dedicated finite-temperature
analysis of the electroweak phase transition
(which is beyond the scope of this work and
left for future studies) we adopt
as a starting point a benchmark point from the
literature in the context of the complex
2HDM~(C2HDM) that has been shown to feature a strong
first-order electroweak phase transition and
CP-violation in the Higgs sector, leading to
a baryon asymmetry of the correct order
of magnitude~\cite{Biekotter:2025fjx}.
We emphasize that the C2HDM benchmark point
adopted here represents only one possible
realization of a strong first-order electroweak
phase transition. A variety of different
Higgs-boson mass hierarchies and parameter
configurations have been shown to give rise to
such a transition,
see e.g.~Refs.~\cite{Basler:2017uxn,Basler:2021kgq,
Goncalves:2023svb,Biekotter:2025fjx,Barni:2025ifb}
for recent studies.}

\TB{
A generic feature of the C2HDM is that CP violation
originates from a single independent source in
the scalar potential, which simultaneously controls
the CP-violating effects relevant for electroweak
baryogenesis and induces contributions to
EDMs. As a consequence,
parameter regions that yield a sufficiently large baryon
asymmetry are typically in tension with the stringent
experimental bounds on the eEDM.
This is also the case for the benchmark point adopted
here, which predicts an electron EDM exceeding
the current experimental limit by approximately one
order of magnitude~\cite{Biekotter:2025fjx}.
In the cS2HDM, however, additional and independent
sources of CP violation are present that allow
this tension to be alleviated.
In particular, we show that by introducing a
complex lepton flavour-alignment parameter $\xi_\ell$,
the predicted eEDM can be suppressed
below the experimental bounds while keeping the
Higgs-sector CP violation and the
properties of the electroweak phase transition unchanged.
}

\TB{
The modification of the lepton Yukawa sector
compared to the C2HDM is expected to have negligible
impact on the generation of the baryon asymmetry,
because the dominant contributions to electroweak baryogenesis arise from transport processes
involving the top quark due to its
large Yukawa coupling.
To obtain the correct DM relic abundance, a small
mixing is introduced between $H_1$, 
the SM-like Higgs boson at 125~GeV,
and the singlet state $H_3$ whose mass
is chosen to be $m_{H_3} = 200$~GeV, 
while the DM mass is set to $m_\chi = 80$~GeV. 
The small mixing between $H_1$ and $H_3$,
introduced by setting $\alpha_2 = 0.02$,
generates $H_{1,3}\chi\chi$ couplings of
the appropriate size to predict the correct
DM relic abundance.
We verified that the mixing is sufficiently small
to not be in tension with LHC Higgs boson
cross section measurements and LHC searches
for additional Higgs bosons.
Moreover, $\alpha_2$ is small enough to
induce only negligible modifications to the
scalar potential, thus preserving the strength
of the electroweak phase transition as it
was found in the C2HDM.
}

\begin{table}
\centering
\begin{tabular}{ll}
\hline\hline
\multicolumn{2}{c}{\textbf{Input parameters}} \\
\hline
$m_{H_1}$ & $125.09$ \\
$m_{H_2}$ & $420$ \\
$m_{H_3}$ & $200$ \\
$m_{H_4}$ & $660$ \\
$m_{H^\pm}$ & $660$ \\
$m_\chi$ & $80$ \\
$\tan\beta$ & $2.960654$ \\
$\alpha_1$ & $\beta$ \\
$\alpha_2$ & $0.02$ \\
$\alpha_3$ & $0.0$ \\
$\alpha_4$ & $0.0525$ \\
$\alpha_5$ & $0.0$ \\
$\alpha_6$ & $-0.024708$ \\
$M$ & $m_{H_2}$ \\
$v_S$ & $100$ \\
$\lambda_5^{\rm Im}$ & $0.563345$ \\
$\lambda_{6,7,11}^{\rm Re}$ & $0.0$ \\
$\xi_{u,d,\ell}^{\rm Re}$ & $0.0$ \\
$\xi_{u,d}^{\rm Im}$ & $0.0$ \\
$\xi_\ell^{\rm Im}$ & $[-2,\,1]$ \\
\hline
\hline
\multicolumn{2}{c}{\textbf{Predictions}} \\
\hline
$\xi_n$ (transition strength) & $2.4$ \\
$\eta_s$ (baryon asymmetry) & $9.5 \cdot 10^{-11}$\\
$\Omega_\chi h^2$ (DM relic abundance) & $0.11$ \\
$\Delta \chi^2$ (\texttt{HiggsSignals}) & $<3.14$ \\
\hline
\hline
\multicolumn{2}{c}{\textbf{Main branching ratios of BSM Higgs bosons}} \\
\hline
$H_2 \to t\bar t$ & $98\%$ \\
$H_2 \to H_1 Z$ & $1.3\%$ \\
$H_3 \to \chi\chi$ & $100\%$ \\
$H_4 \to t\bar t$ & $18\%$ \\
$H_4 \to H_2 Z$ & $81\%$ \\
$H^\pm \to tb$ & $17\%$ \\
$H^\pm \to H_2 W^\pm$ & $83\%$ \\
\hline\hline
\end{tabular}
\caption{Benchmark scenario for the cS2HDM
adapted from a C2HDM parameter point reproducing
the observed baryon-to-entropy ratio of the
universe $\eta_s$.
The top section shows the input parameters
of the cS2HDM, the middle section a few
model predictions for
the electroweak phase transition, DM, and Higgs phenomenology, and the bottom section lists the dominant branching ratios of the BSM Higgs bosons.
Masses are in GeV.}
\label{tab:benchmark_full}
\end{table}

\TB{
In \cref{tab:benchmark_full} we state the values of
the free parameters used for this benchmark scenario.
The masses of $H_2$ and $H_4$ are taken over from
the C2HDM parameter point, as well as the values
of the mass parameter $M$, $\tan\beta$
and $\lambda_5^{\rm Im}$.
The mass difference between $H_2$ and $H_4$ is responsible
for generating sizable quartic scalar couplings
that facilitate an electroweak phase transition
with a strength of $\xi_n = 2.4$.\footnote{\TB{The
strength of the transition is defined as the
ratio between the electroweak symmetry breaking
vacuum expectation value and the transition
temperature. To allow for electroweak baryogenesis,
the value of $\xi_n$ should be larger than one.}}
The values of the mixing angles $\alpha_1$,
$\alpha_4$ and $\alpha_6$ are set to coincide
with the corresponding values in
the C2HDM. The values of $\alpha_2 = 0.02$ and
$v_S = 100$~GeV are chosen
to obtain the correct DM relic abundance,
and the remaining mixing
angles $\alpha_3$ and $\alpha_5$ are set
to zero.\footnote{\TB{To match a C2HDM parameter point
to a cS2HDM parameter point, and using the most
common
notation for the mixing angles in the C2HDM,
the cS2HDM mixing angles should be chosen
according to: $\alpha_1 = \alpha_1^{\rm C2HDM}$,
$\alpha_2 = 0$, $\alpha_3 = 0$, $\alpha_4 =
-\alpha_{3}^{\rm C2HDM}$, $\alpha_5 = 0$ and
$\alpha_6 = -\alpha_{2}^{\rm C2HDM}$.}}
Finally, the flavour alignment parameters are set to
$\xi_{u,d,\ell} = 0$ to resemble the Yukawa type~I,
except for the imaginary part of $\xi_\ell$, which
was varied in the range $-2 \leq \xi_\ell^{\rm Im}
\leq 1$.
We verified that this benchmark scenario satisfies
all theoretical and experimental constraints
discussed in \cref{sec:constraints}.}

\begin{figure}
    \centering
    \includegraphics[width=0.6\linewidth]{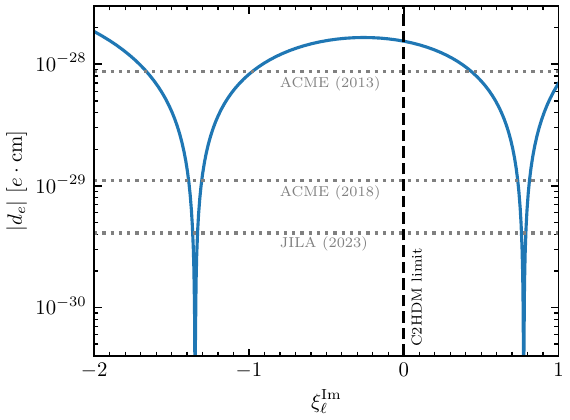}
    \caption{\TB{Predicted values of the eEDM
    as a function of $\xi_\ell^{\rm Im}$. The horizontal black lines represent the experimental upper limits from the ACME~\cite{ACME:2013pal,ACME:2018yjb} and JILA~\cite{Roussy:2022cmp} collaborations.
    The black dashed vertical line indicates
    the limit $\xi_\ell^{\rm Im}$ in which the
    benchmark scenario resembles the C2HDM
    parameter point (see main text for details).}}
    \label{fig:plot_EDM_bau}
\end{figure}

\TB{In Ref.~\cite{Biekotter:2025fjx}, the prediction
for the baryon-to-entropy
ratio $\eta_s$ for the C2HDM parameter point was
estimated to be $\eta_s = 9.5 \cdot 10^{-11}$,
which roughly agrees with the observed
value $\eta_s^{\rm obs} = 8.7 \cdot 10^{-11}$, thus
(considering the sizable theory uncertainties
in the prediction of the baryon asymmetry
generated during the transition)
successfully realizing electroweak
baryogenesis.\footnote{\TB{Predictions for the resulting
baryon asymmetry are subject to various different
sources of theoretical uncertainties from
the modeling of the expanding bubbles of true
vacuum in the early universe and their
interactions with the thermal plasma.
Depending on the properties of the phase transition,
the relative uncertainty on the prediction for $\eta_s$
can be of an order of magnitude or more,
see e.g.~Refs.~\cite{Dorsch:2021ubz,
vandeVis:2025efm,Barni:2025ifb}.}}
As discussed above, by appropriately choosing
the parameters of the cS2HDM,
the prediction for $\eta_s$ can
be taken over here since $\xi_\ell^{\rm Im}$ does
not have an important impact.
The eEDM in the C2HDM was found to be
$|d_e| = 1.5 \cdot 10^{-28}\,e$~cm,
which is about
two orders of magnitude above the current
experimental limits.
Hence, the non-observation of the eEDM precludes
electroweak baryogenesis in this parameter
point of the C2HDM.
However, as can be seen in \cref{fig:plot_EDM_bau},
in the cS2HDM we can use non-zero values of
$\xi_\ell^{\rm Im}$ in order to suppress the eEDM,
making it compatible with the experimental limits
for $\xi_\ell^{\rm Im} \approx -1.45$ and
$\xi_\ell^{\rm Im} \approx 0.78$.
As a result, we demonstrate here with
a concrete parameter point that the cS2HDM
is capable of predicting the correct baryon asymmetry
of the universe via electroweak baryogenesis with
an eEDM that falls well below the current experimental
limits. At the same time, the DM relic abundance is
predicted in agreement with the measured value
via standard thermal freeze-out of Higgs-portal
DM while satisfying the constraints from
DM direct detection experiments.
However, we stress again that in the blind spots
in which the Barr-Zee-type contributions to
the eEDM are strongly suppressed by means of
large accidental cancellations,
additional contributions to the eEDM (not taken
into account here) become relevant and might
be of the order of the current experimental bounds.
}

\TB{
This benchmark scenario can be
tested at the LHC, for instance via the detection
of the heavier Higgs bosons. 
In particular, decays of a heavier scalar into
a lighter Higgs boson plus a gauge boson, 
such as $H_4 \to H_2 Z$ or $H^\pm \to H_2 W^\pm$,
have sizable branching ratios and provide
promising search channels. 
Decays into pairs of DM particles, $\chi\chi$,
would give signatures with missing transverse
energy and could provide a mechanism to distinguish
the cS2HDM from the C2HDM or more general 2HDMs. 
However, for this particular benchmark point,
chosen deliberately close to the C2HDM,
such signatures are not promising: 
The singlet-like state $H_3$ decays exclusively into
$\chi\chi$ but has a very small direct production
cross section due to its singlet nature,
and it is also not produced in decays of
the heavier states. At the same time,
the branching ratios of the doublet-like
BSM scalars $H_2$ and $H_4$ into $\chi\chi$
are negligible.
To make the signal with missing energy more promising,
one would have to go to parameter space regions
of the cS2HDM that depart from C2HDM-like configurations,
where the decays into $H_i \to \chi\chi$ can
become relevant. 
To assess the potential for electroweak baryogenesis
in such scenarios, however, a
dedicated computation of the dynamics of the
electroweak phase transition
and the resulting baryon asymmetry in the cS2HDM
is required, as the results cannot be simply
inferred from the C2HDM.
We leave such an analysis for future work.
}

\section{Conclusions and Outlook}
\label{sec:conclu}

In this paper, we have introduced the CP-violating
singlet-extended 2-Higgs-Doublet Model~(cS2HDM) as a unified framework
that can simultaenously address two important open questions
in particle physics: the unknown nature of Dark Matter~(DM) and
the unknown origin of the baryon asymmetry of the universe~(BAU).
In order to predict a viable Higgs-portal DM
candidate,
our model extends the Standard Model~(SM) by incorporating
a complex scalar singlet field that is charged under a global
$U(1)$ symmetry. The imaginary component of the singlet field
acts as a pseudo-Nambu-Golstdone~(pNG)
DM particle which is naturally faint
to DM direct detection experiments, while the real component
gives rise to an additional Higgs boson contributing to
the portal interactions between the visible and the dark sector.
Additionally, we augment the Higgs sector with a second
$SU(2)$ Higgs doublet, whose presence allows for additional sources
of CP-violation in the Higgs sector, and the additional scalar
interactions can facilitate a strong first-order electroweak~(EW)
phase transition.
In total, the Higgs sector of the cS2HDM predicts
four neutral Higgs bosons, a pair of charged
Higgs bosons, and a spin-0 DM state that interacts
with the fermions exclusively via the exchange
of Higgs bosons.
Instead of imposing natural flavour conservation via
a discrete $\mathbb{Z}_2$ symmetry
under which the two Higgs doublets have opposite parity,
we make use of flavour alignment in the Yukawa sector,
which suppresses flavour-changing neutral currents
while permitting additional sources of CP-violation in the
Yukawa sector that would otherwise be forbidden.
Both the new sources of CP-violation and
the presence of a strong EW phase transition are
necessary ingredients to explain the BAU via EW baryogenesis.

To put the cS2HDM into context, we briefly compare its
features with those of other popular Higgs-portal DM models.
The SM extended only by a singlet scalar can in principle
provide both a DM candidate and a strong first-order
EW phase transition, but it lacks additional CP-violating sources
required for EW baryogenesis, and the predicted DM state
is in strong tension with direct detection bounds except
in the narrow Higgs-funnel region or for heavy DM masses
in the multi-TeV range, making it an unappealing target for
DM searches at the LHC.
The 2-Higgs-Doublet Model (2HDM) offers complementary features: 
in its variant in which the additional Higgs doublet is inert
(inert 2HDM), the model contains a stable DM candidate and
can support a strong first-order EW phase transition. However, 
CP-violation in the Higgs sector is forbidden and the parameter
space predicting the observed DM relic abundance is largely
excluded by direct detection experiments.
On the other hand, in the generic 2HDM in which both
doublets take part in EW symmetry breaking,
CP-violation is possible and EW baryogenesis can be realized,
but no stable DM candidate is present.
The 2HDM+a, which extends the 2HDM by a gauge singlet pseudoscalar
field and a DM particle (typically a Dirac fermion),
and which often serves as benchmark model for LHC DM searches,
predicts a viable DM candidate with suppressed scattering rates
at direct-detection experiments and can accommodate a strong first-order 
EW phase transition. However, its attractive cancellation 
mechaism suppressing
spin-independent DM-nulceon scattering is in general spoiled
once explicit CP-violation is introduced in the Higgs sector.
By contrast, the cS2HDM retains the natural suppression of direct-detection
rates associated with the pNG DM state under the presence
of CP-violation in the Higgs sector.
As a result, the cS2HDM can realize the
standard thermal freeze-out picture of Higgs-portal
DM, and it simultaneously provides a Higgs sector capable of supporting
EW baryogenesis. Moreover, it offers several independent sources
of CP-violation that can be arranged to evade the stringent
experimental limits from the non-observation of the electron
electric dipole moment~(eEDM), in particular through cancellations
between CP-violating sources in the scalar and the Yukawa sector.
As a result, 
the cS2HDM stands out as a particularly attractive benchmark
model for LHC DM searches.

After defining the model, we confronted it with a
broad set of theoretical and experimental constraints.
On the theory side, we imposed vacuum stability
and perturbative unitarity bounds on the parameters
of the scalar potential. On the phenomenological side,
we computed predictions for Higgs-boson branching ratios and
production cross sections at the LHC, enabling us to study
the collider phenomenology of the model at the LHC.
We calculated the DM relic abundance assuming thermal
freeze-out and evaluated one-loop induced DM-nucleon scattering
rates to assess the impact of direct detection experiments. We further investigated the consequences of EW precision observables
by calculating at the one-loop level
the EW $\rho$-parameter that is sensitive to
mass splittings of different isospoin component of scalars
due to the breaking of the custodial symmetry.
Finally, we studied the predicted eEDM, which provides the most stringent probe of new
CP-violation in the Higgs sector through recent upper bounds from ACME and JILA.

A novel aspect of this work concerns the treatment of DM
direct detection in the presence of CP-violating portal
interactions, which in the cS2HDM can be present both
in the coupling of the DM state $\chi$ with the Higgs
bosons $H_i$, as well as in the Yukawa couplings of the Higgs bosons
$H_i$ with the quarks $q$.
By including these CP-violating interactions,
we performed a matching of the CP-even and CP-odd
operators describing the $\chi q \to \chi q$ scattering
in the UV theory onto the non-relativistic effective
field theory operators that describe the DM-nucleon
scattering at low energies.
A consequence of the CP-violating interactions is
that the matching not only generates the usual
spin-independent interactions but also spin-dependent
and momentum-dependent operators.
By studying these contributions numerically,
we demonstrated that the dominant effect of the new
CP-odd components of the portal couplings is a
quantitative modification of the predicted spin-independent
scattering cross sections, while the induced spin-dependent
rates are far too small to be experimentally accessible.

Combining these ingredients, we performed a random parameter scan
in which all theoretical, LHC, DM, eEDM,  and precision constraints were
imposed, allowing us to identify viable benchmark points.
We analyzed in detail the severity of eEDM limits,
and how they are related to DM direct detection bounds.
We payed special attention to the impact of the
new sources of CP-violation contained in the cS2HDM.
In addition, we investigated in examplary parameter
planes the sensitivity of current LHC Higgs data and
direct searches for  additional Higgs bosons to CP-violating couplings
between the Higgs bosons and fermions, highlighting their
complementarity  with the stringent eEDM bounds.

With this paper, we provide a public software tool that
incorporates all the ingredients discussed above and
enables future studies of the cS2HDM.
Our framework combines a modified version of \texttt{HDECAY}
written in Fortran to calculate scalar branching ratios,
an interface to
the C++ package \texttt{HiggsTools} to obtain cross sections from
effective couplings and confront the model with LHC Higgs data and limits from BSM Higgs boson searches
at LEP and the LHC, as well as 
\texttt{micrOMEGAs} to compute the DM relic abundance.
All other calculations discussed in our paper,
including the one-loop predictions for DM scattering
and the matching onto the non-relativistic effective field
theory describing the interactions between $\chi$ and nucleons,
the EW $\rho$-parameter, and the electron EDM at the two-loop
level considering Barr-Zee-type contributions,
are implemented directly in our tool.
Our goal is that the cS2HDM can serve as a benchmark
model both for LHC DM searches and for scenarios motivated
by EW baryogenesis, for which it is essential to provide the necessary tools, which we make available with this publication
under the following link:
\begin{center}
  \href{https://gitlab.com/thomas.biekoetter/cs2hdmtools}
    {https://gitlab.com/thomas.biekoetter/cs2hdmtools}
\end{center}

As an outlook, we plan to extend this work by
performing global scans of the high-dimensional
parameter space of the cS2HDM using more sophisticated
sampling techniques beyond random scanning,
e.g.~applying genetic algorithms using the
public Fortran package
\texttt{evortran}~\cite{Biekotter:2025gkp}.
This will allow us to identify benchmark scenarios that
are suitable to
simultaneously realize EW baryogenesis and predict the
observed DM relic abundance.
Such studies will include a detailed analysis of
the thermal history to determine regions with a
sufficiently strong first-order EW phase transition
by implementing the model into the public
program \texttt{BSMPT}~\cite{Basler:2018cwe,Basler:2020nrq,Basler:2024aaf},
as well as an exploration of
the characteristic collider signatures of such
scenarios model at the LHC.

\section*{Acknowledgements}

RS and PG are supported by FCT under Contracts: UID/00618/2025 and 2024.03328.CERN (\url{https://doi.org/10.54499/2024.03328.CERN}).
PG is also supported by FCT under the contract 2022.11377.BD (\url{https://doi.org/10.54499/2022.11377.BD}).
MM acknowledges support by the Deutsche Forschungsgemeinschaft (DFG, German Research Foundation) under grant 396021762 - TRR 257.
TB acknowledges the support of the Spanish Agencia
Estatal de Investigaci\'on through the grant
``IFT Centro de Excelencia Severo Ochoa CEX2020-001007-S''.
The project that gave rise to these
results received the support of a
fellowship from the ``la Caixa''
Foundation (ID 100010434). The
fellowship code is LCF/BQ/PI24/12040018.

\bibliographystyle{JHEP}
\bibliography{lit}

\end{document}